\documentclass[a4paper,11pt]{book}
\usepackage{titlesec}
\usepackage{siunitx}
\usepackage[version=4]{mhchem}
\usepackage{amsmath}
\usepackage{amssymb}

\usepackage{import}

\usepackage[pdftex]{graphicx}
\usepackage{multirow}
\usepackage[margin=1.25in]{geometry}
\usepackage{scalerel}

\usepackage{url}
\usepackage{upgreek}
\usepackage{subfigure}
\usepackage{lipsum}
\usepackage[colorlinks = true,
            linkcolor = blue,
            urlcolor  = blue,
            citecolor = blue,
            anchorcolor = blue]{hyperref}
\usepackage{xcolor}
\usepackage{subfiles} % Best loaded last in the preamble

\def\Title#1{\begin{center} {\LARGE #1 } \end{center}}

\newcommand\snowmass{\begin{center}\rule[-0.2in]{\hsize}{0.01in}\\\rule{\hsize}{0.01in}\\
\vskip 0.1in Submitted to the  Proceedings of the US Community Study\\ 
on the Future of Particle Physics (Snowmass 2021)\\ 
\rule{\hsize}{0.01in}\\\rule[+0.2in]{\hsize}{0.01in} \end{center}}

\usepackage[utf8]{inputenc}

\usepackage[noblocks]{authblk}
\makeatletter

\renewcommand{\maketitle}{
 {\noindent\@author\par}
}
\makeatother

\begin{document}
\Title{\Huge{MPGDs for tracking and Muon detection at future high energy physics colliders}}
%\vspace{1mm}
\Title{Snowmass Instrumentation Frontier: MPGD White paper 5}

\author[1]{K.~Black~(Coordinator)}
\author[2]{A.~Colaleo~(Coordinator)}
\author[3]{C.~Aimè}
\author[4]{M.~Alviggi}
\author[2]{C.~Aruta}
\author[5]{M.~Bianco}
\author[6]{I.~Balossino}
\author[7]{G.~Bencivenni}
\author[7]{M.~Bertani}
\author[8]{A.~Braghieri}
\author[9]{V.~Cafaro}
\author[8]{S.~Calzaferri}
\author[10,5]{M.~T.~Camerlingo}
\author[4]{V.~Canale}
\author[6]{G.~Cibinetto}
\author[5]{M.~Corbetta}
\author[10]{V.~D'amico}
\author[7]{E.~De~Lucia}
\author[4]{M.~Della~Pietra}
\author[4]{C.~Di~Donato}
\author[10]{R.~Di~Nardo}
\author[7]{D.~Domenici}
\author[2]{F.~Errico}
\author[1]{P.~Everaerts}
\author[11]{F.~Fallavollita}
\author[6]{R.~Farinelli}
\author[7]{G.~Felici}
\author[8]{D.~Fiorina}
\author[6]{I.~Garzia}
\author[7]{M.~Gatta}
\author[9]{P.~Giacomelli}
\author[7]{M.~Giovannetti}  
\author[6]{S.~Gramigna}
\author[5]{R.~Guida}
\author[12]{M.~Hohlmann}
\author[13,5]{P.~Iengo}
\author[14]{M.~Iodice}
\author[15]{L.~Lavezzi}
\author[16]{M.~Maggi}
\author[5]{B.~Mandelli}
\author[6]{M.~Melchiorri}
\author[17]{J.~A.~Merlin}
\author[6]{G.~Mezzadri}
\author[3]{P.~Montagna}
\author[7]{G.~Morello}
\author[7]{G.~Papalino}
\author[2]{A.~Pellecchia}
\author[10]{F.~Petrucci}
\author[7]{M.~Poli~Lener}
\author[2]{R.~Radogna}
\author[3]{C.~Riccardi}
\author[5]{M.~G.~Rigoletti}
\author[8]{P.~Salvini}
\author[6]{M.~Scodeggio}
\author[13]{G.~Sekhniaidze}
\author[10]{M.~Sessa}
\author[16]{F.M.~Simone}
\author[5]{A.~Sharma}
\author[2]{A.~Stamerra}
\author[8]{I.~Vai}
\author[2]{R.~Venditti}
\author[16]{P.~Verwilligen}
\author[3]{P.~Vitulo}
\author[2]{A.~Zaza}

\affil[1]{University of Wisconsin-Madison, Madison, US}
\affil[2]{University and INFN Sez. Bari, Bari, IT}
\affil[3]{University and INFN Sez. Pavia, Pavia, IT}
\affil[4]{University and INFN Sez. Napoli, Naples, IT}
\affil[5]{CERN, Geneva, CH}
\affil[6]{INFN Sez. Ferrara, Ferrara, IT}
\affil[7]{Laboratori Nazionali di Frascati - INFN, Frascati (RM), IT}
\affil[8]{INFN Sez. Pavia, Pavia, IT}
\affil[9]{INFN Sez. Bologna, Bologna, IT}
\affil[10]{University and INFN Sez. Roma Tre, Rome, IT}
\affil[11]{Institute of Physics, Johannes Gutenberg University Mainz, GE}
\affil[12]{Florida Institute of Technology, Melbourne, US}
\affil[13]{INFN Sez. Napoli, Naples, IT}
\affil[14]{INFN Sez. Roma Tre, Rome, IT}
\affil[15]{INFN Sez. Torino, Torino, IT}
\affil[16]{INFN Sez. Bari, Bari, IT}
\affil[17]{University of Seoul, Seoul, KR}

\maketitle

{\it \noindent  E-mail:
    \href{mailto:anna.colaleo@uniba.it}{anna.colaleo@uniba.it},
    \href{mailto:paolo.iengo@cern.ch}{paolo.iengo@cern.ch},
    \href{mailto:jeremie.alexandre.merlin@cern.ch}{jeremie.alexandre.merlin@cern.ch},\\
    \href{mailto:antonello.pellecchia@ba.infn.it}{antonello.pellecchia@ba.infn.it},
    \href{mailto:giovanni.bencivenni@lnf.infn.it}{giovanni.bencivenni@lnf.infn.it},
    \href{mailto:beatrice.mandelli@cern.ch}{beatrice.mandelli@cern.ch}
    }
\snowmass

\begingroup
\let\cleardoublepage\clearpage
\tableofcontents
\endgroup

\begingroup
\let\cleardoublepage\clearpage
\endgroup

\let\cleardoublepage=\clearpage

\titleformat{\chapter}[display]
  {\normalfont\bfseries}{}{0pt}{\Huge}
  
%\addtocontents{toc}{\protect\setcounter{tocdepth}{1}}

%\bigskip 

%\medskip
\label{chap:intro}

%\medskip

%\Address{}

%\medskip

 %\begin{Abstract}
%\noindent 
%\en%d{Abstract}

%\snowmass

\def\thefootnote{\fnsymbol{footnote}}
\setcounter{footnote}{0}
\chapter*{Executive summary}
\addcontentsline{toc}{chapter}{Executive summary}
\par Gaseous Detectors (GDs) are the primary choice for cost effective instrumentation of large areas, with high detection efficiency in a high background and hostile radiation environment, needed for muon triggering and tracking at future facilities.
They can provide a precise standalone momentum measurement or be combined with inner detector tracks resulting in even greater precision. Adding precise timing information ($\mathcal{O}$(ns)) allows control of uncorrelated background, mitigates pile-up and allows detection of extremely long lived particles that behave like slow muons propagating through the  detector volume over a time as long as a few bunch crossings.

\par In the last decades, GDs (in particular wire chambers and Resistive Plate Chambers (RPC)) have proven to be a versatile and cost effective technology for large volume muon spectrometers.  Their role in particle physics experiments remains central, as testified by their use in the trigger and muon systems of all major LHC experiments (ALICE~\cite{Alice_gen}, ATLAS~\cite{ATLAS_gen}, CMS~\cite{CMS_gen}, LHCb~\cite{LHCB_gen}), largely based on those GD technologies. However their performance can be heavily affected by the presence of the intense background and the operation in such conditions could cause detector aging that would compromise long term performance stability. With the invention and evolution of \textbf{Micro Pattern Gaseous Detectors (MPGDs)}~\cite{MPGD} during the last twenty years, GDs improved significantly in spatial resolution and rate capability. 
MPGDs, based on modern photo-lithographic technologies, allow stable operation at very high background particle flux with high efficiency and excellent spatial resolution. These features determine the main applications of these detectors in particle physics experiments as precise muon tracking in high radiation environment as well as muon tagger and trigger in general purpose detectors at HEP colliders. Among the most prominent MPGD technologies, the Gas Electron Multiplier (GEM~\cite{GEM}) and MicroMegas (MM~\cite{MM}) have been successfully operated in many different experiments, such as Compass~\cite{Compass}, LHCb~\cite{LHCb}, TOTEM~\cite{Totem}.
In addition, the low material budget and the flexibility of the base material makes MPGD suitable for the development of very light, full cylindrical fine tracking inner trackers at lepton colliders such as KLOE-2~\cite{Kloe2} at DAFNE (Frascati, IT) and BESIII~\cite{CGEM-BES3} at BEPCII (Beijing, CN).

A big step in the direction of large-size applications has been obtained both with conceptual consolidation and industrial and cost-effective manufacturing of MPGDs by developing new fabrication techniques: resistive Micromegas~\cite{res-MM} (to suppress destructive sparks in hadron environments) and single-mask and self-stretching GEM techniques~\cite{GEMTDR} (to enable  production  of  large-size  foils  and  significantly reduce detector  assembly time). Scaling up of MPGDs to very large single unit detectors of $\mathcal{O}$(m$^2$), has facilitated their use in Muon systems in the LHC upgrades. 
Major developments in the MPGD technology have been introduced for the ATLAS and CMS muon system upgrades, towards establishing technology goals, and addressing engineering and integration challenges. Micromegas and GEM have been recently installed  in the ATLAS New Small Wheel~\cite{ATLAS-MM}, CMS GE1/1 station respectively~\cite{CMS-GEM}, for operation from Run 3 onward, as precise tracking systems. Those radiation hard detectors, able to cope with the expected increased particle rates, exhibit good spatial resolution (O(100 $\mu$m)) and have a time resolution of 5–10 ns. 
%On the algorithm side, 
In the CMS muon system additional stations, GE2/1 and ME0  ~\cite{butalla2021investigation, muon_ph2_tdr}, based on GEMs with high granularity and spatial segmentation, will be installed to ensure efficient matching of muon stubs to offline pixel tracks at large pseudo-rapidities during HL-LHC operation. 
Several solutions 
($\mu$-RWELL~\cite{uRwell}, Micro Pixel Chamber ($\mu$-PIC)~\cite{Yamane_nima951} and small-pad resistive Micromegas~\cite{Paddy}) were also considered for the very forward muon tagger in the ATLAS Phase-II Upgrade Muon TDR proposal~\cite{ATLAS-CERN-LHCC-2017-017}. Here, the main challenges are discharge protection and miniaturization of readout elements~\cite{PaddyAll}, which can profit from the ongoing developments on Diamond-Like Carbon (DLC) technology~\cite{DLC}. The $\mu$-RWELL is the baseline option for the Phase-II Upgrade of the innermost regions of the Muon System of the LHCb experiment (beyond LHC LS4)~\cite{HR-layouts}.

The new era of Particle Physics experiments is moving towards the upgrade of present accelerators and the design of new facilities operating at extremely high intensities and particle energies such as the \textbf{Future Circular Colliders}~\cite{Abada:2019lih} and the \textbf{Muon Collider}~\cite{muon_collider}.
Cost effective, high efficiency particle detection in a high background and high radiation environment is fundamental to accomplish their physics program.
Different critical aspects such as the high particle rates, discharge probabilities and accumulated doses expected at future colliders must be taken into account. Modifications or new detector configurations are to be investigated by relying  on innovative technological solutions.
Muon systems at future lepton colliders, \textbf{e$\mathrm{^{+}}$e$\mathrm{^{-}}$ colliders} (ILC\cite{ILC-ILD}, CLIC\cite{CLIC}, CepC~\cite{cepc:cdr-2018}, FCC-ee, SCTF~\cite{SCTF}) or \textbf{LHeC}~\cite{LHC-e}, do not pose significant challenges in terms of particle fluxes and the radiation environment. Therefore many existing MPGD technologies are suitable for building future large muon detection systems. For example the $\mu$RWELL technology is envisaged to be utilized for the muon detection system and the preshower detector of the IDEA detector concept~\cite{Abada:2019lih,IDEA:test-beam} that is proposed for the FCC-ee and CepC future large circular leptonic colliders. In addition $\mu$RWELL are candidates for the inner tracking system at future high luminosity tau-charm factories, STCF in Russia and SCTF in China.
Generally, background rates in LHeC muon detector, which are based on the updated design of ATLAS Phase-II Muon spectrometer, are lower than in $pp$ colliders. %CITATION NEEDED
On the other hand, the expected particle rates for the muon tracking and triggering at future \textbf{hadron colliders, such as the FCC-hh\cite{fcc-hh}}, make the existing technologies adequate in most regions of the spectrometers, but require a major R\&D  for the very forward endcap region.
In a \textbf{multi-TeV muon collider}, the effect of the background induced by the muon beam decays is extremely important, since it can contaminate the Interaction Region (IR) from a distance that varies with the beam energy, the collider optics and the superconducting magnets. Therefore, the rate of background is particularly relevant in the forward region. Tracking and triggering can be obtained with multi-layer structures, for an efficient local muon segment reconstruction. A new generation Fast Timing MPGD (FTM\cite{rui_ftm}, Picosec\cite{picosec}) is considered to mitigate the Beam Induced Background (BIB), thanks the rejection of hits uncorrelated in time.

MPGDs offer a diversity of technologies that allow them to meet the required performance challenges at future facilities and in various applications, thanks to the specific advantages that each technology provides. On-going R\&D should focus on pushing the detector performance to the limits of  each technology by overcoming the related technological challenges. 

\textbf{Common future R\&D} should focus on stable operation of large area coverage, including precision timing information to ensure the correct track-event association, and on the ability to cope with large particle fluxes, while guaranteeing detector longevity using environmentally friendly gas mixtures and optimized gas consumption (gas recirculating and recuperation system). 
Strong constraints on response stability, discharge probability and space charge accumulation require  innovative technological solutions and novel detector configurations. Considering the high rate exposure of the detectors and the radiation hazards at future colliders, very strong restrictions to access the detector for reparations and replacement are expected. In this scenario long term operation requirements have to be guarantee also in term of mechanical and electronics robustness. The main challenges include gas tightness, over-pressure operation and electronics cooling.
Integration aspects have also to be optimised for easy accessibility and replaceability in complex installations. The assembly of a large scale detector components will require engineering effort to ensure mechanical precision. 
MPGDs require dedicated front-end electronics (FEE) development, both discrete and integrated (ASIC), focused on specific applications, while meeting a large set of challenging requirements such as: fast timing, large input capacitance, low noise, input discharge protection, cross-talk reduction, pixel size, compactness, low power consumption and detector integration~\cite{electronics}.

%\pubblock

%\Title{High granularity resistive Micromegas for high rates}

\chapter{\centering High granularity resistive Micromegas for high rates}
\label{ch:2}

%\addcontentsline{toc}{chapter}{High granularity resistive Micromegas for high rates}
\bigskip 

%\Author{M. Alviggi$^1$, M.T. Camerlingo$^2$, V. Canale$^1$, V. D'amico$^3$, M. Della Petra$^1$, C. Di Donato$^1$, R. Di Nardo$^3$, P. Iengo$^{4*}$, M. Iodice$^5$, F. Petrucci$^3$, G. Sekhniaidze$^6$, M. Sessa$^3$}

\medskip

%{\it{\noindent$^1$University and INFN Sez. Napoli, Naples, IT;
%\newline $^2$University and INFN Sez. Roma Tre, Rome, IT and CERN, Geneva, CH;
%\newline $^3$University and INFN Sez. Roma Tre, Rome, IT;
%\newline $^4$INFN Sez. Napoli, Naples, IT, and CERN, Geneva, CH;
%\newline $^5$INFN Sez. Roma Tre, Rome, IT;
%\newline $^6$INFN Sez. Napoli, Naples, IT;}}

%\noindent $^*$Corresponding author: paolo.iengo@cern.ch

\medskip

\def\thefootnote{\fnsymbol{footnote}}
\setcounter{footnote}{0}

\section{Introduction}
\label{ch:2_sec:intro}
The aim of this project is the development of MPGDs based on the resistive Micromegas (MM) technology able to efficiently operate at particle rates up to 10~MHz/cm$^2$ and beyond.

Resistive Micromegas have been developed for the ATLAS Experiment~\cite{ATLAS-MM} in the last decade to greatly suppress the intensity of discharges to which the non-resistive Micromegas are prone.
Since then, many resistive MPGD have been proposed. In our research work, we are pushing forward the resistive Micromegas technology with the final goal to develop a detector offering precise tracking and rate capability hundred times higher than the Micromegas for ATLAS, and mechanically sound to be considered for large experiments.
Previous works on Micromegas detectors with pad-shaped readout can be found in~\cite{MaxChef}.

\subsection{Detector description}

The basic concept of our detector is a single amplification stage device with a resistive scheme able to guarantee at a time stable operation (spark suppression) and fast charge evacuation at high particle rate. 
The readout electrodes must be small enough (high granularity) to reduce the occupancy and provide precise tracking performance.

We have built and tested detectors with many different resistive schemes, their detailed description is reported in Section \ref{ch:2_sec:state} together with the main results.
Here we describe the main features common to all the detectors.

\begin{figure}
\begin{center}
\includegraphics[width=0.8\textwidth]{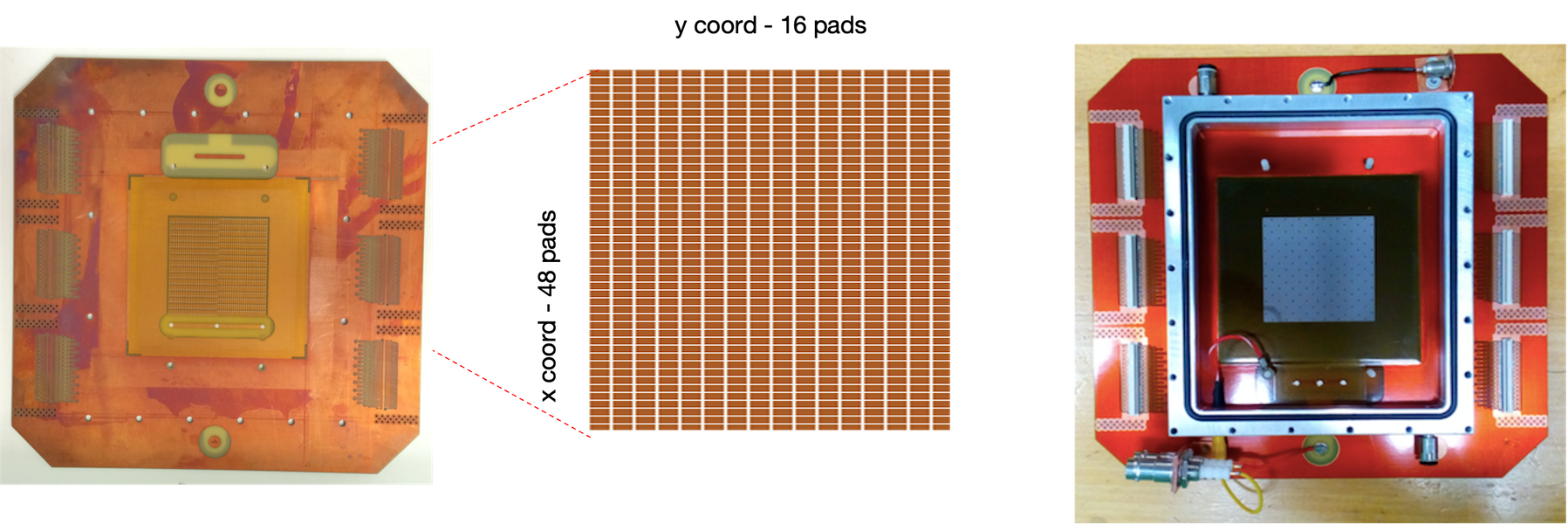}
\end{center}
\caption{Picture of the detector anode plane (left) with an expanded view of the pad structure (center).
Right: picture of the detector after the {\it{bulk}} process with the frame defining the gas enclosure.}
\label{fig:fig1}
\end{figure}

The anode plane is segmented in 48$\times$16 readout pads of rectangular shape of 0.8$\times$2.8 mm$^2$, with a pitch of 1 and 3 mm in the two coordinates x and y, respectively.
The active surface is 4.8$\times$4.8 cm$^2$ with a total number of 768 channels and a
density of about 33 readout elements per cm$^2$, routed to connectors placed at the periphery of the detector board.
A picture of the anode plane can be seen in Figure~\ref{fig:fig1}. 
On top of this anode plane the spark protection resistive layers have been implemented and will be discussed later.
All MM detectors are assembled using the bulk Micromegas process~\cite{MMbulk}, with cylindrical spacer pillars, 128~$\mu$m high and with a diameter of 300~$\mu$m (unless differently stated), supporting the stainless steel micro-mesh\footnote{The mesh wires have a diameter of 18$\mu$m and a pitch of 45$\mu$m and the mesh underwent the calendering process.}. The actual amplification gap is about 100$\mu$m.
The cathode is made by a copper foil placed 5~mm above the anode plane, defining the conversion (or drift) gap.

%%%%%%%%%%%%%%%%%%%%%%%%%%%%%%%%%%%%%%%%%%%%%%%%%%%%%%%%%%%%%%%%%%%%%%%%%
%%%%%%%%%%%%%%%%%%%%%%%%%%%%%%%%%%%%%%%%%%%%%%%%%%%%%%%%%%%%%%%%%%%%%%%%%%%

\section{State of the art}
\label{ch:2_sec:state}

\subsection{Resistive layouts}

The resistive layout is a crucial element of our Micromegas detectors. The detector performance greatly depends on its characteristics. 
We have implemented and studied several concepts of the spark protection resistive layers, described in the following.

\noindent {\bf{Pad-Patterned embedded resistors layout (PAD-P)}}

\begin{figure}
\begin{center}
\includegraphics[width=1.0\textwidth]{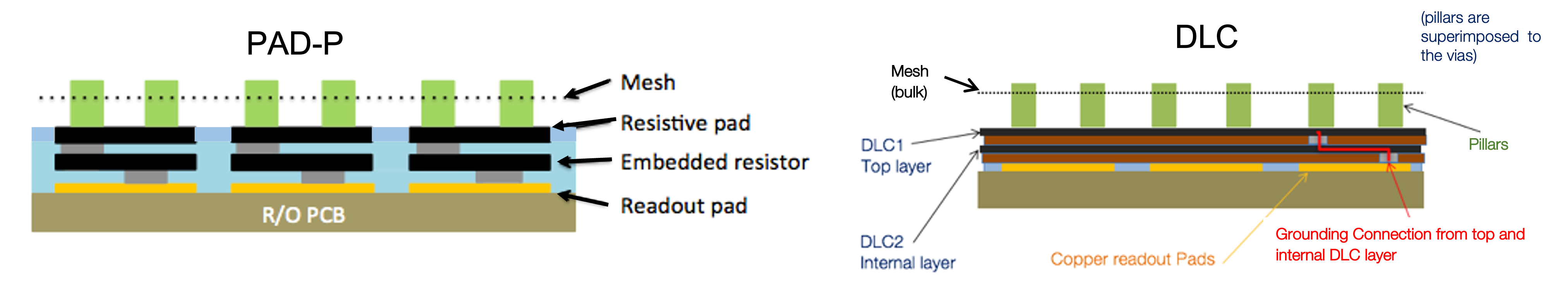}
\end{center}
\caption{Sketch of the pad-patterned embedded resistor layout (left) and of the Diamond-Like Carbon layout (right).}
\label{fig:fig2}
\end{figure}

This solution adopts a pad-patterned resistive scheme~\cite{Paddy}. Resistive pads, with the same dimension of the copper anode pads, overlay the anode pads and are interconnected to them by intermediate resistors, as shown in Figure~\ref{fig:fig2} (left). In this scheme, each resistive pad is connected to the corresponding anode pad, underneath of it.
The total resistance between the resistive and anode pads is in the range 3-7 M$\Omega$. The main characteristic of this detector is that each pad is totally separated from the others, for the anode, as well as for the resistive part. 
A double layer of resistors, with staggered connection vias, is necessary to guarantee an almost uniform resistance to the anode pads, independent of the impact position of the electron avalanche. With a single layer, avalanches close to vias would see very low resistance.
%%%%%%%%%%%%%%%%%%%%%%%%%%%%%%%%%%%%%%%%%%%%%%%%%%%%%%%%%%%%%%%%%%%%%%%%%
%%%%%%%%%%%%%%%%%%%%%%%%%%%%%%%%%%%%%%%%%%%%%%%%%%%%%%%%%%%%%%%%%%%%%%%%%%%

\noindent {\bf{Diamond-Like Carbon (DLC) layout}}

The scheme uses two continuous resistive layers of Diamond Like Carbon structures (DLC), deposited by sputtering on kapton foils and glued on the anode. The two resistive layers are interconnected with the readout pads with a network of conducting vias (filled with flexible polymer silver conductive paste) with a few mm pitch to evacuate the charge, as sketched in Figure~\ref{fig:fig2} (right). 
The smooth, uniform and continuous surface of the DLC foil grants a more homogeneous detector response.

The detector active plane is divided in two halves for testing purposes, with a different pitch of the conducting vias through the DLC layers: 6~mm and 12~mm respectively. The spark protection mechanism with a double DLC layer was inspired by the technique used for the development of $\mu$RWELL detectors~\cite{uRwell}. Two detectors have been built with the standard DLC technique with different resistivity: the first one with average resistivity of about 50~M$\Omega/\Box$, and the other with foils with about 20~M$\Omega/\Box$ referred to as DLC50 and DLC20, respectively. In order to distinguish the two regions with 6~mm and 12~mm vias pitch, the suffixes “6~mm” and “12~mm” are added to the corresponding name.
%%%%%%%%%%%%%%%%%%%%%%%%%%%%%%%%%%%%%%%%%%%%%%%%%%%%%%%%%%%%%%%%%%%%%%%%%
\begin{figure}
\begin{center}
\includegraphics[width=0.6\textwidth]{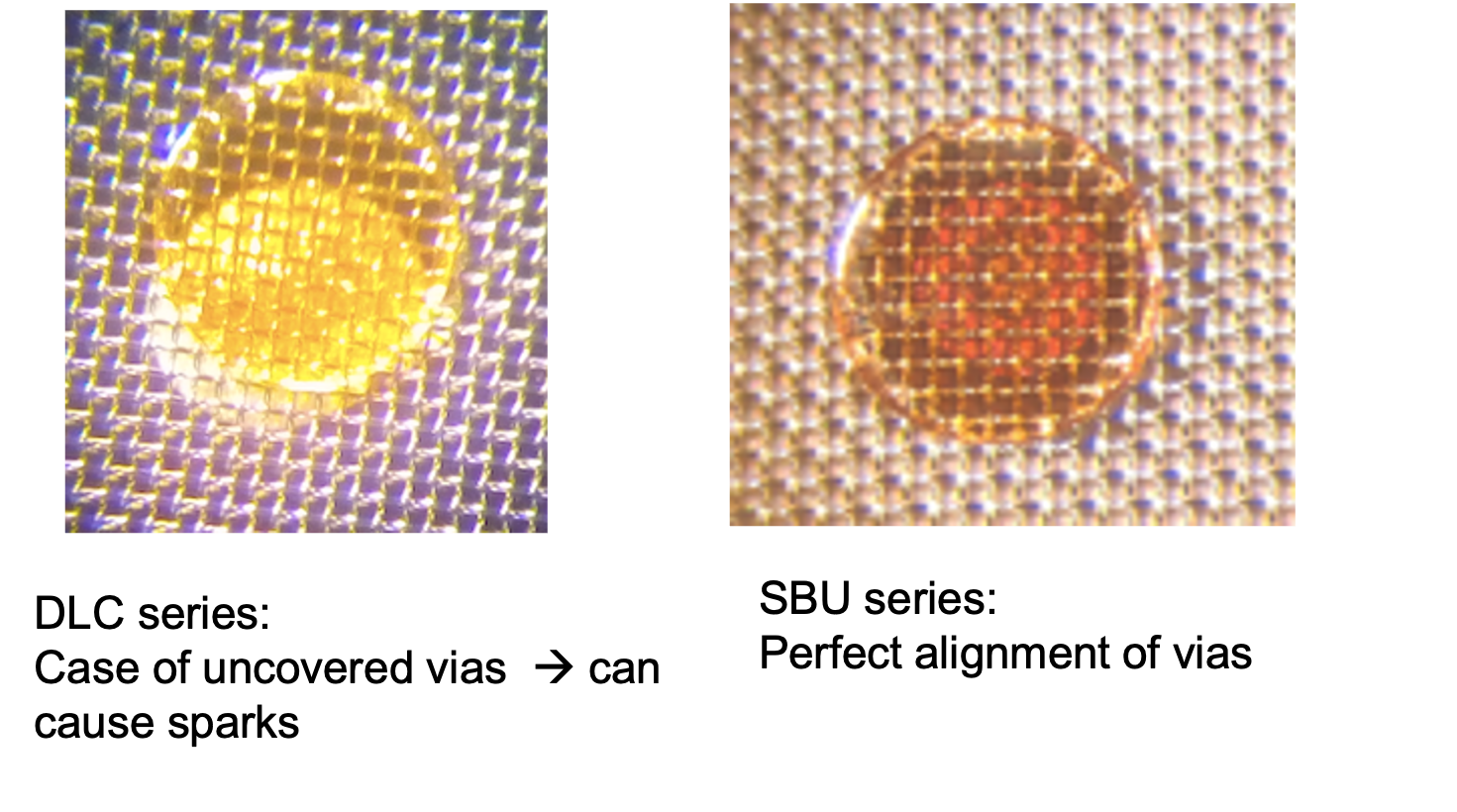}
\end{center}
\caption{Example of a misaligned via as found in a DLC detector (left) and of a via well centered below the pillar in a SBU detector.}
\label{fig:fig3}
\end{figure}
%%%%%%%%%%%%%%%%%%%%%%%%%%%%%%%%%%%%%%%%%%%%%%%%%%%%%%%%%%%%%%%%%%%%%%%%%%%

\noindent {\bf{DLC with the sequential build-up technique (SBU)}}

This detector exploits the copper clad DLC foils to improve the construction technique, making use of the {\it{sequential build up}} (SBU) process~\cite{SBU}. 
This technique has two main advantages.
The first one is that it allows to use photolithography (applied after removing the copper everywhere except at the vias positions) to precisely locate the conductive vias and align them below the pillars, as shown in Figure~\ref{fig:fig3} right in comparison with a misaligned via-pillar pair observed in one of the first standard DLC prototypes (Figure~\ref{fig:fig3} left).
This solution prevents sparks in regions where the conductive vias can be misplaced and partially exposed to the gas gap. 
The second advantage is that it is fully compatible with standard PCB processes (for example the vias can be created with standard plating techniques), significantly facilitating the technological transfer of the production. 

Two prototypes have been built with the SBU technique, referred to as SBU1 and SBU2. In both cases, the configuration with the 6~mm pitch grounding vias is adopted in the full area. Both detectors were built with a 35~M$\Omega/\Box$ inner resistive layer (closest to the anode pads) and the outer layer with resistivity of 5~M$\Omega/\Box$. A third prototype, SBU3, has both DLC foils with an average surface resistivity of 30~M$\Omega/\Box$, and the readout pads in-between the two DLC layers.

\noindent {\bf{Hybrid layout (PAD-H)}}

\begin{figure}
\begin{center}
\includegraphics[width=0.80\textwidth]{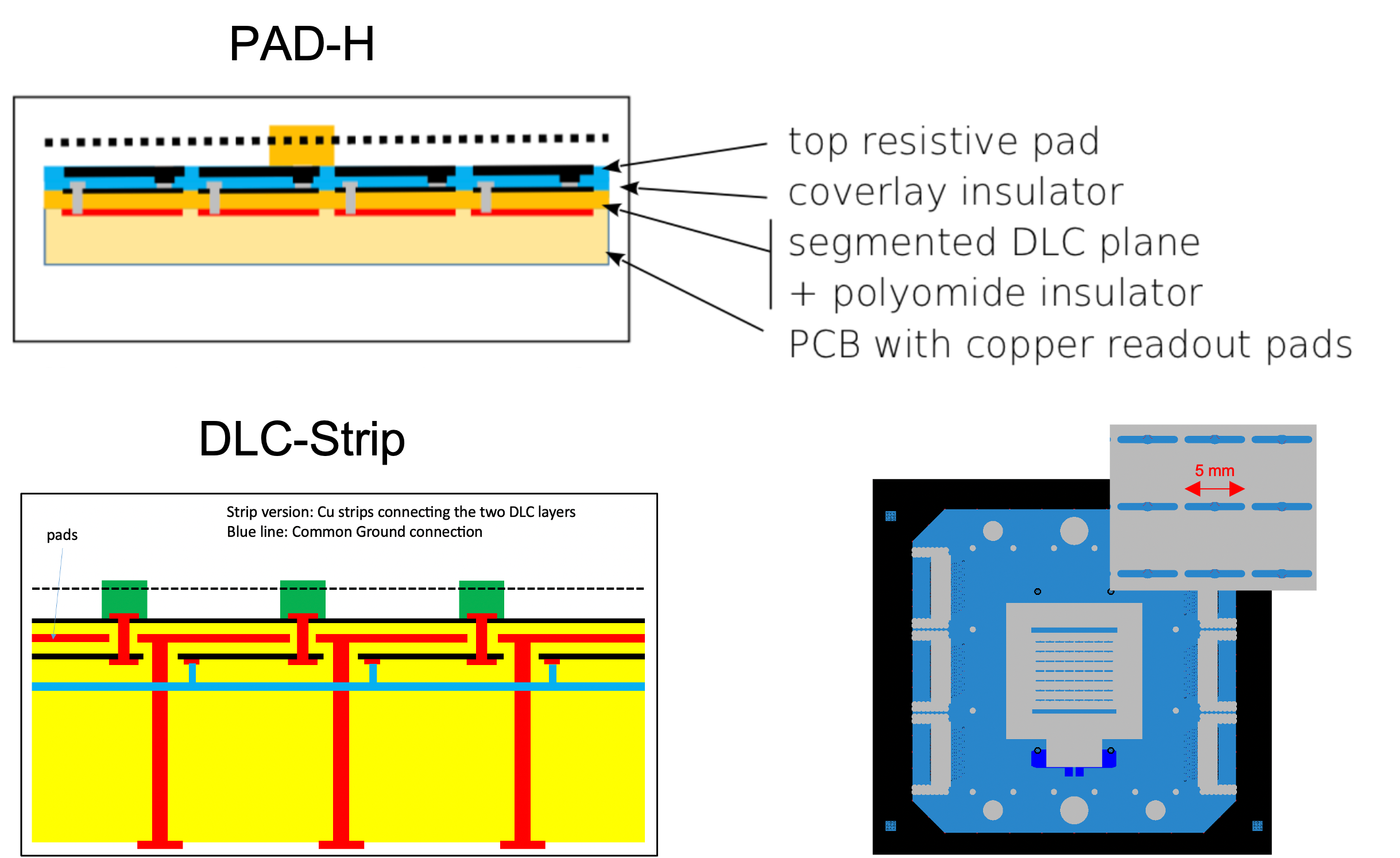}
\end{center}
\caption{Schematic view of the PAD-H (top) and the DLC-Strip (bottom) detectors. The bottom right plot shows the layout of the insulating layer of the DLC-Strip detector, with the characteristic elongated pillars.}
\label{fig:fig4}
\end{figure}

This configuration uses a DLC layer for the inner layer and screen-printed resistive pads for the outher layer. Differently from the DLC and the SBU schemes, in the PAD-H configuration the carbon layer of the DLC foil is patterned in pads. The schematic cross-section of the detector is shown in Figure~\ref{fig:fig4} top, with the indication of the components of the stack.  

The size of the DLC pads can be equal to the one of the screen-printed pads to maximise the number of charge evacuation paths, or to a multiple of them to simplify the construction at the cost of longer evacuation paths. In our prototype we opted for the first solution, leading to the same connection scheme, between the inner and the outer layers, of the PAD-P detectors.

%%%%%%%%%%%%%%%%%%%%%%%%%%%%%%%%%%%%%%%%%%%%%%%%%%%%%%%%%%%%%%%%%%%%%%%%%
%%%%%%%%%%%%%%%%%%%%%%%%%%%%%%%%%%%%%%%%%%%%%%%%%%%%%%%%%%%%%%%%%%%%%%%%%%%

\noindent {\bf{DLC-Strip layout}}

In the DLC-Strip detector the readout pads are located in-between the two resistive layers (as in SBU-3). The latter are uniform DLC for the outer and segmented DLC for the inner one (same as the inner layer of PAD-H).
A schematic view of the detector structure is shown in Figure~\ref{fig:fig4} bottom.
This configuration is expected to improve the capacitive coupling for signal induction. 

The connection between the DLC layers can't be reliably realised with silver-loaded polymer, thus metal connection strips are used. 
The connection strips divides the detector surface into several separated sectors, rendering the detector performance independent form the irradiated area.
The top surface of these connections must be completely insulated by the gas gap, to prevent the development of intense discharges. 
As for the insulation of the vias for the PAD detectors, for the DLC-Strip the insulation is obtained by covering the exposed conductive material with pillars. Following the specific layour of the detector, in this case the pillars need to be elongated to a lenght of 5~mm, as shown in Figure~\ref{fig:fig4} bottom right.

\subsection{Performance}

All the detectors underwent a number of test to assess their performance in terms of charge up behavior, gain, energy resolution, rate capability, efficiency and spatial resolution. Tests have been conducted at CERN with $^{55}$Fe radioactive sources, with X-rays from copper, and with particle beams at the SPS at CERN and at PSI. 
At the test beams the detector signals have been acquired with APV25 hybrids read out with the SRS system~\cite{APVSRS}. Unless differently stated, all tests have have been conducted with Ar:CO$_2$ gas mixture in the fraction 93:7~vol, at a gas flow of few renewal per hour, for a tpycal gas volume of 0.2~l.

Detailed results can be found here~\cite{PaddyAll}; in this section a summary of the most relevant results is reported.

\subsection{Charging up effect}

The variation of the gain in MPGD when exposed to intense particle fluxes is a well known effect, observed by many authors, owing to the charging up of the dielectric material in the detector structure. 
It can lead either to a reduction or to an increase of the gain, depending on the field configuration. Often both effects are present in the same structure although with different time scales. For resitive detectors the resistivity plays an important role too.

%%%%%%%%%%%%%%%%%%%%%%%%%%%%%%%%%%%%%%%%%%%%%%%%%%%%%%%%%%%%%%%%%%%%%%%%%
\begin{figure}
\begin{center}
\includegraphics[width=1.0\textwidth]{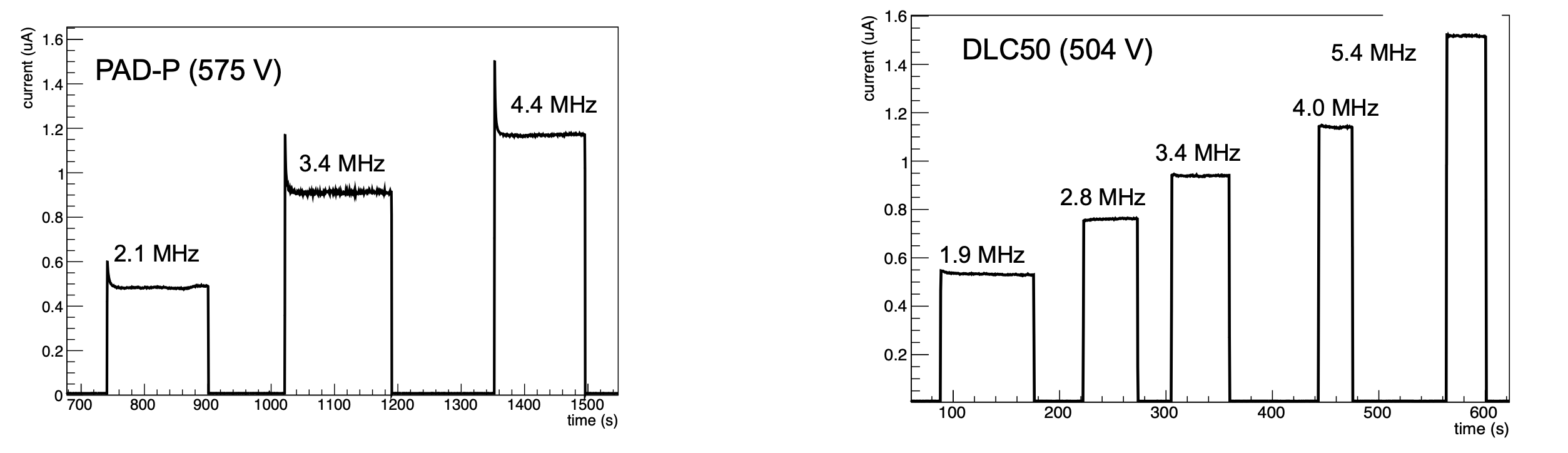}
\end{center}
\caption{Charge up for the PAD-P (left) and the DLC (right) detectors irradiated with X-rays.}
\label{fig:fig5}
\end{figure}
%%%%%%%%%%%%%%%%%%%%%%%%%%%%%%%%%%%%%%%%%%%%%%%%%%%%%%%%%%%%%%%%%%%%%%%%%%%

We have found two different behaviors for detectors with the upper resistive layer segmented in pads (like the PAD-P series) or with uniform DLC layer (as the DLC series).
As an example, Figure~\ref{fig:fig5} shows the current as function of the time of a PAD-P and a DLC detectors when irradiated with X-rays at different irradiation rates of the order of MHz/cm$^2$.
The PAD-P detector shows a characteristic fast gain reduction of the order of 15-20\%. For the DLC detector the charging up effect at short time is much smaller, with a gain reduction of less than few percent. A behavior similar to the DLC have been observed for the SBU detector, while the PAD-H showed a charging up effect similar to the PAD-P series.

\begin{figure}
\begin{center}
\includegraphics[width=1.0\textwidth]{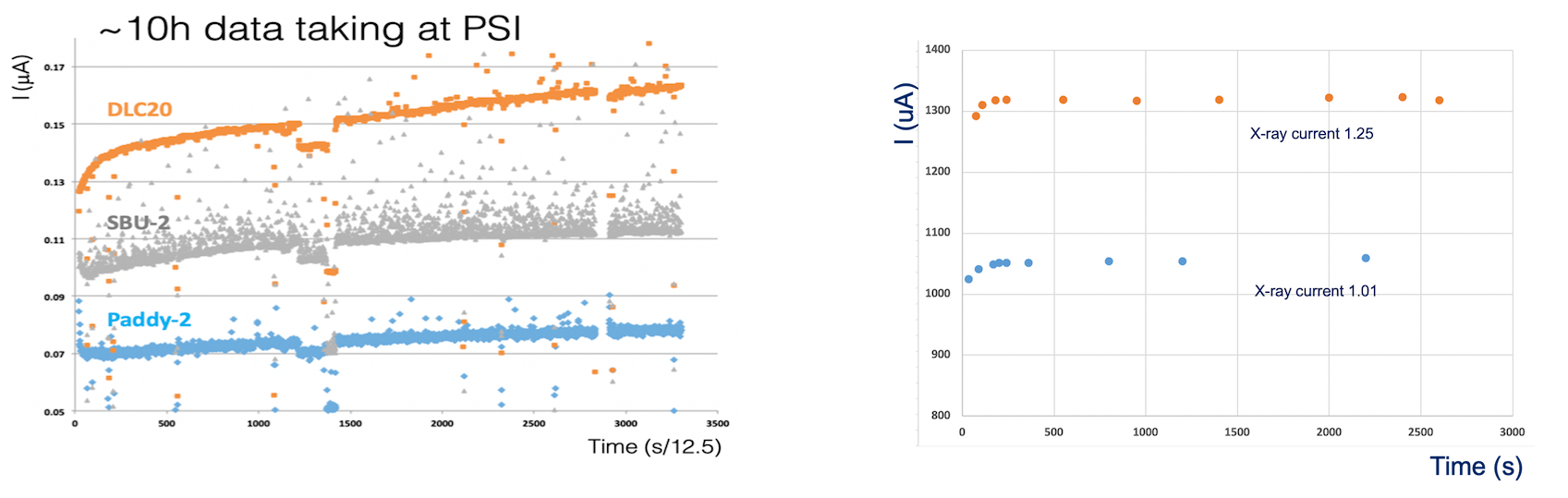}
\end{center}
\caption{Charge up for the PAD-P, DLC and SBU at PSI pion beam (left) and of the DLC-Strip (right) irradiated with X-rays.}
\label{fig:fig6}
\end{figure}

It is interesting to notice that during long-term irradiation a gain increase has been observed for all the detectors. This is shown in Figure~\ref{fig:fig6} left, obtained with a high intensity pion beam at PSI for a measurement of about 10~h. After an initial period where the PAD detector behaves differently for the DLC and SBU, all the detector showed a slow increase of the gain with time.

The DLC-Strip detector showed a peculiar charging-up: no measurable gain drop at short time scale (similar to DLC) but a fast gain increase, as illustrated in Figure~\ref{fig:fig6} right, where the detector current has been measured under X-ray irradiation at a hit rate of few hundreds kHz/cm$^2$. A gain increase of about 3\% has been measured in 200~s.

%%%%%%%%%%%%%%%%%%%%%%%%%%%%%%%%%%%%%%%%%%%%%%%%%%%%%%%%%%%%%%%%%%%%%%%%%
%%%%%%%%%%%%%%%%%%%%%%%%%%%%%%%%%%%%%%%%%%%%%%%%%%%%%%%%%%%%%%%%%%%%%%%%%%%

\subsection{Gain and energy resolution}

The amplification gain of the detectors has been measured in several conditions; here we only report few highlights.
In what follows the drift field is usually kept at 600~V/cm, unless differently stated.

\begin{figure}
\begin{center}
\includegraphics[width=1.0\textwidth]{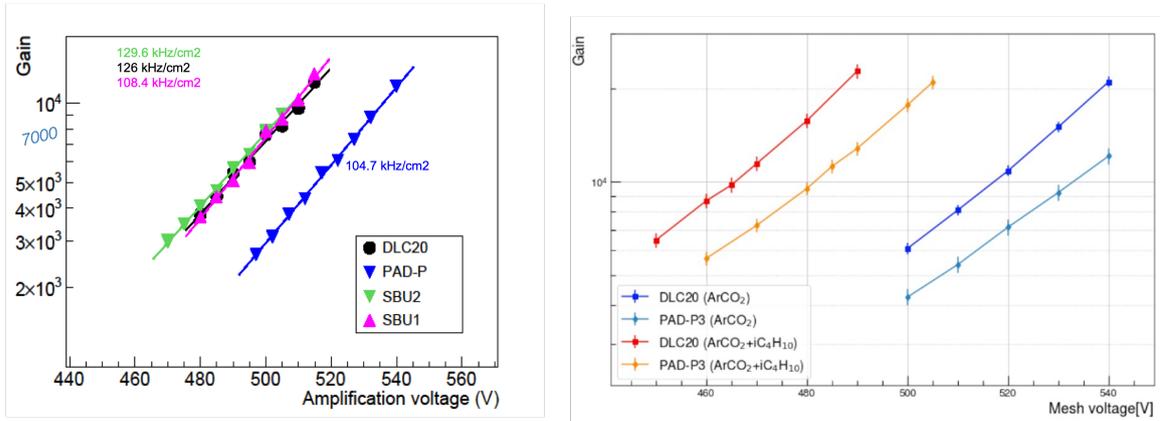}
\end{center}
\caption{Gain vs amplification voltage for the PAD-P, DLC and SBU detectors in Ar:CO$_2$ (left) and for PAD-H and DLC in Ar:CO$_2$ and Ar:CO$_2$:iC$_4$H$_{10}$.}
\label{fig:fig7}
\end{figure}

Figure~\ref{fig:fig7} left shows the gas gain as function of the amplification voltage of a PAD-P, a DLC and two SBU detectors. 
DLC and SBU, all exploiting an external uniform DLC resistive layer, show the same gain demonstrating the high level of uniformity reached in the production process.  
The PAD-P detector shows a lower gain, by about a factor 2, only partially justified by the larger charge-up. The main difference is attributed to the different field configuration as consequence of the different layout of the external resistive layers.
%%%%%%%%%%%%%%%%%%%%%%%%%%%%%%%%%%%%%%%%%%%%%%%%%%%%%%%%%%%%%%%%%%%%%%%%%
%%%%%%%%%%%%%%%%%%%%%%%%%%%%%%%%%%%%%%%%%%%%%%%%%%%%%%%%%%%%%%%%%%%%%%%%%%%

The same difference was observed comparing the DLC20 detector with the PAD-H, as shown in  Figure~\ref{fig:fig7} right (where PAD-H is indicated as PAD-P3 for historical reasons). The same plot shows the gain of the two detectors measured with the Ar:CO$_2$:iC$_4$H$_{10}$ mixture in the fraction 93:5:2 vol. The introduction of 2\% of isobutane lead to a gain increase of a factor about 4 with respect to the Ar:CO$_2$ mixture, owing to the higher Penning transfer. The mixture with the addition of 2\% of isobutane allows then to operate the detector at lower voltage to reach the same gain. Moreover, it is worth to mention that the mixture is still not flammable and can be safely used for application in experiments.
%%%%%%%%%%%%%%%%%%%%%%%%%%%%%%%%%%%%%%%%%%%%%%%%%%%%%%%%%%%%%%%%%%%%%%%%%
\begin{figure}
\begin{center}
\includegraphics[width=1.0\textwidth]{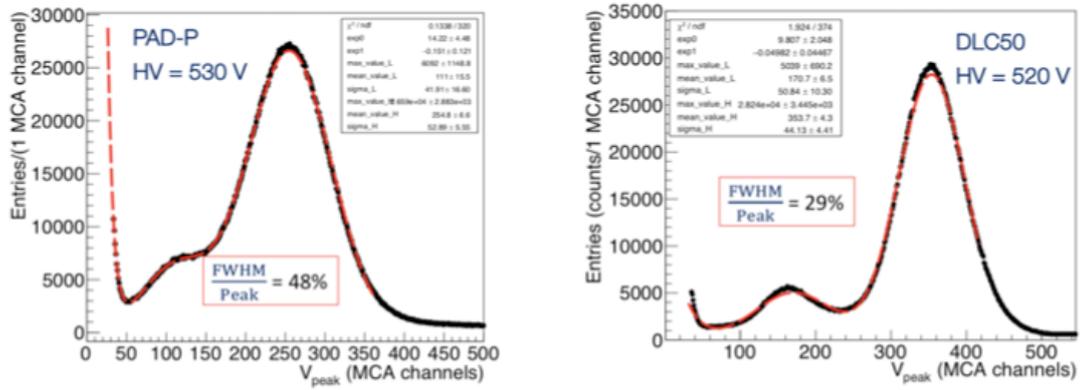}
\end{center}
\caption{Energy spectra of $\gamma$s from an $^{55}$Fe source for the PAD-P (left) and the DLC (right) detectors. The main peak of each distribution is fitted with a gaussian and the energy resolution is computed as FWHM/peak value, as indicated in the figures.}
\label{fig:fig8}
\end{figure}
%%%%%%%%%%%%%%%%%%%%%%%%%%%%%%%%%%%%%%%%%%%%%%%%%%%%%%%%%%%%%%%%%%%%%%%%%%%

The energy resolution has been obtained by measuring the energy spectra of the detector response with a multi-channel analyser (MCA) when irradiated with gammas from $^{55}$Fe sources.
Figure~\ref{fig:fig8} shows a typical result obtained for PAD-P (left) and DLC (right) detectors. The smooth surface of the carbon layer in the DLC detector assure a more uniform electric field when compared to the screen-printed pads of the PAD-P one, resulting in a significant improvement in the energy resolution, as indicated in figure. During the construction process of the PAD-H detector (whose outer resistive layer is similar to PAD-P) more care was taken to flatten the resistive pads, resulting in an energy resolution of 38\% (not shown in figure). This result shows that with PAD type detectors one can easily reach an energy resolution below 40\% (FWHM/mean).

\subsection{Rate capability}

\begin{figure}
\begin{center}
\includegraphics[width=1.0\textwidth]{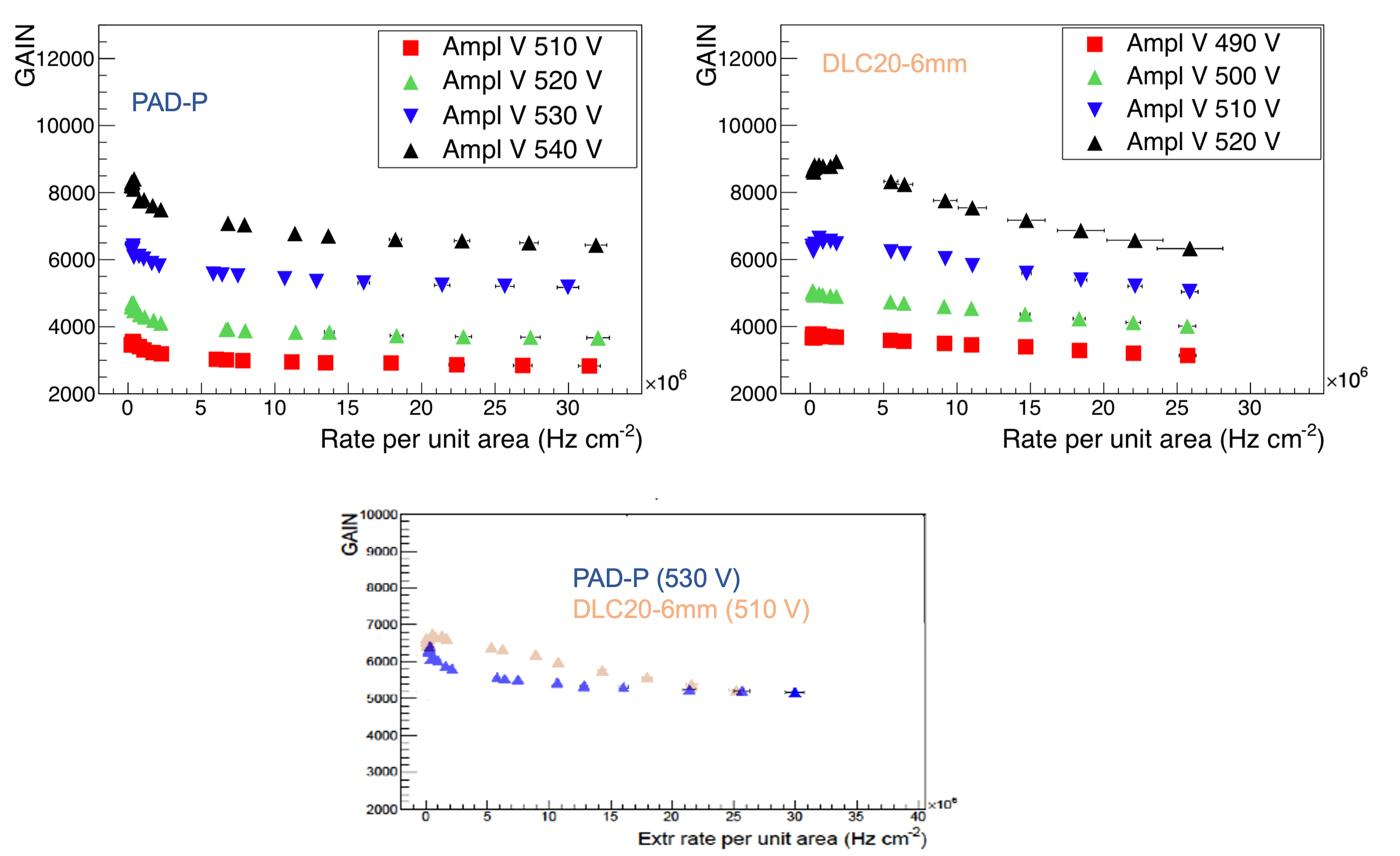}
\end{center}
\caption{Top: gain of the PAD-P (left) and DLC20-6mm (right) in the range of rates up to 30 MHz/cm$^2$ for different values of the amplification voltage, measured with X-rays. Bottom: direct comparison between the two detectors operated at approximately the same initial gain.}
\label{fig:fig9}
\end{figure}

The gain of the PAD-P and DLC-20 detectors in the range of rates up to 30 MHz/cm$^2$, is reported in Figure~\ref{fig:fig9} top, for different values of the amplification voltage, as measured with X-rays. The bottom plot of Figure~\ref{fig:fig9} shows a direct comparison between the two detectors.
PAD-P shows a significant gain drop at lower rates dominated by charging up effect, up to about 20\% at 20 MHz/cm$^2$ at 530~V, while it has a negligible ohmic voltage drop in this range of rates. The DLC-20 detector shows a significant ohmic voltage drop for rates higher than a few MHz/cm$^2$ (with a relative drop of about 20\% at 20 MHz/cm$^2$ at 510~V). It has also been observed that all the DLC series detectors, including the SBU type, have approximately the same gain behaviour, and show systematically a gain higher than PAD-P (at low/moderate rates) for the same value of the amplification voltage owing to the less uniform electric field in the amplification gap of PAD-P, where significant pad-edge effects occur. In order to compare results on the rate capabilities among the different detectors, we have operated them at approximately the same gain (at low rates), around 6500, setting the reference operating conditions at 530 V for the amplification voltage of PAD-P and 510 V for all DLC types. 
%%%%%%%%%%%%%%%%%%%%%%%%%%%%%%%%%%%%%%%%%%%%%%%%%%%%%%%%%%%%%%%%%%%%%%%%%
%%%%%%%%%%%%%%%%%%%%%%%%%%%%%%%%%%%%%%%%%%%%%%%%%%%%%%%%%%%%%%%%%%%%%%%%%%%

\begin{figure}
\begin{center}
\includegraphics[width=1.0\textwidth]{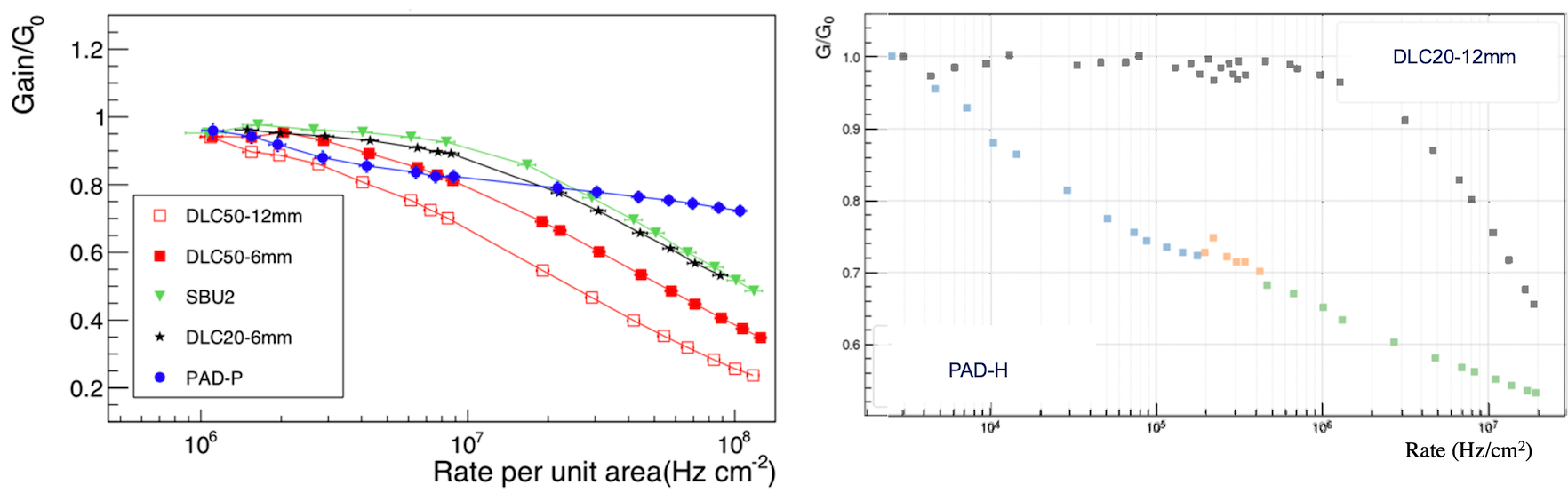}
\end{center}
\caption{Left: Dependence of the gain of the PAD-P, DLC and SBU detectors, normalised to their value at low rates, as a function of the X-Rays hit rates. The amplification voltage was set to have a gain about 6500 at 100~kHz/cm$^2$ for all the detectors. Right: direct comparison between the PAD-H and DLC20 detectors operated at approximately the same initial gain.}
\label{fig:fig10}
\end{figure}

In Figure~\ref{fig:fig10}  left, the dependence of the gain of the PAD-P, DLC and SBU detectors, normalised to their value at low rates, are reported as a function of the hit rates, in the range 1-100 MHz/cm$^2$. 
PAD-P has a very different behaviour from all the DLC types: its gain drop is dominated by the charge up, increasing with the rates and almost saturating at 20 MHz/cm$^2$, where the gain drops by about 20\%. The voltage ohmic drop is contributing only for very high rates (with currents larger than 0.5~$\mu$A per pad) up to a total gain drop of about 30\% at 100~MHz/cm$^2$. In the comparison between the different DLC configurations, at rates above 10~MHz/cm$^2$, the DLC50 prototype is more severely affected by the ohmic voltage drop and the gain is significantly reduced, as expected because of its higher resistivity. It can also be seen that the configuration with 6~mm pitch grounding vias gives better performance. For what concerns the DLC20-6mm and SBU2 detectors, they show a similar behaviour at high rates, with a gain drop similar to PAD-P at about 20~MHz/cm$^2$, further reduced up to about 50\% at 100~MHz/cm$^2$.
%%%%%%%%%%%%%%%%%%%%%%%%%%%%%%%%%%%%%%%%%%%%%%%%%%%%%%%%%%%%%%%%%%%%%%%%%
%%%%%%%%%%%%%%%%%%%%%%%%%%%%%%%%%%%%%%%%%%%%%%%%%%%%%%%%%%%%%%%%%%%%%%%%%%%

Figure~\ref{fig:fig10} right shows the comparison between PAD-H and DLC20 taken as reference. The behaviour of PAD-H is similar to the one observed for the PAD-P detector, with the gain reduction dominated by the charge up and a smaller reduction at higer rates when compared with the DLC detector. The gain reduction of PAD-H results however larger than the one measured for PAD-P.

\begin{figure}
\begin{center}
\includegraphics[width=0.50\textwidth]{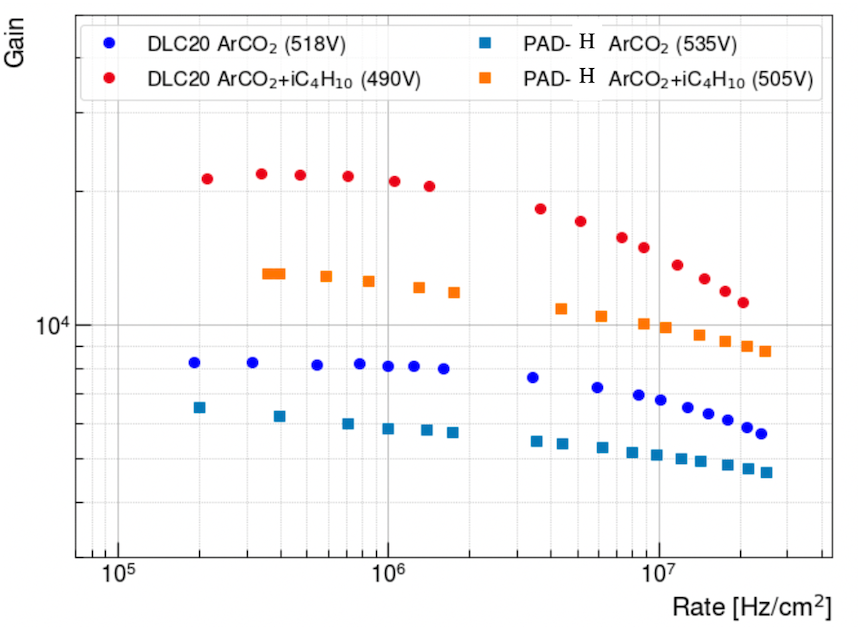}
\end{center}
\caption{Gain as function of the hit rate for the DLC20 and PAD-H detectors operated with Ar:CO$_2$ (93:7) and Ar:CO$_2$:iC$_4$H$_{10}$ (93:5:2) gas mixtures.}
\label{fig:fig11}
\end{figure}

Finally, Figure~\ref{fig:fig11} shows the  gain as function of the hit rate for the DLC20 and PAD-H detectors operated with Ar:CO$_2$ (93:7) and Ar:CO$_2$:iC$_4$H$_{10}$ (93:5:2) gas mixtures. The higher gain of the isobutane-enriched mixture makes possible to stably operate the detectors with gain above 10$^4$ up to extremely high rates of the order of 10MHz/cm$^2$ and more. This result is very encouraging in view of applications where very high particle fluxes are expected.
%%%%%%%%%%%%%%%%%%%%%%%%%%%%%%%%%%%%%%%%%%%%%%%%%%%%%%%%%%%%%%%%%%%%%%%%%
%%%%%%%%%%%%%%%%%%%%%%%%%%%%%%%%%%%%%%%%%%%%%%%%%%%%%%%%%%%%%%%%%%%%%%%%%%%

\begin{figure}
\begin{center}
\includegraphics[width=0.9\textwidth]{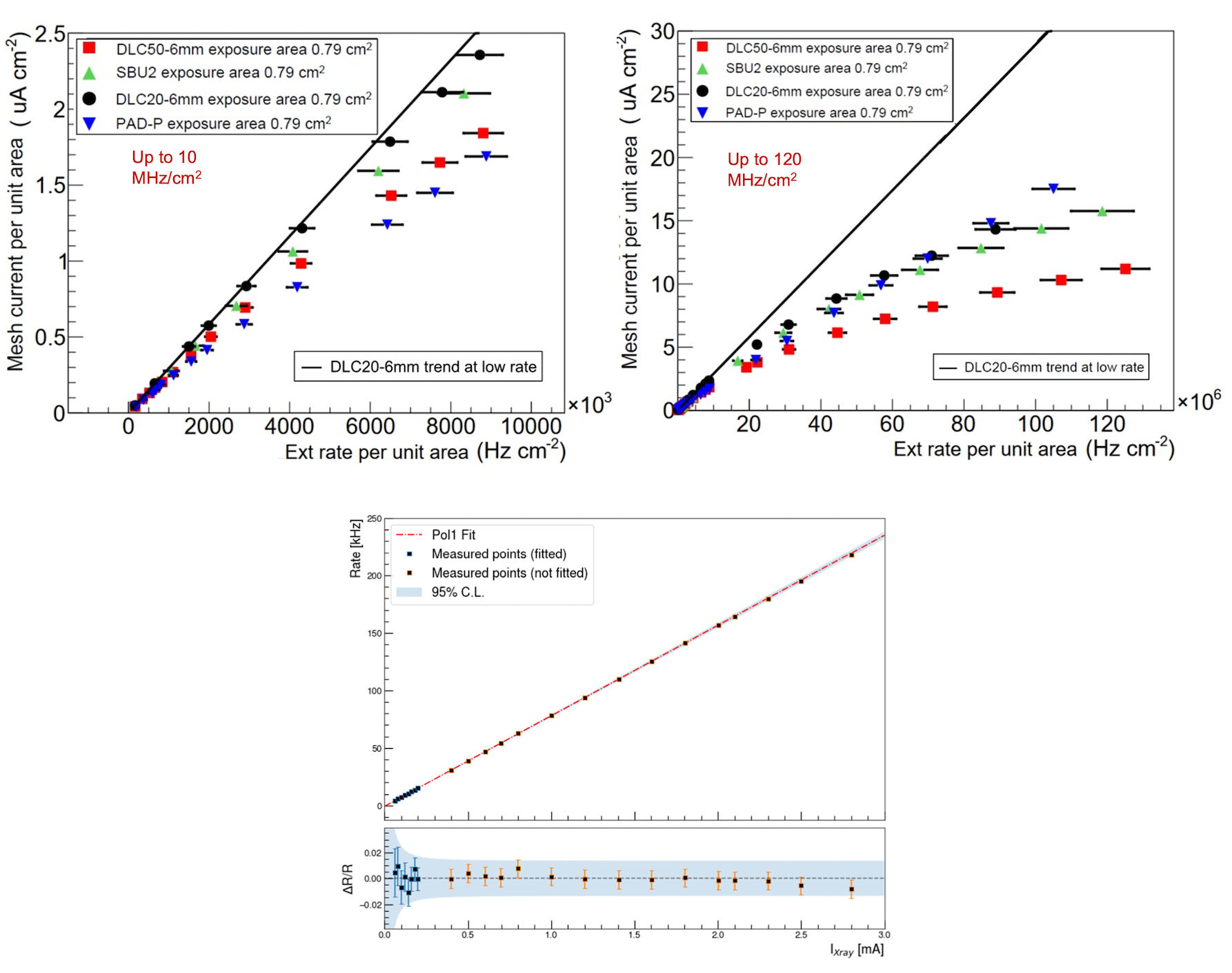}
\end{center}
\caption{Top: detector response (current) as function of the particle rates for the PAD-P, DLC20, DLC50 and SBU detectors. Left and right plots refer to different rate ranges. Bottom: PAD-H detector response as function of the current of the X-ray gun.}
\label{fig:fig12}
\end{figure}

Another way to present the rate capability is to look to the linearity of the response as function of the hit rate.
This is shown in Figure~\ref{fig:fig12} top for the PAD-P, DLC20, DLC50 and SBU detectors for irradiation with X-rays. The left plot shows the behavior at rates up to 10~MHz/cm$^2$ and the right plot up to 120~MHz/cm$^2$. At lower rates the charge up of the PAD-P detector dominates the gain reduction while at high rates the most relevant factor is the ohmic voltage drop along the charge evacuation path. The detector experiencing the larger gain reduction is DLC50, with the highest surface resistivity. The other detectors have a similar behavior with a gain reduction of about a factor two at rates of 10$^8$~Hz/cm$^2$.
Figure~\ref{fig:fig12} bottom shows the linearity of the PAD-H detector, as the correlation between the mesh current and the current of the X-ray cannon. The point at highest I$_{Xray}$ corresponds to about 200~kHz/cm$^2$, where the detector response linearity is still excellent. 
%%%%%%%%%%%%%%%%%%%%%%%%%%%%%%%%%%%%%%%%%%%%%%%%%%%%%%%%%%%%%%%%%%%%%%%%%
%%%%%%%%%%%%%%%%%%%%%%%%%%%%%%%%%%%%%%%%%%%%%%%%%%%%%%%%%%%%%%%%%%%%%%%%%%%

\subsection{Spatial resolution and efficiency}

\begin{figure}
\begin{center}
\includegraphics[width=1.0\textwidth]{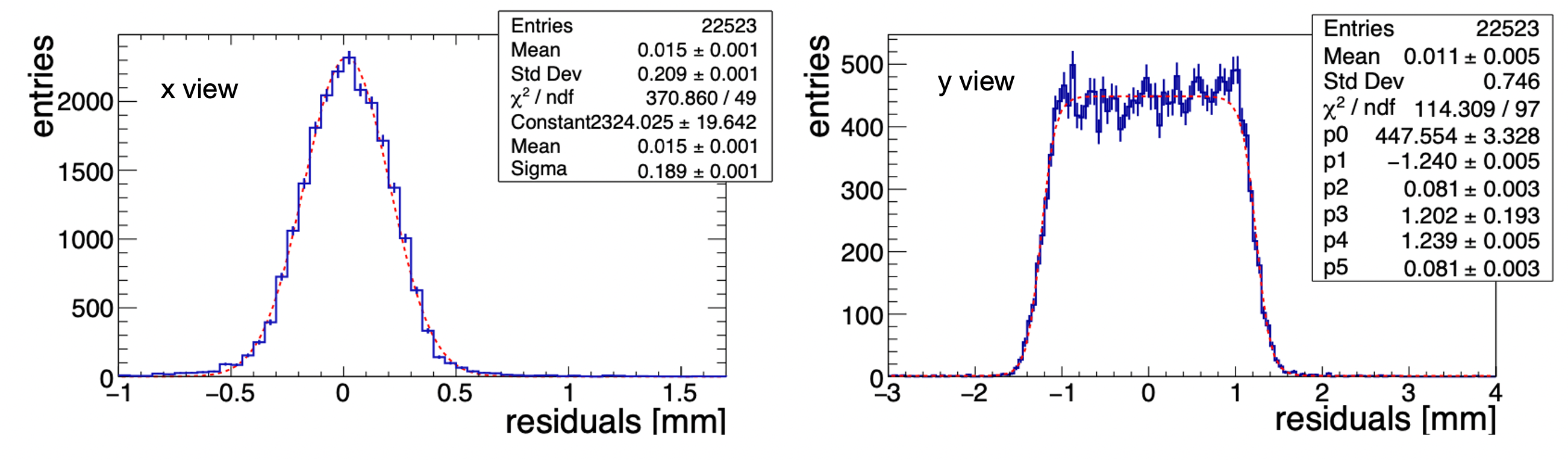}
\end{center}
\caption{Distribution of the residuals between the reconstructed cluster position on the PAD-P detector and the extrapolated track position measured with an external tracker, for the x- (left) and y-coordinate (right).}
\label{fig:fig13}
\end{figure}

The spatial resolution and efficiency  of the detectors have been measured during several test beam campaigns with particle beams at the CERN SPS H4 line and at PSI, and with muon at the CERN GIF++ facility~\cite{GIF} where a muon beam is available together with 662~keV photon background from a $^{137}$Cs source of 13.9~TBq activity.
All the presented results have been obtained with muon or pion beams perpendicular to the detector surface.
As an example, Figure~\ref{fig:fig13} shows the residual distribution of the reconstructed cluster position on the PAD-P detector under test and the extrapolated track position measured with an external tracker based on 2D resistive bulk-Micromegas. 
The cluster position is obtained as the pad charge weighted centroid of the fired pads with a signal passing minimal quality cuts.
The spatial resolution measured with CERN pion beam in the precise coordinate (x, with pad size of 0.8~mm and a pitch of 1~mm) is 190~$\mu$m. In the y-coordinate, with a pad readout pitch of 3~mm the residual distribution has a box-like shape with a FWHM of about 2.2~mm, owing to the large pad size which gives a reconstructed cluster with a single pad in most of the cases.   
%%%%%%%%%%%%%%%%%%%%%%%%%%%%%%%%%%%%%%%%%%%%%%%%%%%%%%%%%%%%%%%%%%%%%%%%%
%%%%%%%%%%%%%%%%%%%%%%%%%%%%%%%%%%%%%%%%%%%%%%%%%%%%%%%%%%%%%%%%%%%%%%%%%%%

\begin{figure}
\begin{center}
\includegraphics[width=1.0\textwidth]{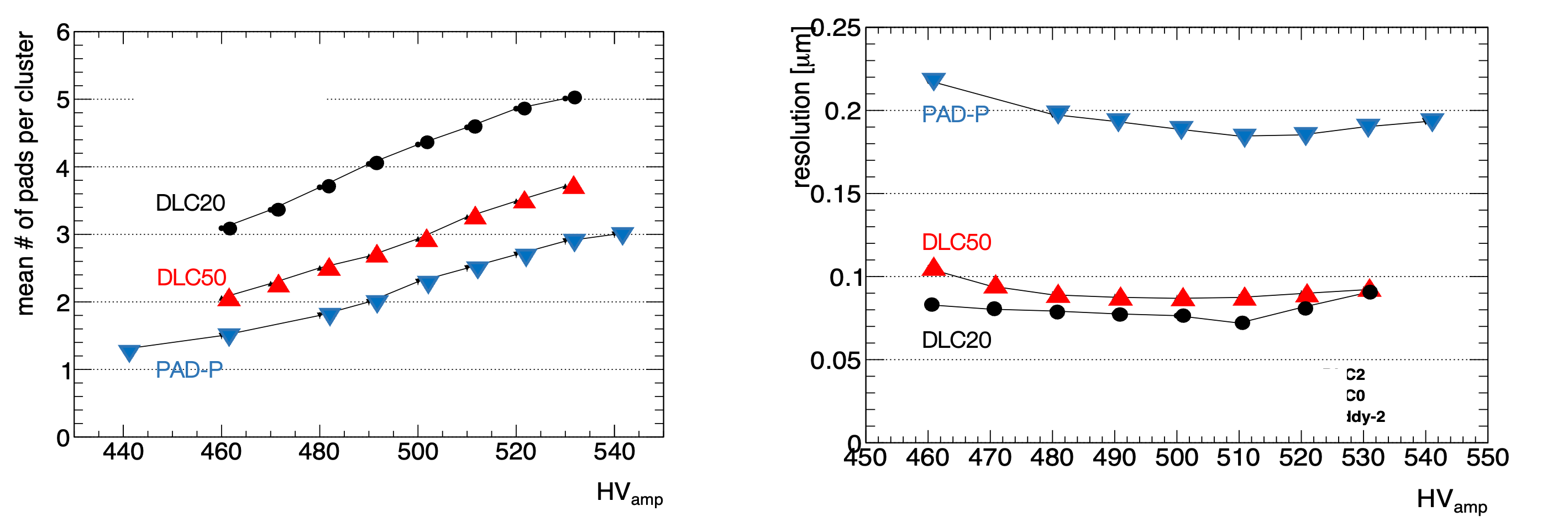}
\end{center}
\caption{Cluster dimension (left) and spatial resolution (right) for the PAD-P, DLC20 and DLC50 detectors measured with pion beam at the CERN SPS.}
\label{fig:fig14}
\end{figure}

The main parameters affecting the spatial resolution are the readout pad dimension, equal for all the detectors, and the configuration of the resistive layer. The latter affects the size of the induced signal. In detector with uniform layers (DLC, SBU) the induced signal spreads over more pads, leading to larger average dimension of the reconstructed cluster and a more precise centroid reconstruction. A lower resistivity of the external carbon layer goes in the same direction.
All that is shown in Figure~\ref{fig:fig14} where the cluster size (left) and the spatial resolution (right) are shown as function of the amplification voltage for the PAD-P, DLC20 and DLC50 detectors.
As expected PAD-P, with segmented resistive pad in the external layer, has smaller cluster dimensions and correspondingly a slightly worst spatial resolution, while the DLC detectors show a larger cluster size and a spatial resolution better than 100~$\mu$m in the x-coordinate. 
In the comparison between the two DLC detectors, DLC20 behaves slightly better in terms of spatial resolution because of its smaller surface resistivity.
%%%%%%%%%%%%%%%%%%%%%%%%%%%%%%%%%%%%%%%%%%%%%%%%%%%%%%%%%%%%%%%%%%%%%%%%%
%%%%%%%%%%%%%%%%%%%%%%%%%%%%%%%%%%%%%%%%%%%%%%%%%%%%%%%%%%%%%%%%%%%%%%%%%%%

\begin{figure}
\begin{center}
\includegraphics[width=1.0\textwidth]{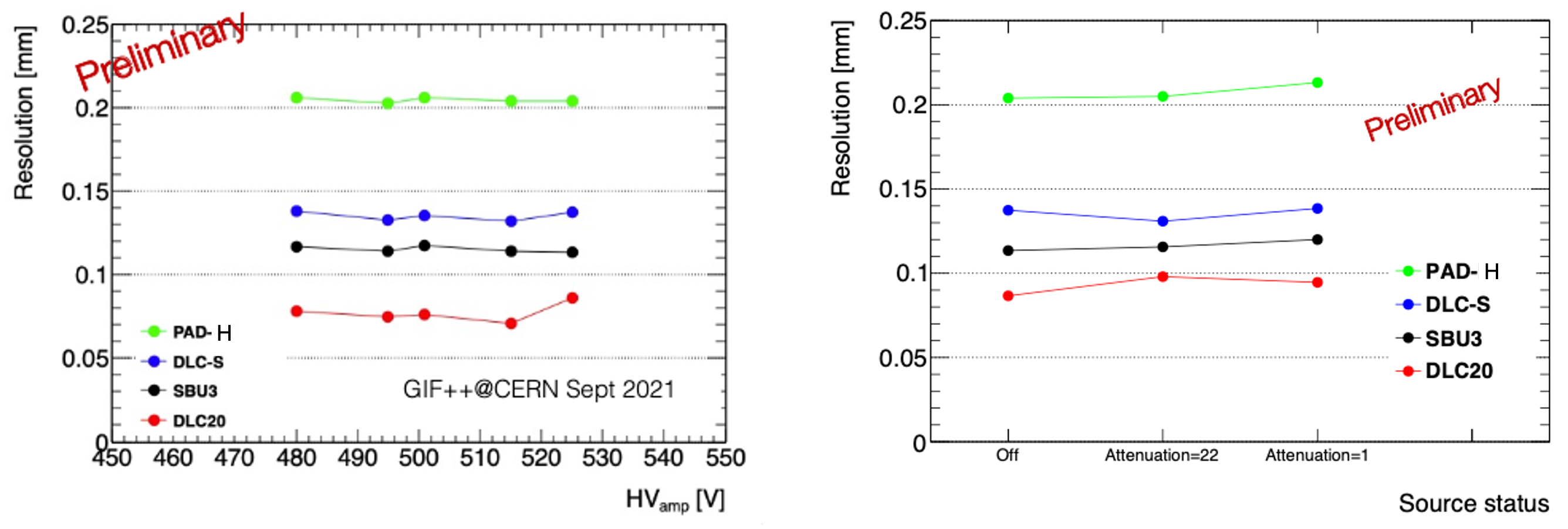}
\end{center}
\caption{Spatial resolution measured at the GIF++ facility at CERN as function of the amplification voltage (left) and of the $^{137}$Cs source status (right), for the PAD-H, DLC20, DLC-Strip and SBU detectors.}
\label{fig:fig15}
\end{figure}

We have recently tested some detectors at the GIF++ facility at CERN with muon beam from the SPS. 
The preliminary results for the PAD-H, the DLC-Strip, an SBU (the third of the series) and the DLC20 are shown in Figure~\ref{fig:fig15}.
The left plot shows the spatial resolution as function of the amplification voltage, with the expected results: the DLC detector with uniform layer of lowest resistivity (DLC20) has a resolution of 80~$\mu$m, in agreement with older measurements, while the pad-patterned device (PAD-H) shows a resolution of about 200~$\mu$m, of the same order of PAD-P.
%%%%%%%%%%%%%%%%%%%%%%%%%%%%%%%%%%%%%%%%%%%%%%%%%%%%%%%%%%%%%%%%%%%%%%%%%
%%%%%%%%%%%%%%%%%%%%%%%%%%%%%%%%%%%%%%%%%%%%%%%%%%%%%%%%%%%%%%%%%%%%%%%%%%%
The spatial resolution was also measured with increasing photon background. The result is shown in Figure~\ref{fig:fig15} right where the x-axis reports the status of the GIF++ gamma source: Off means that during the measurement no photon flux was present; Attenuation=1 means full source, corresponding to approximately few tens of kHz/cm$^2$; Attenuation=22 means that the measurement was performed with a set of absorbers in front of the GIF++ source providing a total photon flux reduced by about a factor 22 with respect to Attenuation=1.
No degradation of the tracking performance have been observed for any of the tested detectors for the full range of the GIF++ photon flux. This result confirms, once more, the suitability of these detector as tracking devices in conditions with high background.

\begin{figure}
\begin{center}
\includegraphics[width=0.50\textwidth]{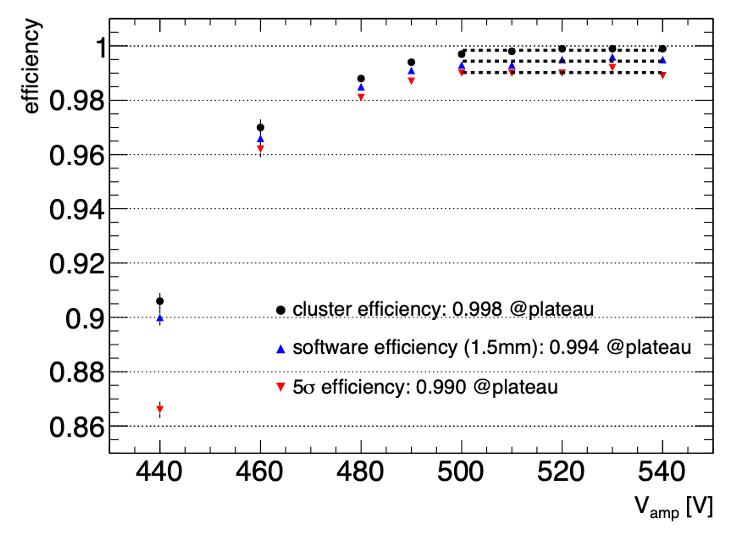}
\end{center}
\caption{PAD-P efficiency as measured with a high energy muon beam.}
\label{fig:fig16}
\end{figure}

With particle beams the detector efficiency was studied, too. As an example, Figure~\ref{fig:fig16} shows the efficiency as function of the amplification voltage measured for the PAD-P detector.
The efficiency is computed with three difference cuts when looking to the position of a reconstructed cluster with respect to the extrapolated muon track: cluster efficiency (the cluster is anywhere in the detector active area);
software efficiency (the cluster is within 1.5~mm of the extrapolated reference track in the precision coordinate); 5$\sigma$ efficiency (the cluster is within 5$\sigma$, about 1~mm, of the extrapolated reference track in the precision coordinate).
%%%%%%%%%%%%%%%%%%%%%%%%%%%%%%%%%%%%%%%%%%%%%%%%%%%%%%%%%%%%%%%%%%%%%%%%%
%%%%%%%%%%%%%%%%%%%%%%%%%%%%%%%%%%%%%%%%%%%%%%%%%%%%%%%%%%%%%%%%%%%%%%%%%%%
Even with the most stringent requirement, the detector efficiency reaches a value well above 98\%.

\begin{figure}
\begin{center}
\includegraphics[width=1.0\textwidth]{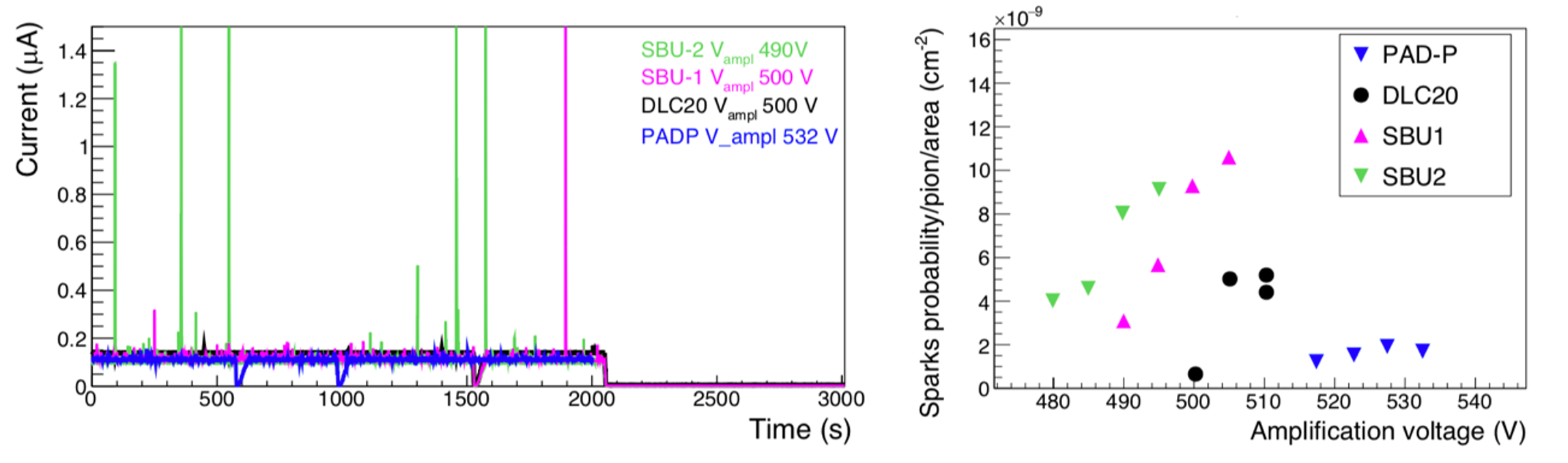}
\end{center}
\caption{Detector stability with a pion beam of 350 MeV/c. Left: current as a function of time under a particle rate of about 100~kHz/cm$^2$. Right: spark probability density per pion.}
\label{fig:fig17}
\end{figure}

Finally, with the high intensity pion beam of 350~MeV/c energy at the PSI, we have measured the spark probability of several of our detectors.
Figure~\ref{fig:fig17} left shows the detector current trend for particle rate of about 100~kHz/cm$^2$. Some discharges (seen as high current peaks) are visible, more frequently on one of the two SBU-type detectors.
The right part of the plot shows the spark probability as function of the amplification HV for the PAD-P, DLC-20 and two SBU detectors.
PAD-P shows a very high stability with a spark probability $<$2$\times$10$^{-9}$/pion/cm$^2$. 
DLC20 shows better stability of the two SBU, explained by the lower resistivity of the external resistive layer for the latter.
%%%%%%%%%%%%%%%%%%%%%%%%%%%%%%%%%%%%%%%%%%%%%%%%%%%%%%%%%%%%%%%%%%%%%%%%%
%%%%%%%%%%%%%%%%%%%%%%%%%%%%%%%%%%%%%%%%%%%%%%%%%%%%%%%%%%%%%%%%%%%%%%%%%%%

\section{Fields of applications}

The proposed detector is suited for multiple application of HEP experiments where good tracking performance is required in presence of high particle rates.
The construction process is well established and the design, notably the readout pad granularity, can be adapted to the needs. 
The different options to implement the resistive layer (pad-patterned, uniform DLC, SBU etc) Provides an additional handle to target the detector technology to the specific application, depending on the expected rate and the desired spatial resolution.
The operation conditions can also be adapted to the application, in particular the choice of the gas mixture must take into account a number of parameters like desired gain, diffusion, velocity, presence of magnetic field, external limitations like safety-related restrictions, possibility for a re-circulation system etc.

The fields of application of this family of detectors is then very large, a comprehensive discussion goes beyond the scope of this paper.
Here we only give a partial list of possible applications in HEP experiments.

\begin{itemize}
    \item Extension of the Muon system of existing experiments. The MM-pad detector can be employed as muon detector in the end-cap region (where the particle rate mostly from background is more relevant) of existing experiments at colliders. An example is the extension of the Muon system in high-$\eta$ region for the ATLAS experiment during LHC LS3.
    \item Muon detector at future experiment. Use in end-cap and forward muon system in experiments at future hadron machines (FCC-hh).
    \item Tracking detector at lepton machines: FCC-ee, ILC, CEPC, Muon Colliders.
    \item Amplification and readout stage for TPC. Micromegas provide an intrinsically large ion back-flow suppression, which can be further improved with a dedicated R\&D on the number and configuration of the micro-mesh.
    \item Muon veto at beam-dump experiments.
    \item Sampling calorimeters and pre-shower for electromagnetic calorimeters.
    \item Detector for beam monitoring.
\end{itemize}

\section{Ongoing development and future work}

The R\&D work on high granularity resistive Micromegas detectors is continuing in various directions, including construction and performance improvements. Here we mention some research lines that are particularly relevant.

\begin{figure}
\begin{center}
\includegraphics[width=1.0\textwidth]{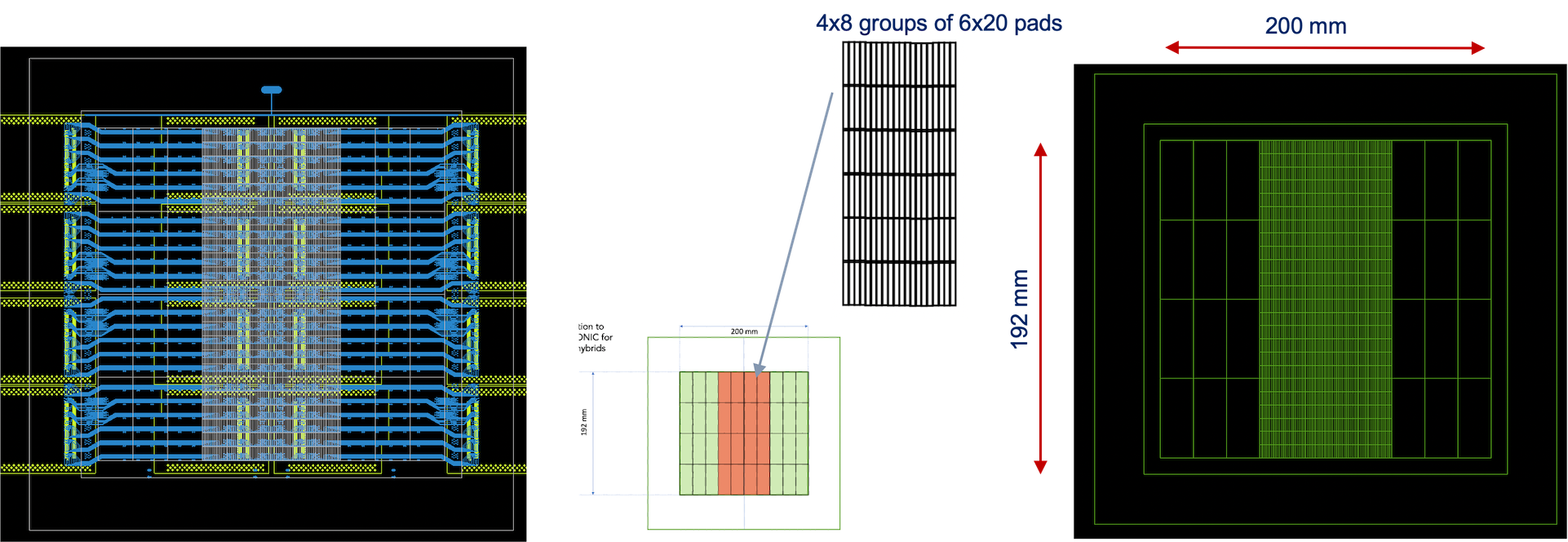}
\end{center}
\caption{Schematic of the large-size demonstrator under construction. The figure includes the zoom showing the readout pad layout.}
\label{fig:fig18}
\end{figure}

The detectors that we have developed have demonstrated to fulfil the needs for a gaseous detector to be employed in environments where high rates are expected and good tracking performance and reliable operations are required.
The performance have been studied so far on detectors with an active surface limited to 48$\times$48~mm$^2$. 
Applications of such devices will require larger detector surface.
In this regard, we are building a demonstrator with an active area of 192$\times$200~mm$^2$ with 4800 readout pads with a pitch of 1$\times$8~mm$^2$ based on the SBU technology.
Figure~\ref{fig:fig18} shows the schematic layout of the large detector.
The high density of copper traces needed to connect all the readout pads to the connectors hosted on the sides of the detector board renders the PCB construction very challenging and introduces a sizeable inactive region around the sensitive area.
For these reasons in the demonstrator only half of the readout pads will be connected.
%%%%%%%%%%%%%%%%%%%%%%%%%%%%%%%%%%%%%%%%%%%%%%%%%%%%%%%%%%%%%%%%%%%%%%%%%
%%%%%%%%%%%%%%%%%%%%%%%%%%%%%%%%%%%%%%%%%%%%%%%%%%%%%%%%%%%%%%%%%%%%%%%%%%%

\begin{figure}
\begin{center}
\includegraphics[width=0.90\textwidth]{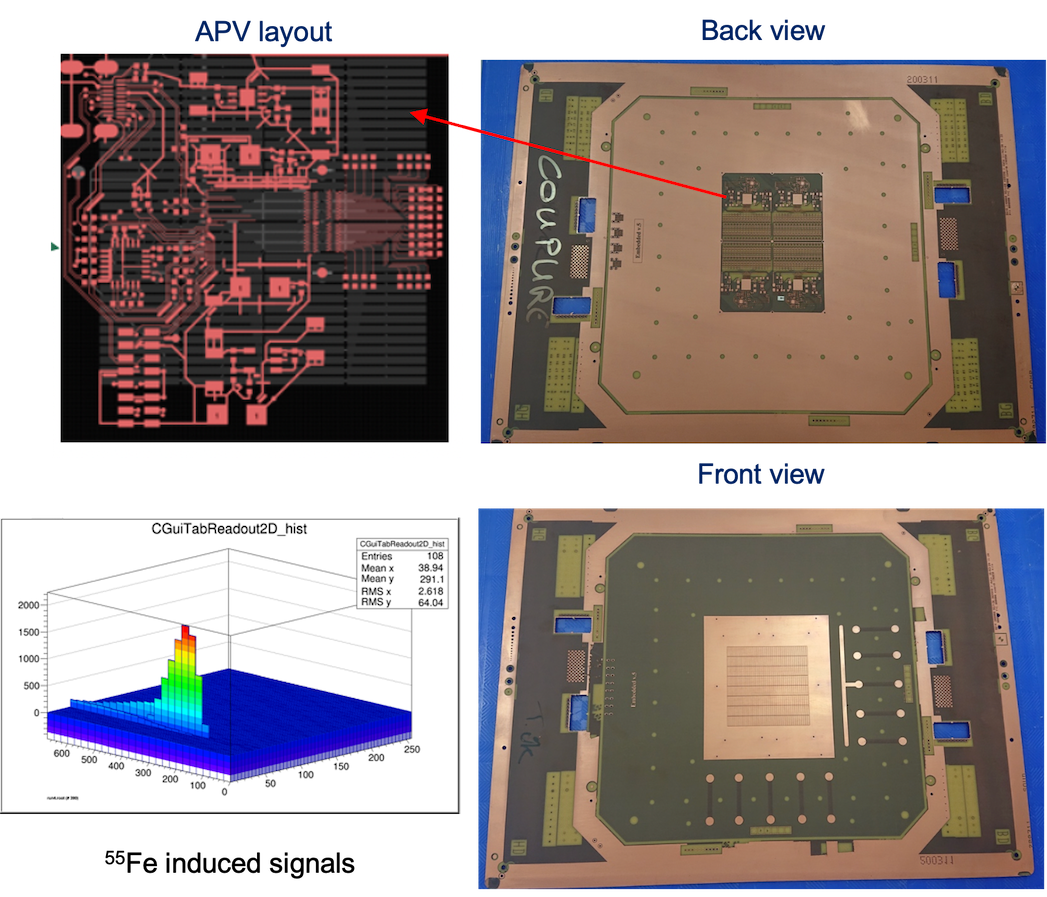}
\end{center}
\caption{Top and bottom right: pictures of the first implementation of the PAD detector with integrated electronics.
Bottom left: representation of a signal induced by a photon form a $^{55}$Fe source acquired with the PAD detector with integrated electronics.}
\label{fig:fig19}
\end{figure}

The issue of the construction complexity of the detector PCB, requiring many layers, is addressed by another ongoing development, consisting in the integration of the readout electronics directly on the back side of the detector board. This solution will drastically reduce the dead area around the active region too.
We have performed first tests with APV chips, obtaining encouraging results. 
Figure~\ref{fig:fig19} shows the picture of a Micromegas pad detector with such implementation and one of the first acquired signals. The work in this direction is still ongoing.

The next step, after the electronics integration, will be the cooling integration if chips requiring continuous cooling has to be used. The idea here is to include a micro-channel cooling loop inside the base-plane material of the detector to have a compact and highly integrated detector system which includes the sensitive device and the electronics with its cooling.
%%%%%%%%%%%%%%%%%%%%%%%%%%%%%%%%%%%%%%%%%%%%%%%%%%%%%%%%%%%%%%%%%%%%%%%%%
%%%%%%%%%%%%%%%%%%%%%%%%%%%%%%%%%%%%%%%%%%%%%%%%%%%%%%%%%%%%%%%%%%%%%%%%%%%

Another future development is the construction of a double-side detector. In this configuration two detectors are places face-to-face with a single gas gap divided in two halves by a metallic mesh playing the role of drift cathode for both detectors. The result is a detector with two anode planes that can be independently segmented and with a very reduced transverse envelope and low material budget.

As for any other MPGD and gaseous detector in general, the choice of the right gas mixture is very relevant and has to be optimised for the envisaged application.
For the early development phase we decided to use a simply, cheap and eco-friendly gas mixture without any flammable component. The used mixture (Ar:CO$_2$) is however sub-optimal for many applications.
The addition of isobutane allows to increase the gas gain at a given voltage, as described in Section \ref{ch:2_sec:state}.
Another component under consideration is CF$_4$ that with its higher drift velocity can improve the time resolution of the detector.
%%%%%%%%%%%%%%%%%%%%%%%%%%%%%%%%%%%%%%%%%%%%%%%%%%%%%%%%%%%%%%%%%%%%%%%%%
\begin{figure}
\begin{center}
\includegraphics[width=0.90\textwidth]{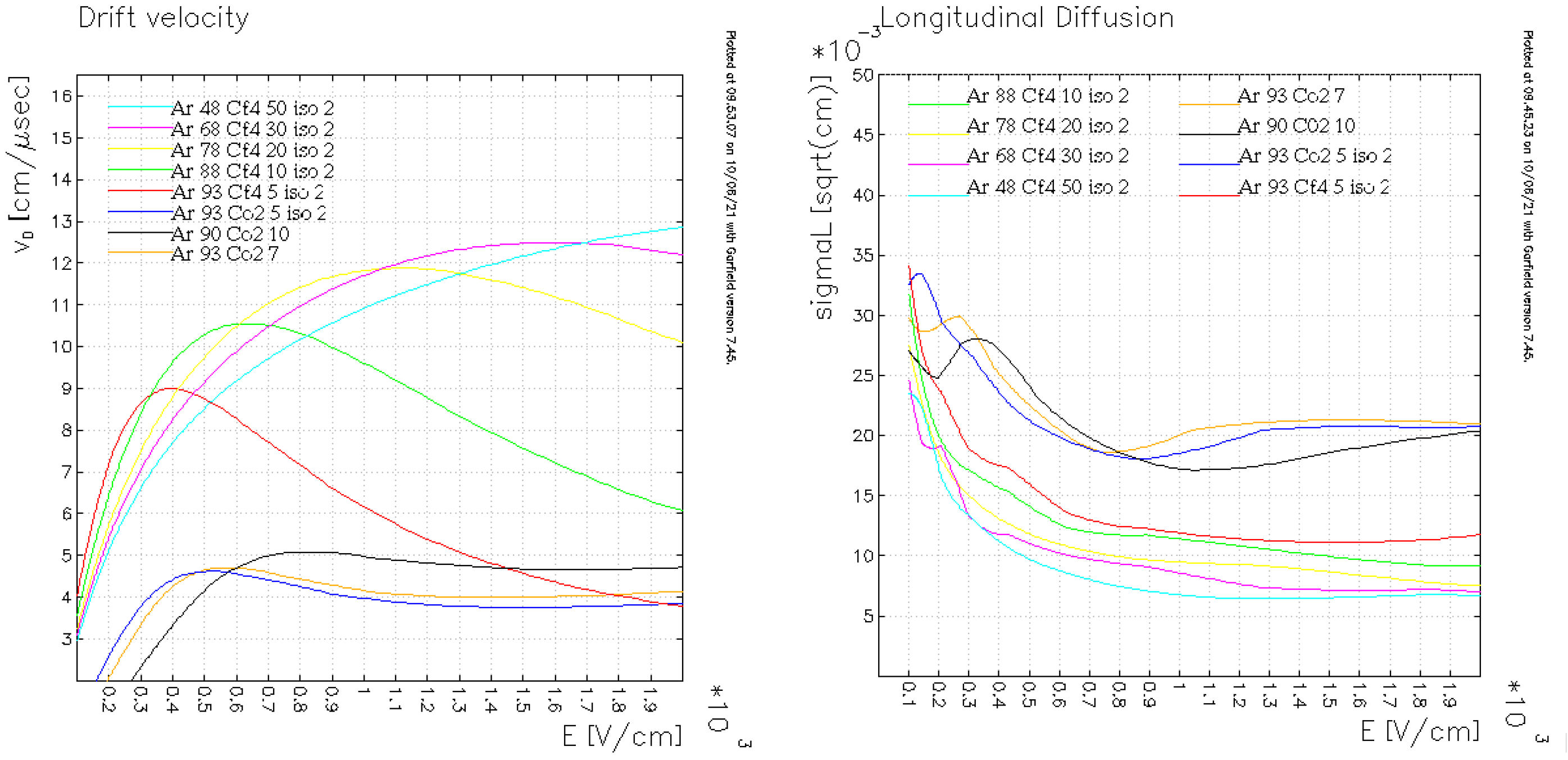}
\end{center}
\caption{Drift velocity (left) and longitudinal diffusion of argon-based gas mixtures simulate with Garfield.}
\label{fig:fig20}
\end{figure}
%%%%%%%%%%%%%%%%%%%%%%%%%%%%%%%%%%%%%%%%%%%%%%%%%%%%%%%%%%%%%%%%%%%%%%%%%%%
This is shown in Figure~\ref{fig:fig20}, displaying the drift velocity (left) and longitudinal diffusion (right) as function of the electric field in the conversion gap of a number of Ar-based mixtures with CO$_2$, iC$_4$H$_{10}$ and CF$_4$ simulated with Garfield~\cite{Garfield}.

\section{Conclusions}
This chapter summarizes the R\&D activity, ongoing since several years, aiming to the development of a new generation of single amplification stage resistive Micromegas. 
Several detectors have been so far built with different construction techniques and fully characterised with test in laboratory and at beam lines.
The developed device, with 3~mm$^2$ readout pads, can be efficiently operated up to particle fluxes of 10~MHz/cm$^2$, have an efficiency to charged particles above 98\% and a spatial resolution ranging between 80 and 200~$\mu$m (depending on the technology) for readout pads with 1~mm wide pitch.
The proposed detector can find a suitable application in many fields of particle detection in future experiments, as muon tracking or tagging detector, calorimeters, Time Projection Chamber, central tracker with low material budget, ad other.

The R\&D of this new technology is still ongoing to consolidate the construction of large-size detectors; to develop a fully integrated system that include the the front-end electronics and the cooling in a single structure; to optimise the performance and the operating conditions to the specific application. 

%\section*{Acknowledgements}
%We are indebted with the CERN MPT workshop (in particular R. de Oliveira and his group) for ideas, discussions and the construction of the detectors; with the RD51 Collaboration for support with the tests at the Gas Detector Development (GDD) Laboratory and for the test-beam at CERN; with the team of the piM1 Beam facility for their support for the test beam at PSI; with the GIF++ facility team for their support during the beam and irradiation test at GIF++.

%We would like to thank the organisers of Snowmass 2021 (US Community Study on the Future of Particle Physics) and in particular the conveners of the withepaper 'MPGD for muon detection at future colliders'.

%%%%%%%%%%%%%%%%%%%%%%%%%%%%%%%%%%%%%%%%%%

%  If you would like to use BibTEX for the bibliography, please feel free to do so.  It is not required.

%  To use BibTeX,

%    1.  uncomment the following two lines, 
%    2.  comment out everything below from  \begin{thebibliography}{99}   to \end{thebibliography).
%    3.  create the file  myreferences.bib, and process this file in the usual way

%\bibliographystyle{JHEP}
%\bibliography{myreferences}  % file myreferences.bib

%%%%%%%%%%%%%%%%%%%%%%%%%%%%%%%%%%%%%%%%%

%\section{Advanced GEM detectors for future collider experiments}

%\pubblock

%\Title{Advanced GEM detectors for future collider experiments}
\chapter{\centering Advanced GEM detectors for future collider experiments}

%\addcontentsline{toc}{chapter}{Advanced GEM detectors for future collider experiments}
\bigskip 

%\Author{A.~Colaleo, A.~Pellecchia, R.~Venditti, P.~Verwilligen}
%\Address{INFN Bari and University of Bari}

%\Author{M. Hohlmann}
%\Address{Florida institute of technology}

%\Author{J. Merlin}
%\Address{University of Seoul}

%\Author{A. Sharma}
%\Address{CERN - European Organization for Nuclear Research}

\medskip
\medskip

% \begin{Abstract}
%\noindent I describe the classification of whales in terms perhaps more familiar to the reader.
%\end{Abstract}

%\section{Introduction}

%\textcolor{blue}{basics of GEM technology, used in HEP experiments, current project and on-going R\&D.}

Since their invention in 1997, detectors based on the Gaseous Electron Multiplier (GEM) technology \cite{Sauli} have risen among the most consolidated classes of micro-pattern gaseous detectors in present-generation experiments. Specifically, among high-energy physics experiments in the last 20 years triple-GEM detectors have provided tracking with a \SI{70}{-\micro\m} space resolution in the high-rate environment (\SI{2.5}{\mega\Hz/\centi\m^2}) of the COMPASS inner tracker \cite{compass} at the SPS and have been operated in the hadron\-blind detector of the PHENIX experiment \cite{phenix} and in the forward tracker of STAR \cite{star}, both at RHIC. At the LHC, triple-GEM detectors have instrumented the T2 telescope of the TOTEM experiment \cite{totem} and the muon system of LHCb \cite{lhcb}, sustaining rates up to \SI{600}{\kilo\Hz/\centi\m^2} and demonstrating a longevity up to \SI{}{0.8 \milli\coulomb/\centi\m^2} while maintaining a good timing performance down to a \SI{4}{\nano\s} time resolution.

\begin{figure}[t]
    \centering
    \includegraphics[width=0.6\textwidth]{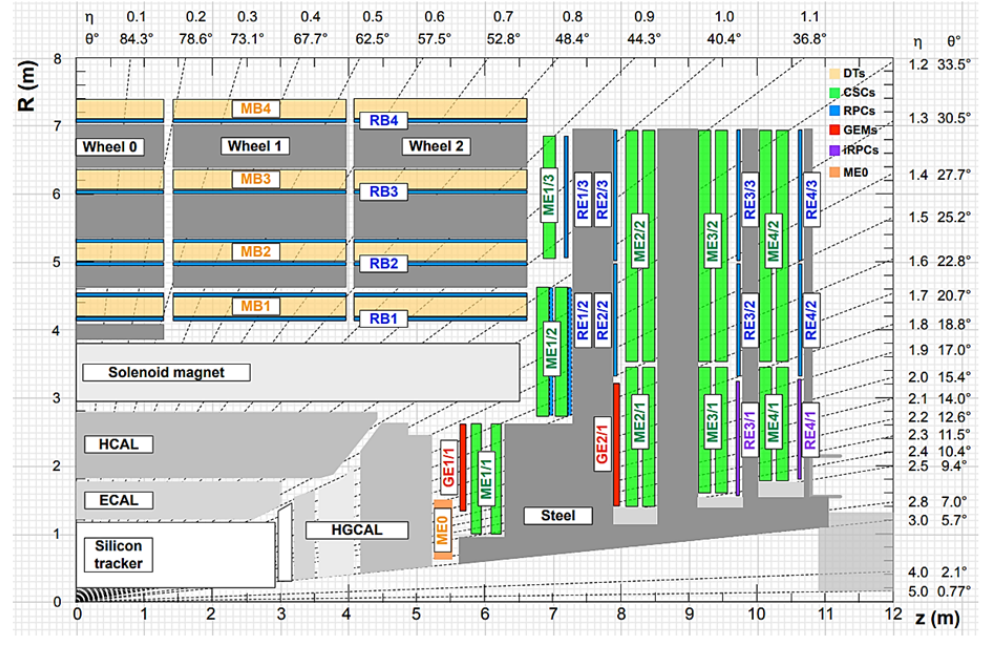}
    \includegraphics[width=0.3\textwidth]{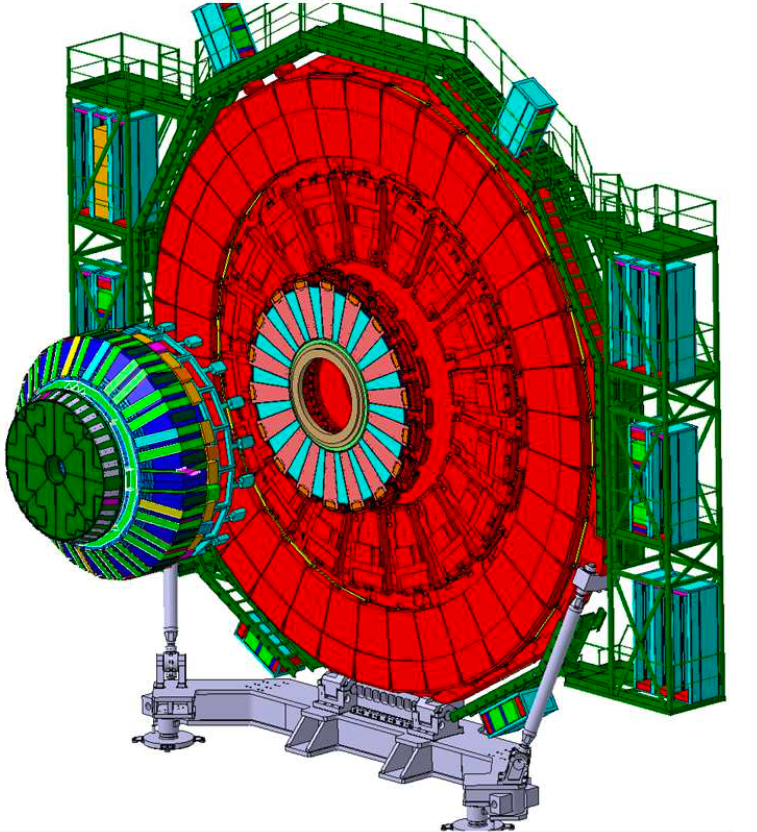}
    \caption{On the left, quadrant of a section of the CMS muon detector after the High-Luminosity LHC upgrade, showing the GE1/1 and GE2/1 stations in red and the ME0 station in orange. On the right, drawing of the GE1/1 station (with GEM chambers in orange and blue) in the first ring of the spectrometer \cite{GEMTDR}.}
    \label{fig:cms_quadrant_gems}
\end{figure}

Coming to present years, three stations of the CMS muon spectrometer (Fig.~\ref{fig:cms_quadrant_gems}) instrumented by triple-GEM detectors are being installed between the second and third LHC long shutdowns (LS2 and LS3) as part of the CMS phase-2 upgrade \cite{muon_ph2_tdr}. The first two GEM stations to be installed (GE1/1 and GE2/1) will complement the existing subdetectors of the muon endcap in providing muon tracking and triggering, allowing CMS to maintain its excellent $p_T$ resolution in the high-luminosity environment of HL-LHC; the third station, ME0, will extend the coverage of the muon system in the very forward pseudorapidity region 2.4$<|\eta|<2.8$.

Despite being based on the same working principle, triple-GEM detectors operated at present-generation experiments have undergone a sequence of design optimization relevant to their different scope of applications, as exemplified by the CMS case: the CMS GEM stations will be the largest GEM detectors among high-energy physics experiment, covering a total surface of over \SI{200}{\meter\squared} divided in modules of area between 0.3 and \SI{0.4}{\meter\squared} each. Therefore, within the CMS experiment, most of the past and ongoing R\&D on GEM detectors has been related to scaling the technology to large areas.

Furthermore, despite the excellent performance of present MPGDs, new applications in high energy physics experiments require the consolidation of the existing technologies and the development of new structures in order to cope with the long-term sustained operation in harsh environments, and to satisfy the more demanding physics programs.

After a succinct introduction to the working principle of GEM detectors, the following sections outline the challenges faced in the most recent applications of GEMs and the areas of improvement towards advanced GEM detectors for the High Luminosity LHC and experiments at future colliders, such as the proposed Muon Collider \cite{muon_collider}. Most of these topics are in a mature development stage and are either a natural adaptation of GEM detectors to required challenges faced by new environments or novel designs for application scenarios yet to be realized.

\section{Basic principle of a GEM detector}

\begin{figure}
    \centering
    \includegraphics[width=0.4\textwidth]{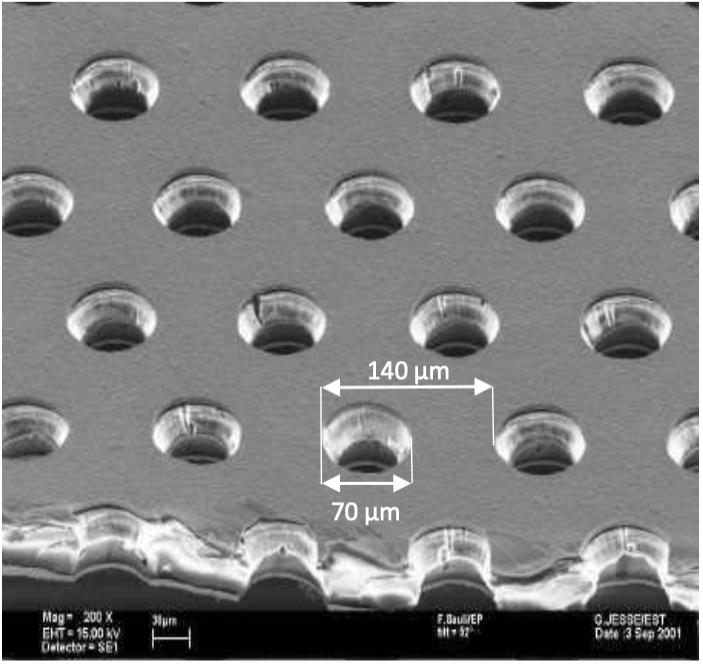}
    \includegraphics[width=0.45\textwidth]{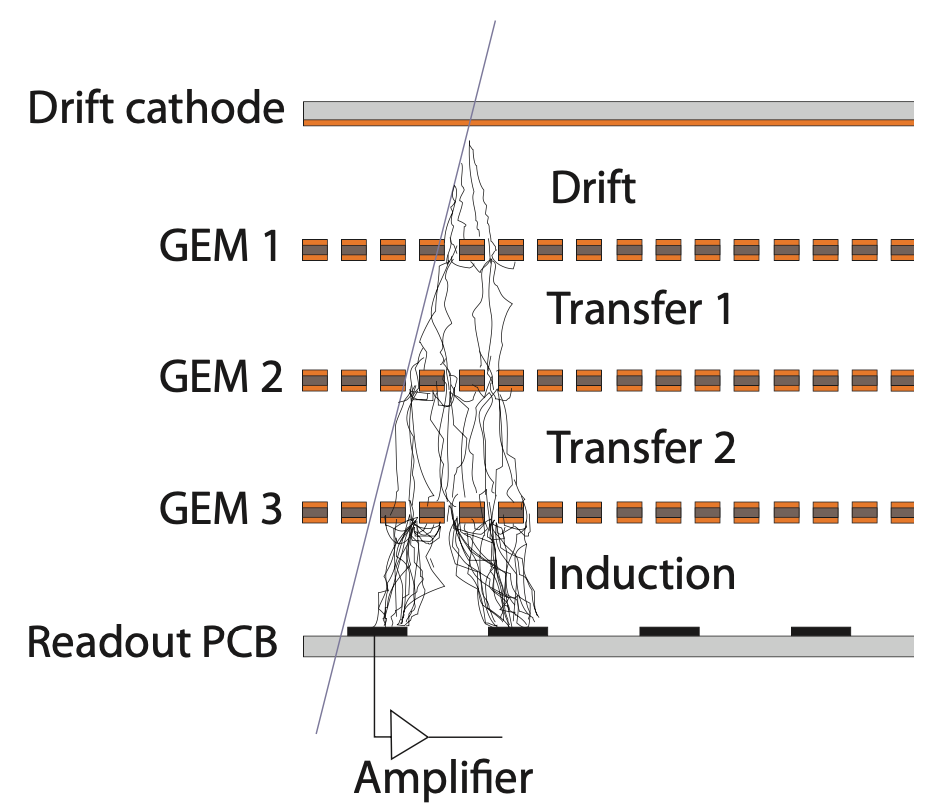}
    \caption{On the left, microscope picture of a GEM foil; on the right, layout of a triple-GEM detector.}
    \label{fig:gem_principle}
\end{figure}

The basic element of a GEM detector is a GEM foil, a flexible sheet made by a polyimide layer (typically \SI{50}{\micro\m} thick) clad on top and bottom by thin copper electrodes (\SI{5}{\micro\m}). The foil presents holes of typical internal diameter between \SI{50} and \SI{70}{\micro\m} and \SI{140}{\micro\m} pitch obtained by chemical etching (Fig.~\ref{fig:gem_principle}). The electric field arising in the holes when a voltage difference is applied to the electrodes allows to sustain a stable charge multiplications up to a factor of a few hundreds in typical counting gas mixtures.

Stacking multiple GEM foils in cascade allows to safely operate each foil as an independent amplification stage \cite{sauli_review}, with electric fields in the gaps guiding the charge in the avalanche from one foil to the next one (Fig.~\ref{fig:gem_principle} right). Detectors instrumented with a three-GEM amplification stack have become a widely adopted solution thanks to their excellent space resolution (of the order of \SI{100}{\micro\m}), an intrinsic rate capability higher than \SI{2}{\mega\Hz/\milli\m^2} over small areas and a good time resolution around 10 ns.

%\textcolor{blue}{GEM is good choice for particle tracking and identification, but also ...}

\section{GEM design optimization for high rate applications}

This section summarizes recent areas of improvements of the GEM technology for high-rate, large-area environments, as the ones prospected for the CMS GEM upgrade. Most challenges in such contexts are related to a characteristic trade-off, well-known for GEM detectors but more broadly driving many developments among MPGDs, between the detector robustness to discharges and the ability to maintain good efficiency under intense irradiation.

\subsection{Rate capability over large areas}

Triple-GEM detectors have been proven capable of locally sustaining extreme particle fluxes, with measurements performed with low energy x-rays on a surface of few \SI{}{\milli\m\squared} not showing significant gain drops up to up to \SI{2}{\mega\Hz/\milli\m\squared} \cite{jeremie_phd}. Such measurements are mostly sensitive to space charge effects, which are strongly suppressed in GEM detectors because of the spreading of the avalanche over several microscopic channels of amplification, defined by the GEM holes, but also thanks to the fast collection of the ions, which allows for the quick recovery of the gas neutrality. Under small-area irradiation above \SI{1}{\mega\Hz/\milli\m\squared}, space charge effects in triple-GEM detectors cause an increase of the effective gain due to a local distortion of the transfer electric fields, which improves the collection efficiency in the gaps \cite{thuiner_rate_capability}.

\begin{figure}[t]
    \centering
    \includegraphics[width=0.5\textwidth]{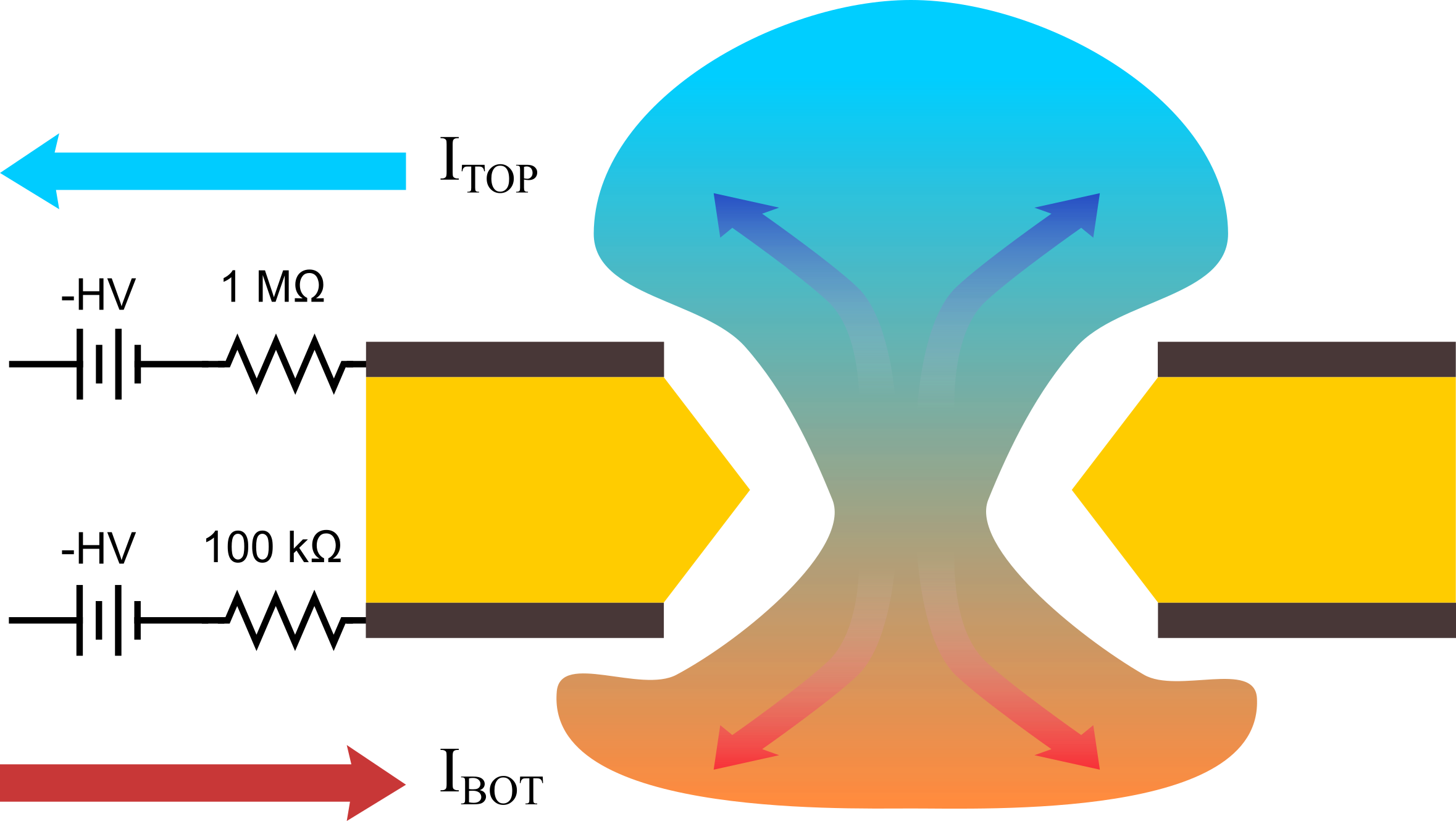}
    \caption{Avalanche in a GEM hole showing the currents induced by the moving charges in the gas flowing through the protection resistors. Such currents are responsible for the voltage drop on the electrodes that lower the effective gain under intense irradiation. In average, the current on the top electrode flows from the electrode to the power supply, as it is induced by ions approaching the electrode, while the bottom current has opposite direction being mostly due to electrons being collected. The actual voltage on the top electrode is then higher than the one applied from the power supply, while the bottom voltage is lower, resulting in a lower voltage difference (and thus a lower amplification) across the GEM foil.}
    \label{fig:gem_rate_capability_avalanche}
\end{figure}

However, in large area detectors the main limit to the rate capability is an \emph{ohmic} effect due to the presence of discharge protection circuits decoupling the GEM electrodes from the high voltage power supply: the moving charges in the avalanche (Fig.~\ref{fig:gem_rate_capability_avalanche}) induce currents on the electrodes that, flowing through the protection resistors, reduce the voltage on the electrodes with respect to the power supply. Such effect is negligible in small-area detectors (or, equivalently, in measurements on large-area detectors with small irradiation area), where the fraction of the charges captured by the GEM electrodes generates a current of the order of some \SI{}{\nano\ampere}; however, being a collective effect over the entire GEM foil it can be observed by irradiating the entire detector surface with moderate particle fluxes (a few tens of \SI{}{\kilo\Hz/\centi\m\squared}) \cite{fallavollita_me0}.

\begin{figure}[t]
    \centering
    \includegraphics[width=0.46\textwidth]{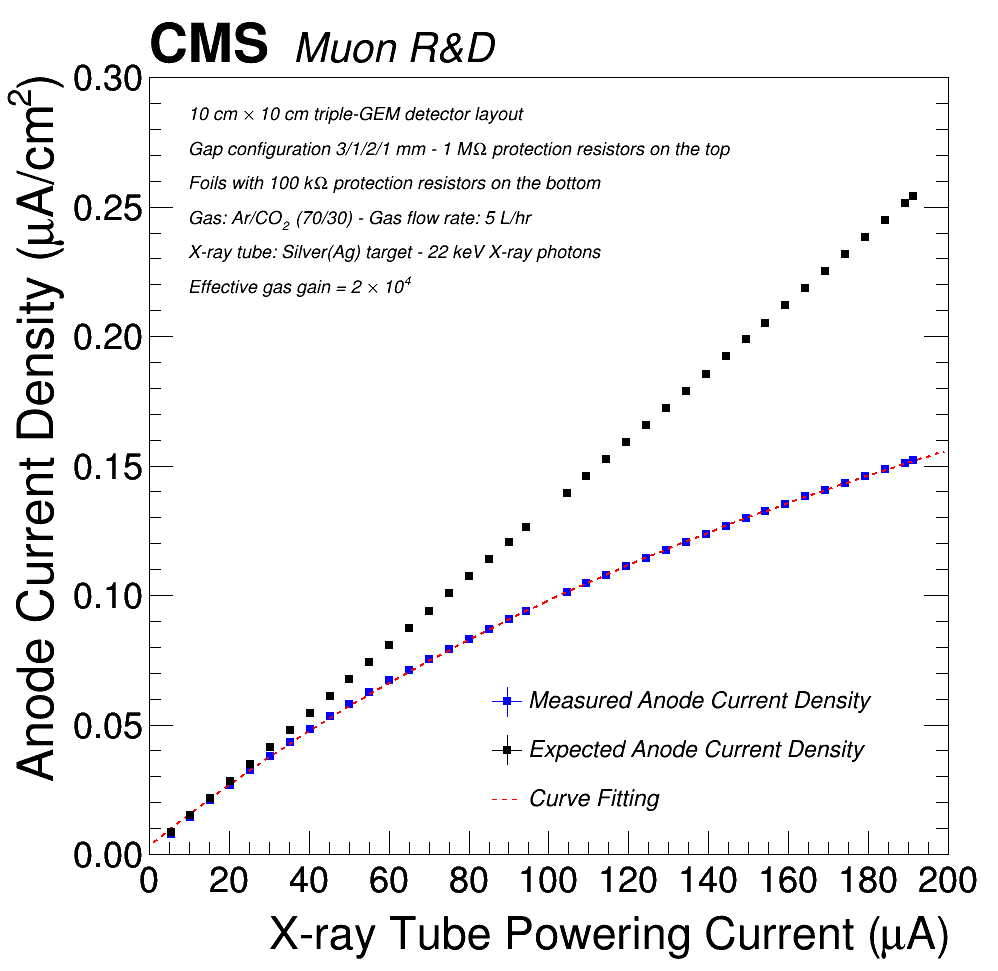}
    \includegraphics[width=0.53\textwidth]{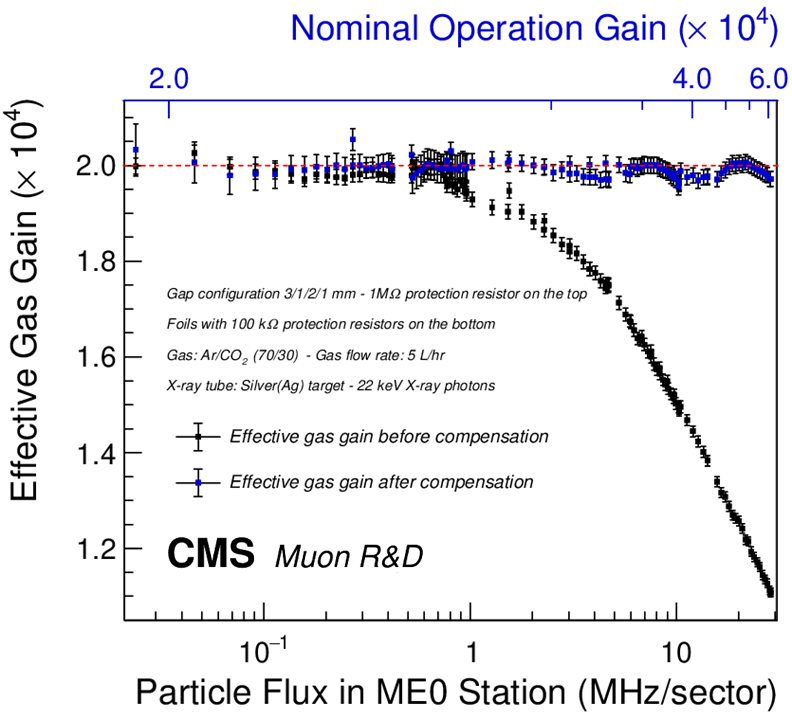}
    \caption{On the left, linearity curve of a $10\times$\SI{10}{\centi\m^2} triple-GEM detector irradiated by an x-ray generator. The anode current bends at increasing radiation flux, showing the gain drop at high particle rates. On the right, rate capability curve of a $10\times$\SI{10}{\centi\m^2} triple-GEM detector irradiated by an x-ray generator (black points). The blue points show the recovered effective gain of the detector (of \SI{2e4}) obtained by operating it under irradiation at increasing bias voltage. The blue axis shows the nominal gain at which the detector has to be operated in order to maintain a fixed gain of \SI{2e4} under irradiation. Image from F. Fallavollita et al. \cite{fallavollita_me0}.}
    \label{fig:cms_gem_rate_capability}
\end{figure}

The ohmic effect has been proven to be dominant in the rate capability of triple-GEM detector in large-area irradiation by a set of independent rate capability measurements \cite{pellecchia_me0}. The linearity curve of a $10\times$\SI{10}{\centi\m\squared} detector shows the gain saturation at increasing incident particle rate (Fig.~\ref{fig:cms_gem_rate_capability} left). The effect is observed to be dependent on the value of the protection resistors and on the average primary charge per background event; for different values of irradiated areas, the gain drop is dependent on the total background particle rate rather than on the particle flux (i.e. rate per unit surface). For a prototype instrumented with \SI{1}{\mega\ohm} protection resistors on each GEM top electrode, the gain drop under irradiation by a low-energy x-ray source is found to be 10\% at a total hit rate of \SI{2}{\mega\Hz} (Fig.~\ref{fig:cms_gem_rate_capability} right).

The same studies have shown that the original gain of the detector can be recovered by operating it at increasing bias voltage on each electrode, compensating the voltage drop due to the irradiation. As a downside for such recovery mechanism, the bias voltage required under irradiation increases more than linearly with the background particle rate, as a consequence of the positive effect loop between the nominal GEM foil amplification and the background current induced on the electrodes. This limits the scope of application of the gain compensation as the only mitigation mechanism in experiments, due to the high risk of detector damage in case of sudden beam loss. An optimization of the detector design is necessary to ensure safe operations while maintaining high efficiency.

The rate capability issue is of particular concern in the case of the ME0 station at the CMS experiment, scheduled for installation during the third LHC long shutdown in 2025. The ME0 detector will be the closest to the LHC beam line among the CMS muon stations (see Fig.~\ref{fig:cms_quadrant_gems}); among the high-rate detectors based on MPGDs at the major LHC experiments (with a background rate of \SI{150}{\kilo\Hz/\centi\m\squared} in the highest pseudorapidity region, Tab.~\ref{tab:lhc_experiments_rate}), ME0 is going to be the station covering the largest area.

\begin{table}[h]
    \centering
    \begin{tabular}{lcccccc}
         \hline
          & LHCb & ATLAS & ALICE & CMS & CMS & CMS  \\ \hline
         \textbf{Station} & M1 & NSW & TPC & GE1/1 & GE2/1 & ME0 \\
         \textbf{Technology} & 3-GEM & MicroMegas & 4-GEM & 3-GEM & 3-GEM & 3-GEM \\
         \textbf{Module area (cm$^2$)} & 480 & 20000 & 3000 & 4000 & 4000 & 3000 \\
         \textbf{Max. rate (kHz/cm$^2$)} & 600 & 15 & 100 & 5 & 1.5 & 150 \\ \hline
    \end{tabular}
    \caption{Comparison of the expected or measured rates in detectors instrumented by different MPGD technologies at the four major LHC experiments, together with the surface of a single detector module.}
    \label{tab:lhc_experiments_rate}
\end{table}

\begin{figure}[h]
    \centering
    \includegraphics[width=0.49\textwidth]{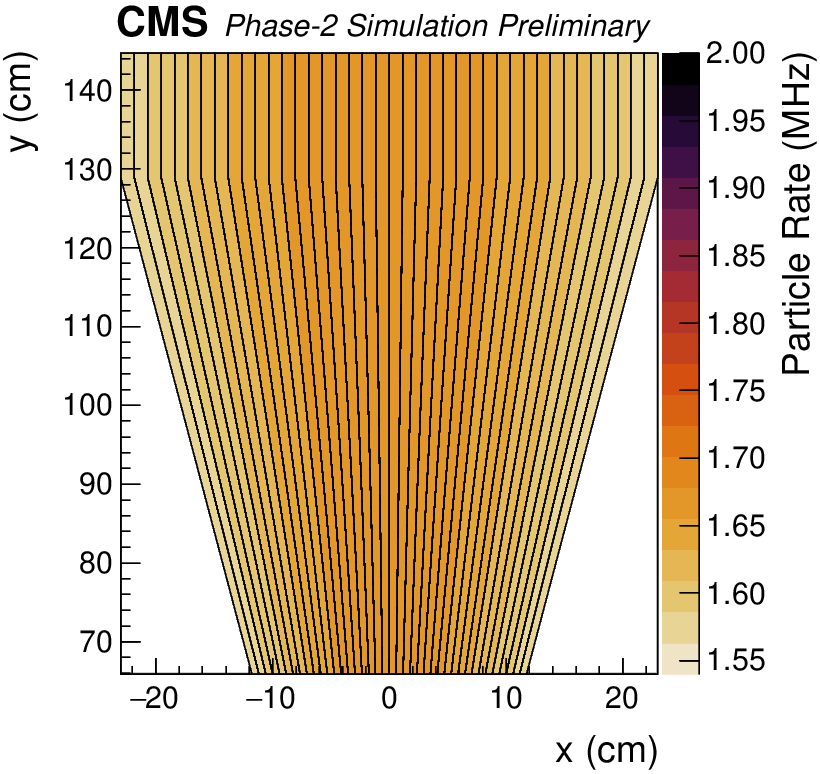}
    \includegraphics[width=0.49\textwidth]{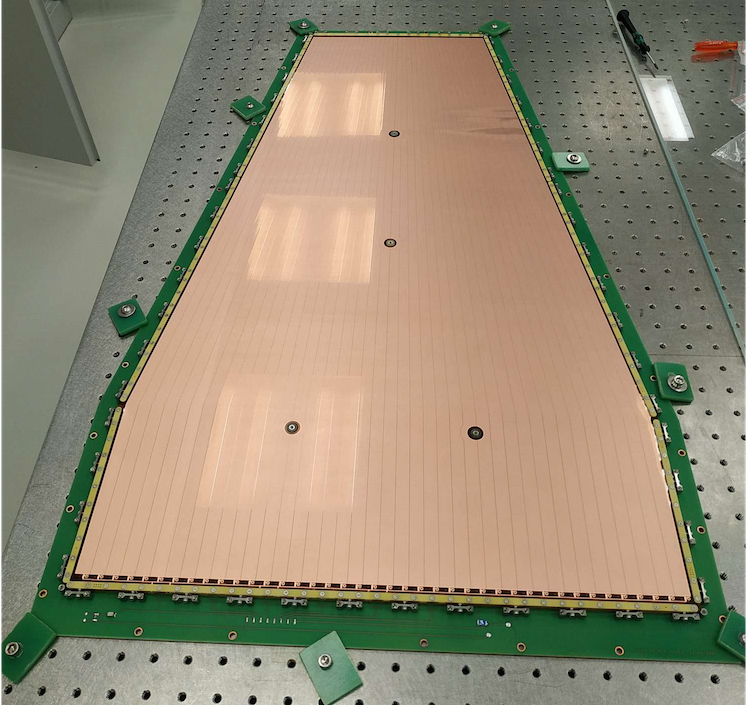}
    \caption{On the left, expected background particle rate per sector in a single ME0 module with a 40-sector azimuthal segmentation. On the right, picture of a stack of GEM foils with azimuthal segmentation during the assembly of an ME0 module.}
    \label{fig:cms_me0_segmentation}
\end{figure}

The chosen mitigation strategy under ongoing investigation for the ME0 rate capability issue consists of dividing the GEM foils in independently powered sectors in the azimuthal direction with respect to the LHC beam line (Fig.~\ref{fig:cms_me0_segmentation}). Such a solution allows to limit the average gain drop across the entire chamber under irradiation to 5\% of the nominal gain of \SI{2e4}. Such design choice has yet to be validated taking into account the additional effect of a resistive high-voltage filter -- included to limit the noise observed by the readout electronics -- which increases the total resistance between the power supply and the GEM electrodes. Besides the HL-LHC applications, the use of GEM-based technologies in future HEP experiments will require the development of new design solutions, detector concepts and new materials to ensure more reliable systems while pushing forward the rate capability.

\subsection{Discharge propagation and long-term protection}

Besides the ability of operating in stable conditions in high background environments, detector are also required to sustain heavy ionising interactions without suffering punctual or long-term degradation. 
In GEM-based detectors for example, the electrical breakdown of the gas between the amplifying electrodes can be induced by an excess of charge carriers in the electron avalanche. The amount of energy released during this process is typically of the order of 1 mJ and can be sufficient to permanently damage the inner geometry of the GEM holes, provoking short-circuits that could prevent the normal operation of the entire detector. This phenomenon, often referred as ”discharges”, remains one of the major limitations of current MPGDs and a possible show-stopper for their use in very high-rate environments.

\subsubsection{Discharge probability and consequences}
 \label{sec:discharge_studies}

The probability of triggering discharges in GEM-based structures was extensively studied by the MPGD community over the past decades \cite{Sauli} \cite{Bashmann}. General recommendations were established in order to minimize this effect by defining optimal design configurations and operation settings. The core protection apparatus, which became standard for most of the large detector systems, essentially consists of three components: the use of quenching molecules in the gas mixture to prevent the formation of secondary avalanches; the segmentation of the GEM top electrode into multiple small regions separated via a protection resistors in order to minimize the discharge energy; the distribution of the gas gain over several GEM layers with asymmetric HV settings to keep the charge density below the critical limit.

Together with the progressive improvement of the GEM production techniques, which offers higher quality GEMs with cleaner geometry, the implementation of such design constraints with specific discharge protection circuits has led to a significant reduction of the discharge probability with respect to the early observations \cite{Bashmann}, as shown in Tab.~\ref{tab:DU_limit} with the comparison of a newly produced small size detector \cite{GEMTDR} and large size chambers developed for the CMS \cite{KCMSW} and ALICE experiments \cite{ALICETDR}\cite{ALICED}.

\begin{table}[h]
    \centering
\begin{tabular}{ |p{3cm}||p{3cm}|p{3cm}|p{3cm}|  }
 \hline
 \multicolumn{4}{|c|}{Discharge Probability with $^{241}Am$  ($E_{\alpha}=5.5 MeV$)} \\
 \hline
 Detector system & $10\times10$ $cm^2$ CMS & CMS GE1/1 & ALICE IROC \\
  & triple-GEM & triple-GEM & quadruple-GEM\\
 \hline
Gas   & $Ar/CO_2$   & $Ar/CO_2$  &   $Ne/CO_2/N_2$ \\
 & (70:30)    & (70:30) &   (90:10:5)\\
 \hline
 Upper limit &   $< 8 \times 10^{-11}$  & $< 9 \times 10^{-10}$   & $< 1.5 \times 10^{-10}$\\
 \hline
\end{tabular}
  \caption{Comparison of the upper limit of the discharge probability with alpha particles for $10\times10$ $cm^2$ triple-GEM \cite{GEMTDR}, CMS GE1/1 triple-GEM \cite{KCMSW} and ALICE IROC quadruple-GEM \cite{ALICETDR}.}
  \label{tab:DU_limit}
\end{table}

More recently, additional studies were conducted in the framework of the CMS GEM upgrade project in order to evaluate more precisely the consequences of discharges at a microscopic level and to determine the critical value for the protection resistors \cite{SingleHole}. The measurements, based on a previous work from the LHCb GEM group \cite{Cardini}, were taken with a single-GEM hole design connected to a custom circuitry in order to control the discharge energy and the protection resistance. The single holes were confronted to multiple discharges until the GEM entered in short circuit.

\begin{figure}[!h]
    \centering
    \includegraphics[width=0.6\textwidth]{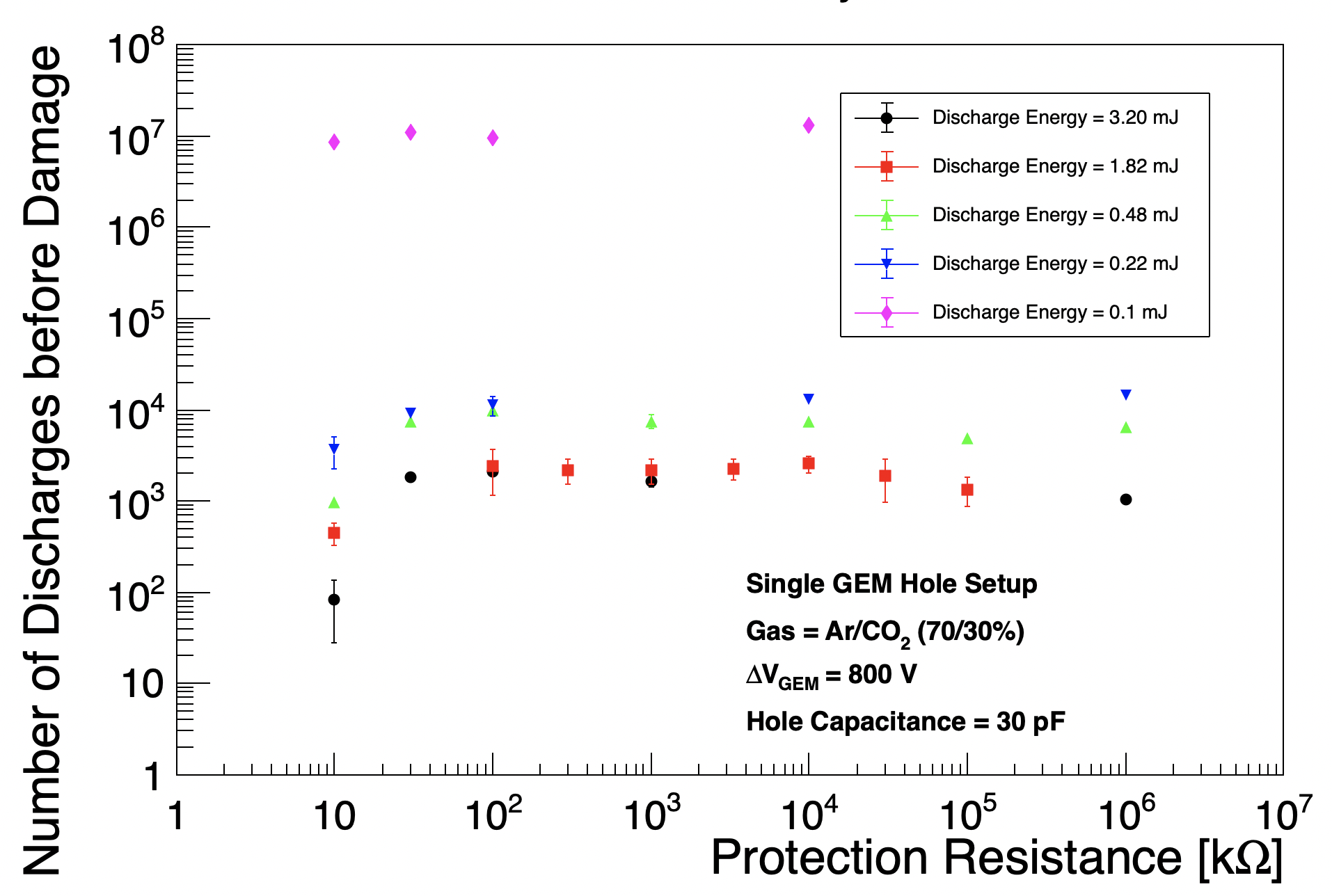}
    \caption{Longevity of a single GEM hole as a function of the protection resistance. The longevity is defined as the number of discharges sustained by the hole before a short circuit is created. The tests were performed in $Ar/CO_2$ (70:30)\cite{SingleHole}.}
    \label{fig:Single_Hole_Resistance}
\end{figure}

The results shown in Fig.~\ref{fig:Single_Hole_Resistance} indicate that a protection resistor of 30 kOhm is sufficient to effectively de-couple the energy stored in the GEM segment from the one stored in the power supply or in the neighbour segments. This gives the possibility to reduce the individual protection resistors in next-generation detectors in order to improve the overall rate capability while maintaining an optimal protection against energetic discharges.

\begin{figure}[!h]
    \centering
    \includegraphics[width=0.5\textwidth]{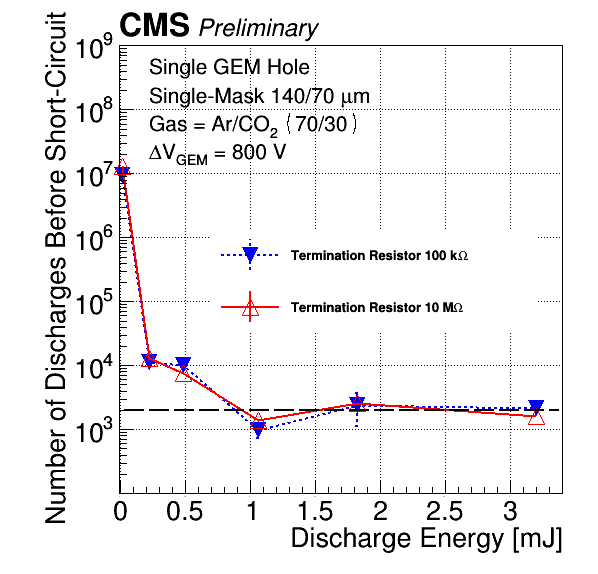}
    \caption{Longevity of a single GEM hole as a function of the discharge energy. The longevity is defined as the number of discharges sustained by the hole before a short circuit is created. The tests were performed in $Ar/CO_2$ (70:30) with two values of protection resistance connecting the top electrode of the GEM and the power supply\cite{SingleHole}.}
    \label{fig:Single_Hole_Energy}
\end{figure}

On Fig.~\ref{fig:Single_Hole_Energy} is reported the average number of discharge accumulated in a single GEM hole before it turns into a short circuit, as a function of the discharge energy. This parameter, which is defined as the discharge longevity of the hole, is significantly improved for energies below 0.5 mJ while it tends to stabilize above 1 mJ at about 2000 discharges.

\begin{figure}[!h]
    \centering
    \includegraphics[width=0.99\textwidth]{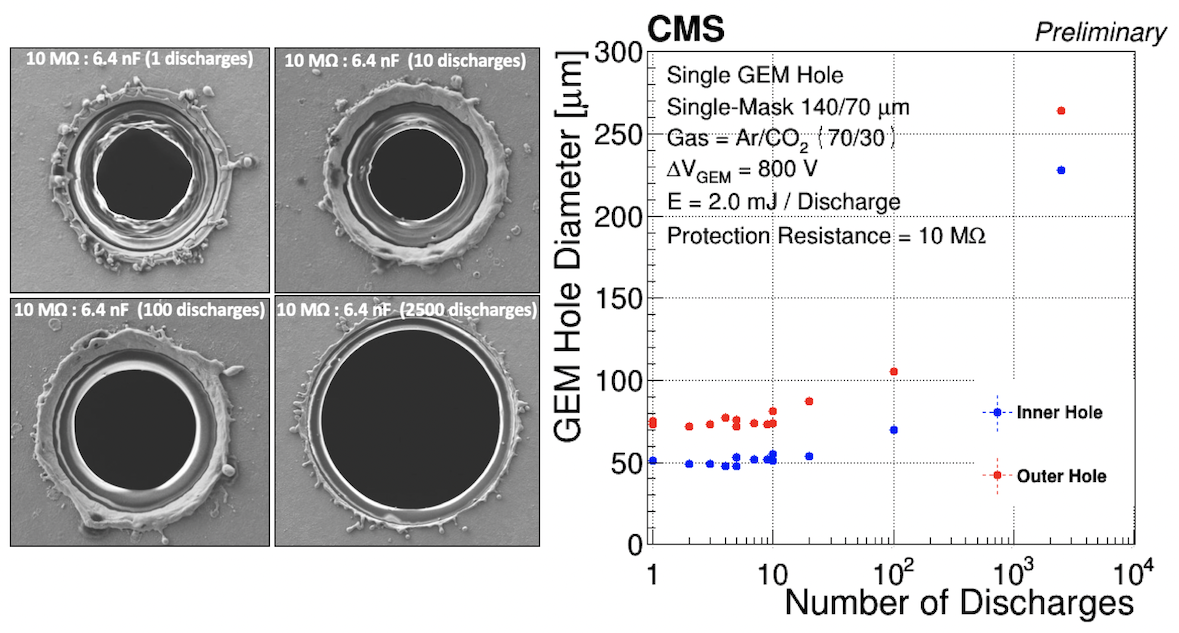}
    \caption{Left: SEM images of single GEM holes subject to 1, 10, 100, and 2500 discharges with an average energy of 2 mJ. Right: Evolution of a GEM hole inner and outer diameters as a function of the number of accumulated discharges~\cite{SingleHole}.}
    \label{fig:Single_Hole_Diameter}
\end{figure}

The GEM samples tested above were also systematically inspected with a SEM microscope. The observations reported on Fig.~\ref{fig:Single_Hole_Diameter} clearly indicate an increase of the hole inner and outer diameters caused by the accumulation of energetic discharges. This effect it due to the plasma etching of the polyimide material together with a vaporization of the copper rim on both sides of the GEM. As a consequence, a GEM foil subject to frequent discharges might show a degradation of it's amplification power and therefore a loss of detection performance with time.

\begin{figure}[!h]
    \centering
    \includegraphics[width=0.6\textwidth]{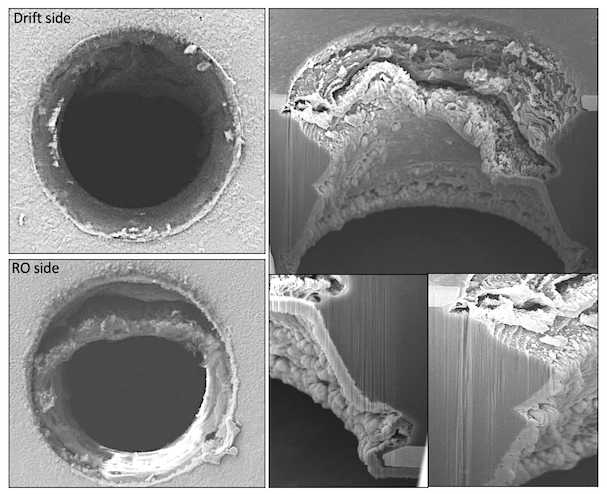}
    \caption{SEM images of a single GEM hole damaged by the accumulation of destructive discharges. The hole in in short-circuit state due to the internal degradation of it's structure and the depositing of carbon-based material on the inside walls~\cite{SingleHole}.}
    \label{fig:Single_Hole_Short}
\end{figure}

Fig.~\ref{fig:Single_Hole_Short} shows an example of a hole damaged by the accumulation of energetic discharges. The internal structure of the polyimide is clearly compromised by the presence of cracks and the uneven deposition of carbon-based material resulting from the energy transfer. The hole is in short-circuit state which make it unusable for particle detection.

The practical experience highlighted by this study brings a new perspective on the process of discharge formation and gives the possibility to further optimize the detector design to reduce the probability of destructive event.

\subsubsection{Discharge propagation}

\begin{figure}[!h]
    \centering
    \includegraphics[width=0.6\textwidth]{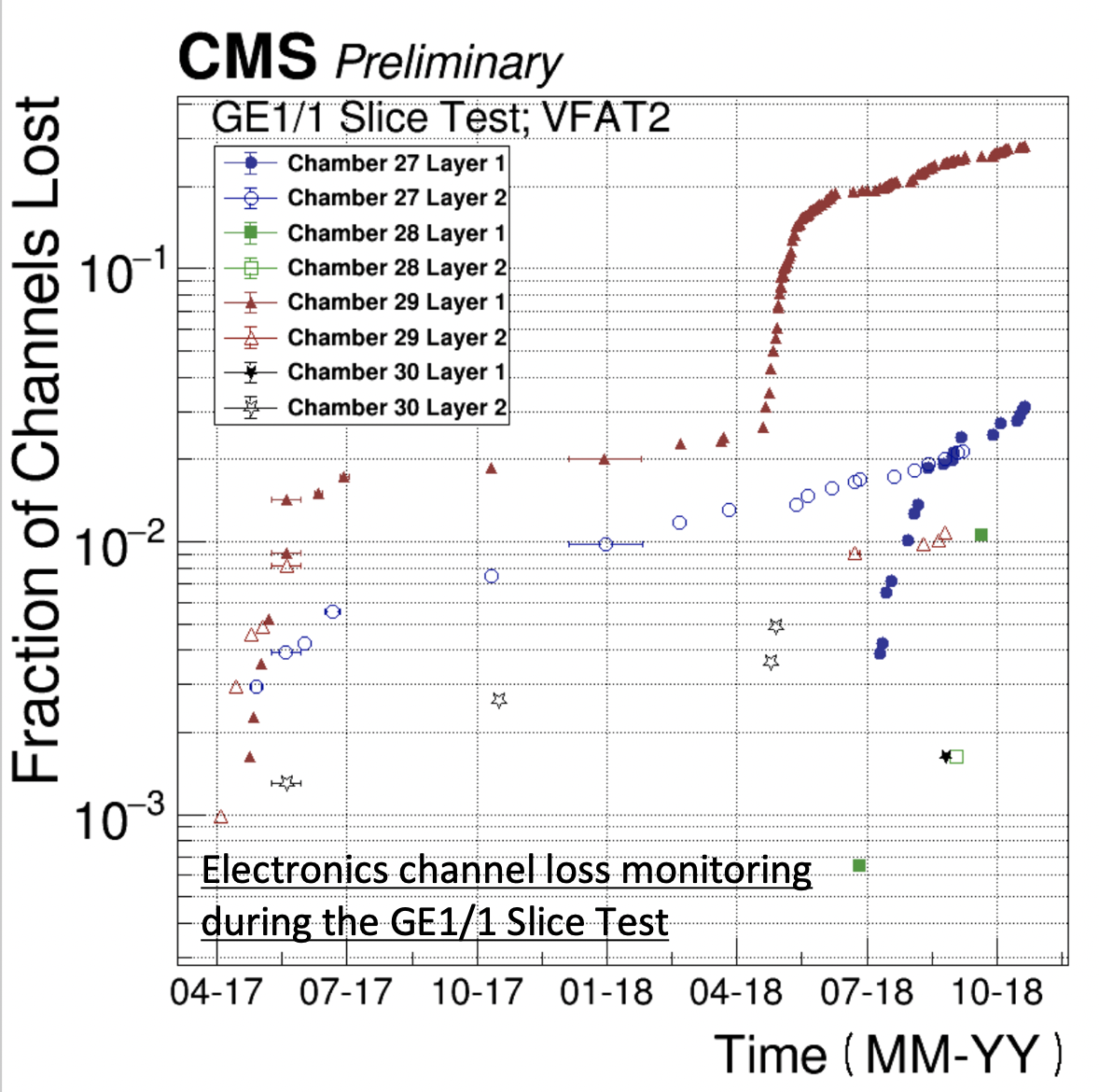}
    \caption{Evolution of the electronics channel loss during the CMS GE1/1 Slice test\cite{SliceTest}.}
    \label{fig:Slice_Test}
\end{figure}

Under certain conditions, a GEM discharge can further propagate between GEM foils and potentially reach the readout board and compromise the electronics integrity. In such cases, the amount of energy transferred to the electronics can exceed the typical values for which the readout protection is designed for and therefore cause permanent damages to the chips. Fig.~\ref{fig:Slice_Test} shows the evolution of the number of dead electronics channels in the eight GE1/1 Slice Test detectors installed in CMS in 2016. Some chambers, subject to frequent discharges, lost up to 30\% of their detection capability because of the propagation of the discharge energy toward the readout board~\cite{SliceTest}. 

The process of discharge propagation, even though observed in the early stage of the GEM development \cite{Bashmann}, was not fully understood until recent investigations conducted jointly by ALICE \cite{Zagreb}, CMS \cite{Propa} and with the support of the RD51 collaboration.

\begin{figure}[!h]
    \centering
    \includegraphics[width=0.9\textwidth]{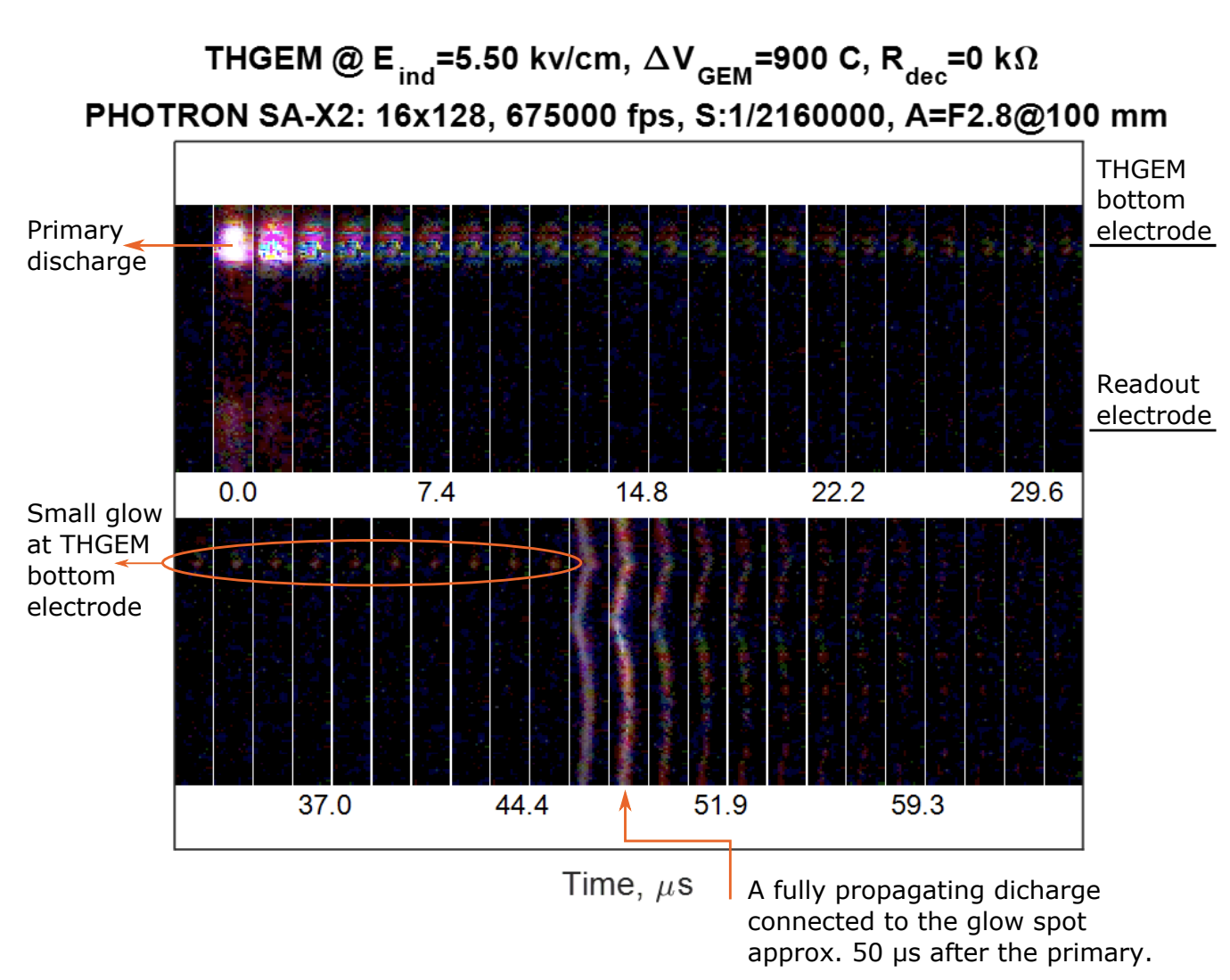}
    \caption{A fast camera recording of the delayed discharge propagation with a THGEM foil at an induction field value of 5.5 kV/cm without a decoupling resistor in $Ne-CO_2-N_2$ (90:10:5)  \cite{Zagreb}.}
    \label{fig:Propa_Glow}
\end{figure}

When a discharge develops across a GEM hole, it creates a plasma that can heat up the copper rims on both top and bottom electrodes. These hot spots would typically take several tens of microseconds to cool down to normal temperature. During this period, we can observe a thermionic emission of electrons in the gas, which generates a current between the GEM copper layers and the neighbour electrodes. This effect is further enhanced by the presence of a strong electric field in the gaps between the GEMs, as described by the Schottky effect \cite{Murphy}. The evidence \cite{Zagreb} of the hot metallic rims is given by the presence of a thermal glow on both top and bottom electrodes after the primary discharge (see Fig.~\ref{fig:Propa_Glow}). This hypothesis is also consistent with the earlier observations of melted copper in the vicinity of the discharging hole (see Fig.~\ref{fig:Single_Hole_Diameter}).

\begin{figure}[!h]
    \centering
    \includegraphics[width=0.7\textwidth]{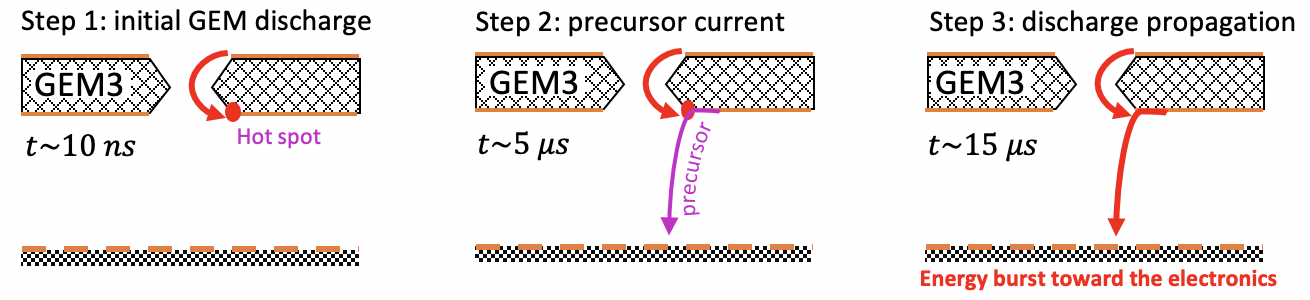}
    \caption{Schematic view of the discharge propagation process showing: 1. the initial formation of the primary discharge; 2. the development of the precursor current; 3. the ignition of the propagated discharge \cite{Propa}}
    \label{fig:Propagation_Small}
\end{figure}

The intensity of the thermionic current strongly depends on the electric field and the capacitance of the gap. In high field environments, typically above 5 - 7 kV/cm, the current tends to grow and convert into a streamer that can trigger a secondary discharge in the gap. This secondary discharge, also called propagated discharge, can itself re-boost the thermionic emission and initiate additional discharges. A schematic overview of the discharge propagation process is shown on Fig.~\ref{fig:Propagation_Small}.

\begin{figure}[!h]
    \centering
    \includegraphics[width=0.99\textwidth]{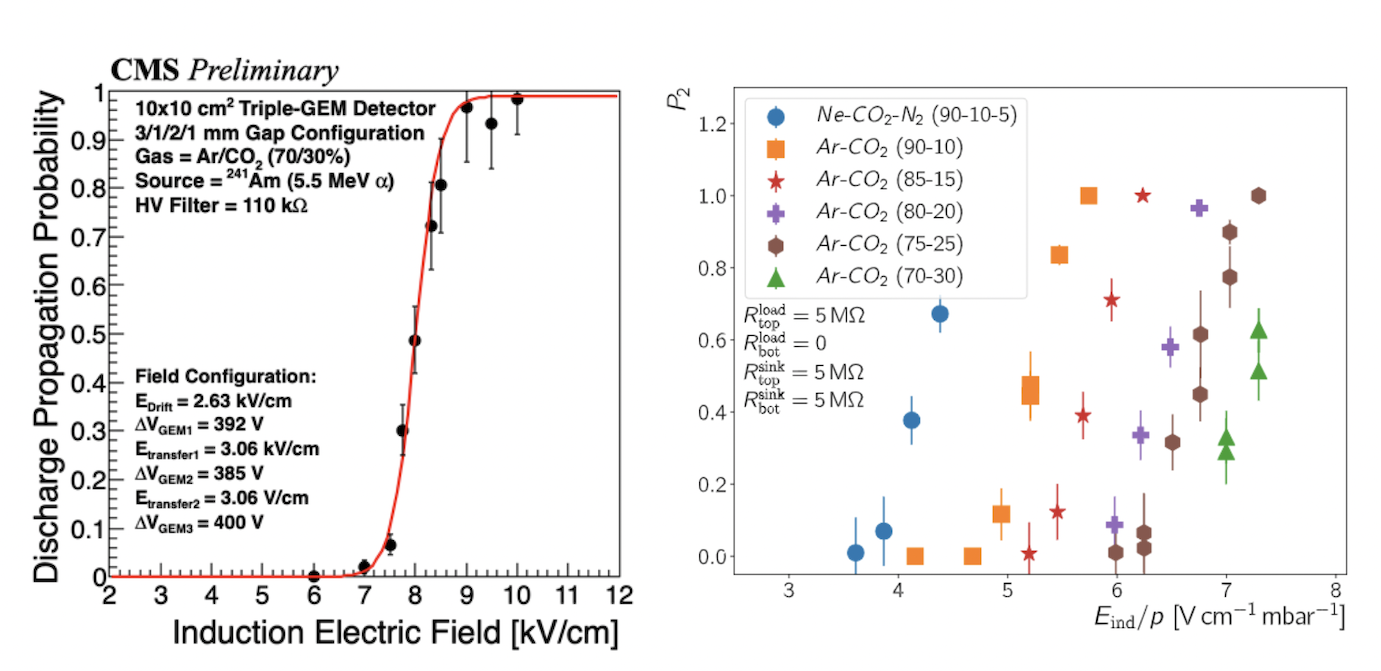}
    \caption{Left: discharge propagation probability in a $10\times10$ $cm^2$ triple-GEM detector as a function of the induction electric field \cite{Propa}. Right: secondary discharge probability $P_2$ as a function of the reduced induction field for different gas mixtures \cite{Deisting}.}
    \label{fig:Propa_Field}
\end{figure}

The probability of the triggering discharge propagation as a function of the induction electric field is reported on Fig.~\ref{fig:Propa_Field}.

\begin{figure}[!h]
    \centering
    \includegraphics[width=0.99\textwidth]{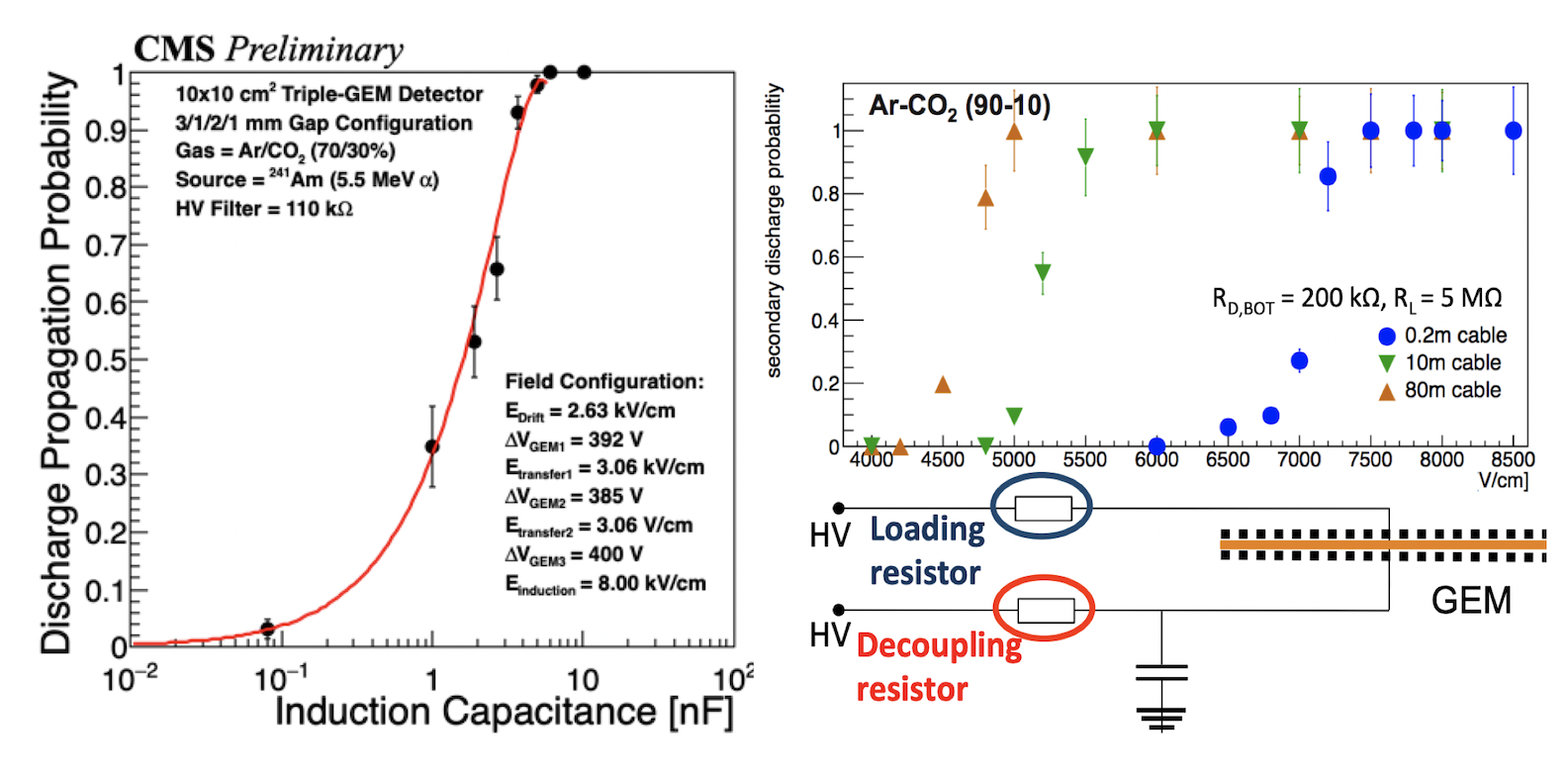}
    \caption{Left: discharge propagation probability in a $10\times10$ $cm^2$ triple-GEM detector as a function of the induction capacitance \cite{Propa}. Right: secondary discharge probability as a function of the induction field for different values of parasitic capacitance in the induction gap \cite{Deisting}.}
    \label{fig:Propa_Capa}
\end{figure}

\begin{figure}[!h]
    \centering
    \includegraphics[width=0.99\textwidth]{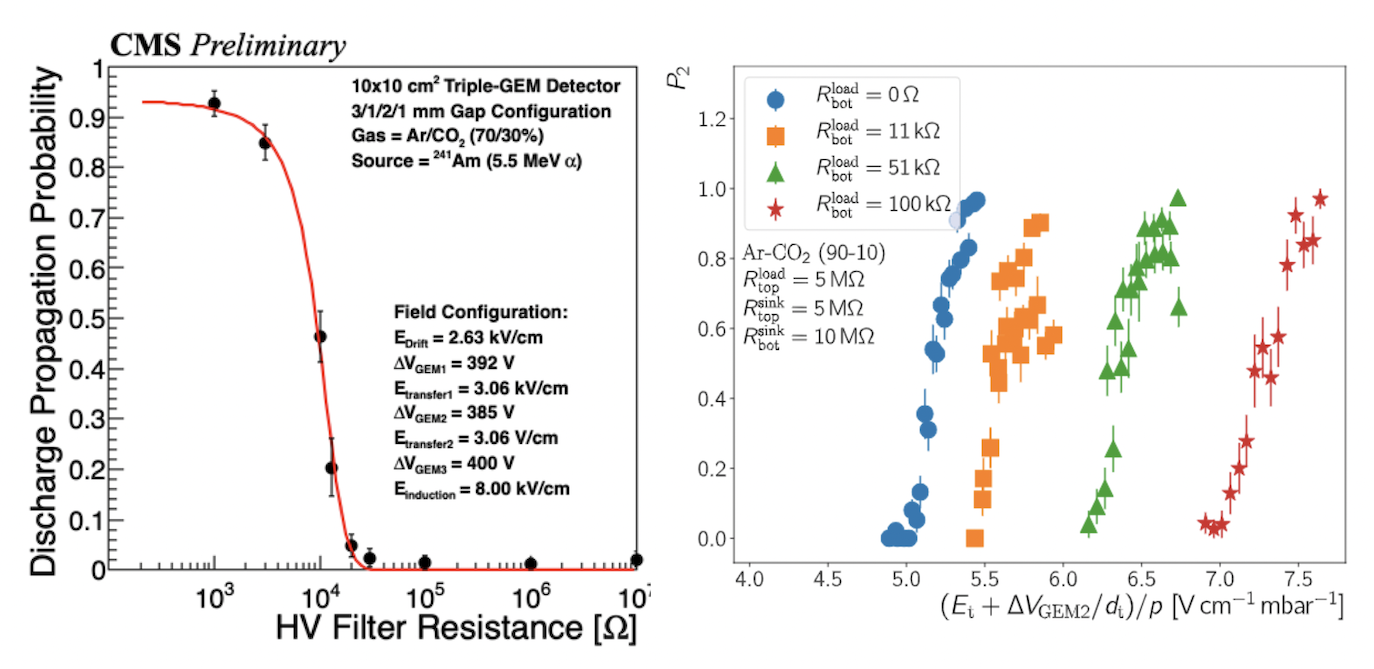}
    \caption{Left: discharge propagation probability in a $10\times10$ $cm^2$ triple-GEM detector as a function of the induction capacitance \cite{Propa}. Right: secondary discharge probability as a function of the reduced induction field for different values of filter (load) resistors \cite{Deisting}.}
    \label{fig:Propa_Filter}
\end{figure}

Additionally, the use of a protection resistor on the foil top electrode helps to limit the energy transfer and therefore prevent the growth of the precursor current. The propagation probability in particular gets significantly reduced for filter values above 30 k$\Omega/$, as reported on Fig.~\ref{fig:Propa_Filter}.

\begin{figure}[!h]
    \centering
    \includegraphics[width=0.99\textwidth]{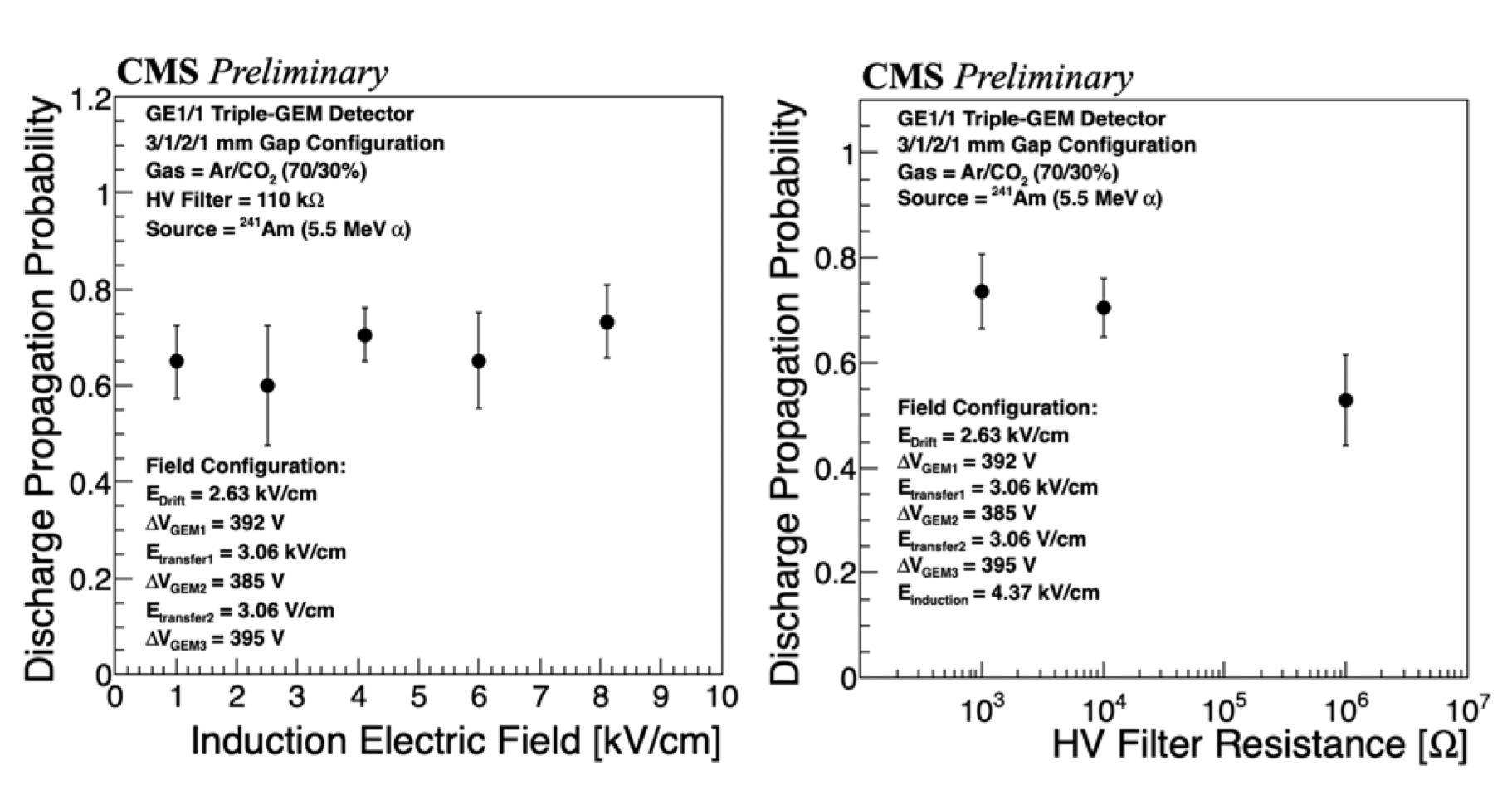}
    \caption{Discharge propagation probability in a GE1/1 CMS detector as a function of the induction capacitance (left) and the filter resistance (right) \cite{Propa}.}
    \label{fig:Propa_Large}
\end{figure}

In large detectors however, the discharge propagation process is significantly enhanced by the large capacitance of the gaps between the GEM foils. The correlation between the propagation probability and the induction capacitance is shown on Fig.~\ref{fig:Propa_Capa}. The size of the foil and the thickness of the gap are critical parameters, but parasitic capacitance introduced by HV filters or HV cables on the GEM bottom electrode could also affect the detector behavior. The example of the discharge propagation behavior in the CMS GE1/1 chambers, given on Fig.~\ref{fig:Propa_Large}, demonstrates that in large detectors the propagation probability is not sensitive anymore to the induction field, and the use of a strong de-coupling filter only has a minor effect.

\begin{figure}[!h]
    \centering
    \includegraphics[width=0.99\textwidth]{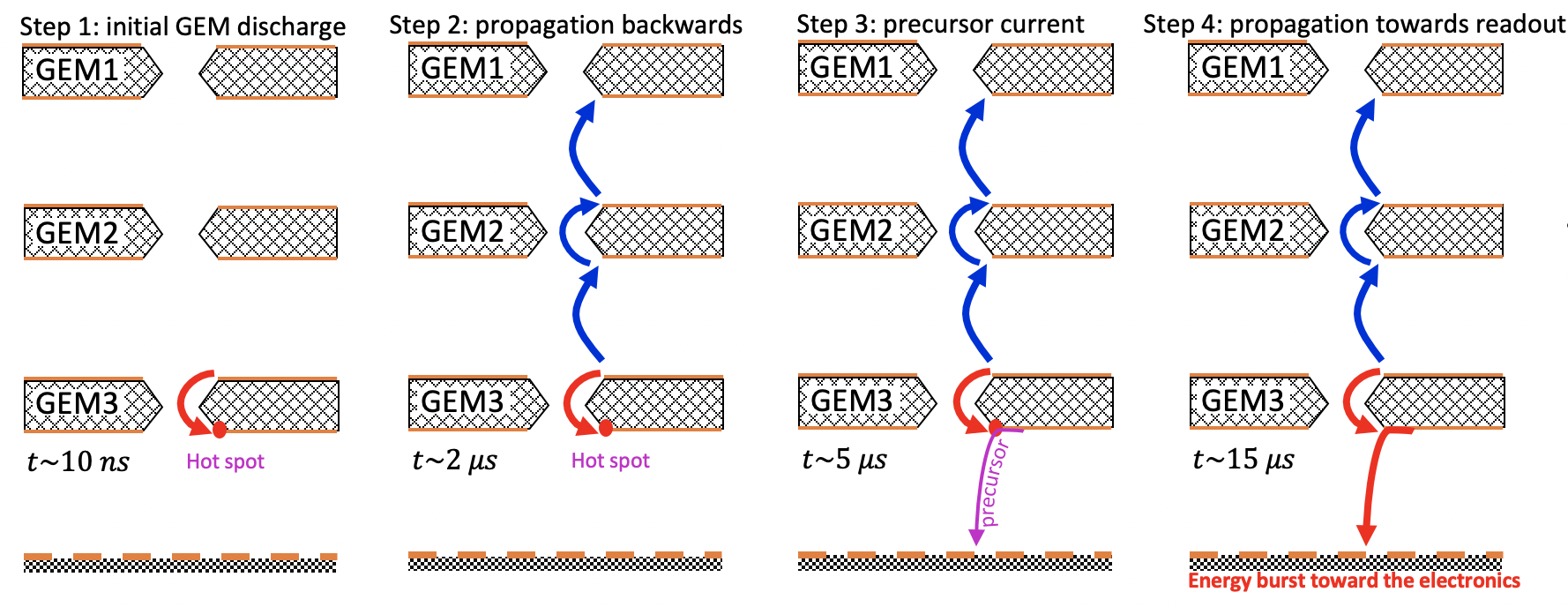}
    \caption{Schematic view of the discharge propagation process in a large GE1/1 CMS detector: 1. the initial formation of the primary discharge in GEM3; 2. backward propagation involving GEM1 and GEM2; 3. formation of the precursor current in the induction gap; 4. propagation of the discharge toward the readout board. \cite{Propa}.}
    \label{fig:Propa_Process_Large}
\end{figure}

Further studies have shown that in this configuration, the discharge propagation involves almost systematically all the three GEM foils, not only the last GEM facing the readout board. The process, described on Fig.~\ref{fig:Propa_Process_Large}, is more complex and problematic than in smaller detectors. In such conditions, the mitigation of the discharge propagation cannot be achieved only by the fine-tuning of the detector configuration, but it requires instead the re-design of the multi-GEM concept. In particular, the implementation of the HV segmentation on both sides of the foils is an effective way for reducing the gap capacitance and for improving the de-coupling of the discharge energy.

\begin{figure}[!h]
    \centering
    \includegraphics[width=0.99\textwidth]{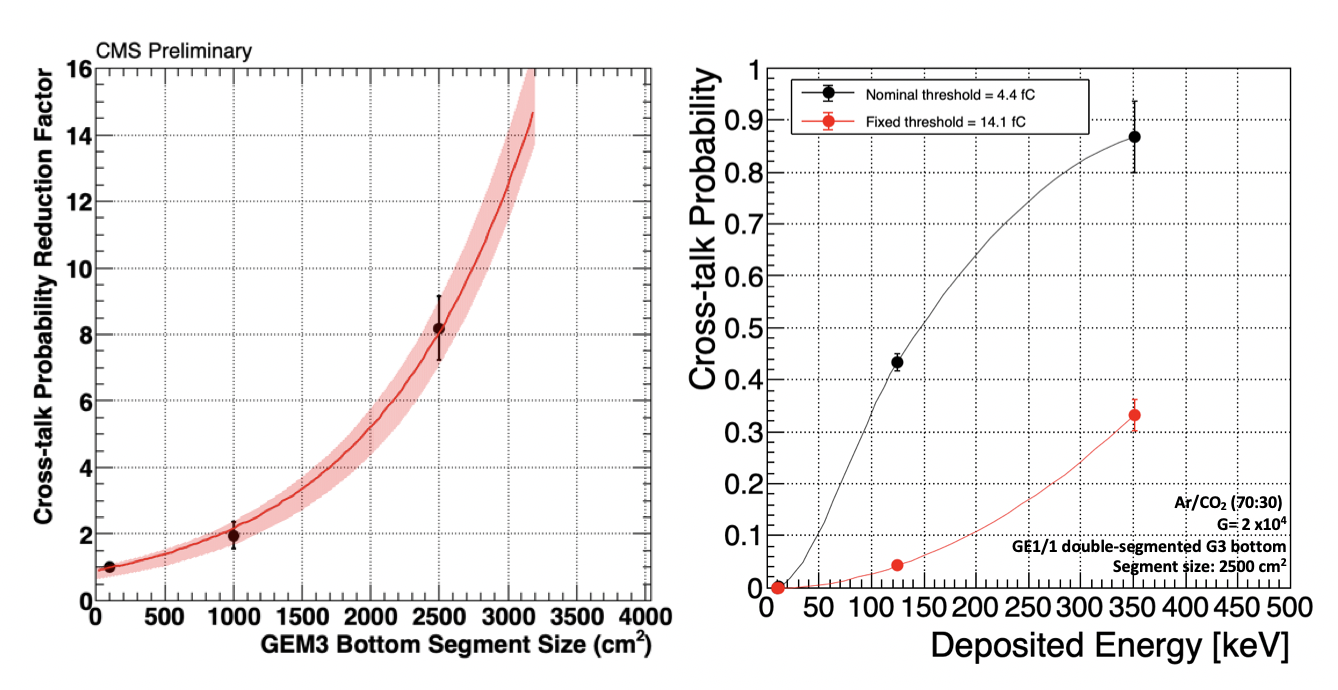}
    \caption{Left: Normalized cross-talk reduction factor as a function of the GEM bottom segment size. Right: Cross-talk probability as a function of the energy deposited in the detector gap for two values of electronics threshold \cite{Propa}.}
    \label{fig:CrossTalk}
\end{figure}

Nevertheless, the use of a fine HV segmentation on the bottom side of the GEM facing the readout strips tends to significantly increase the probability of generating parasitic cross-talk when the detector is subject to heavy ionizing particle (see Fig.~\ref{fig:CrossTalk}).

\begin{figure}[!h]
    \centering
    \includegraphics[width=0.90\textwidth]{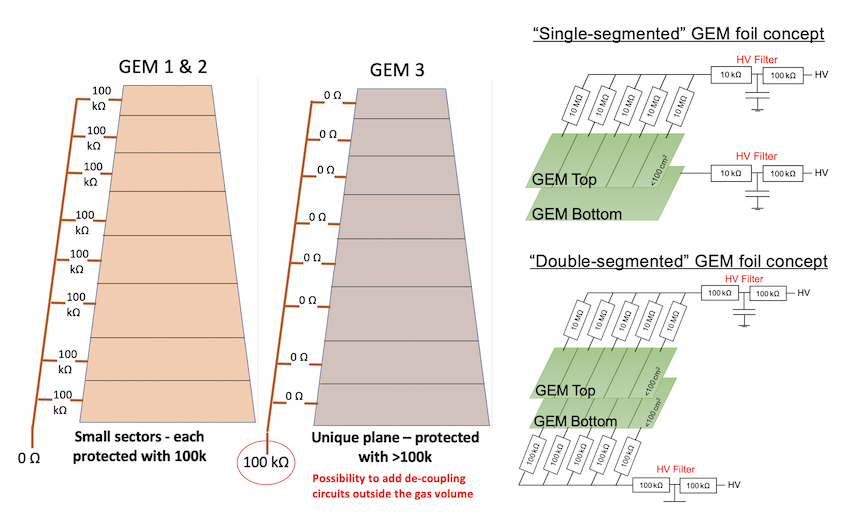}
    \caption{Overview of the "Mixed" design configuration for discharge propagation and cross-talk mitigation in the CMS GE2/1 detectors. \cite{Propa}.}
    \label{fig:MixedDesign}
\end{figure}

\begin{figure}[!h]
    \centering
    \includegraphics[width=0.50\textwidth]{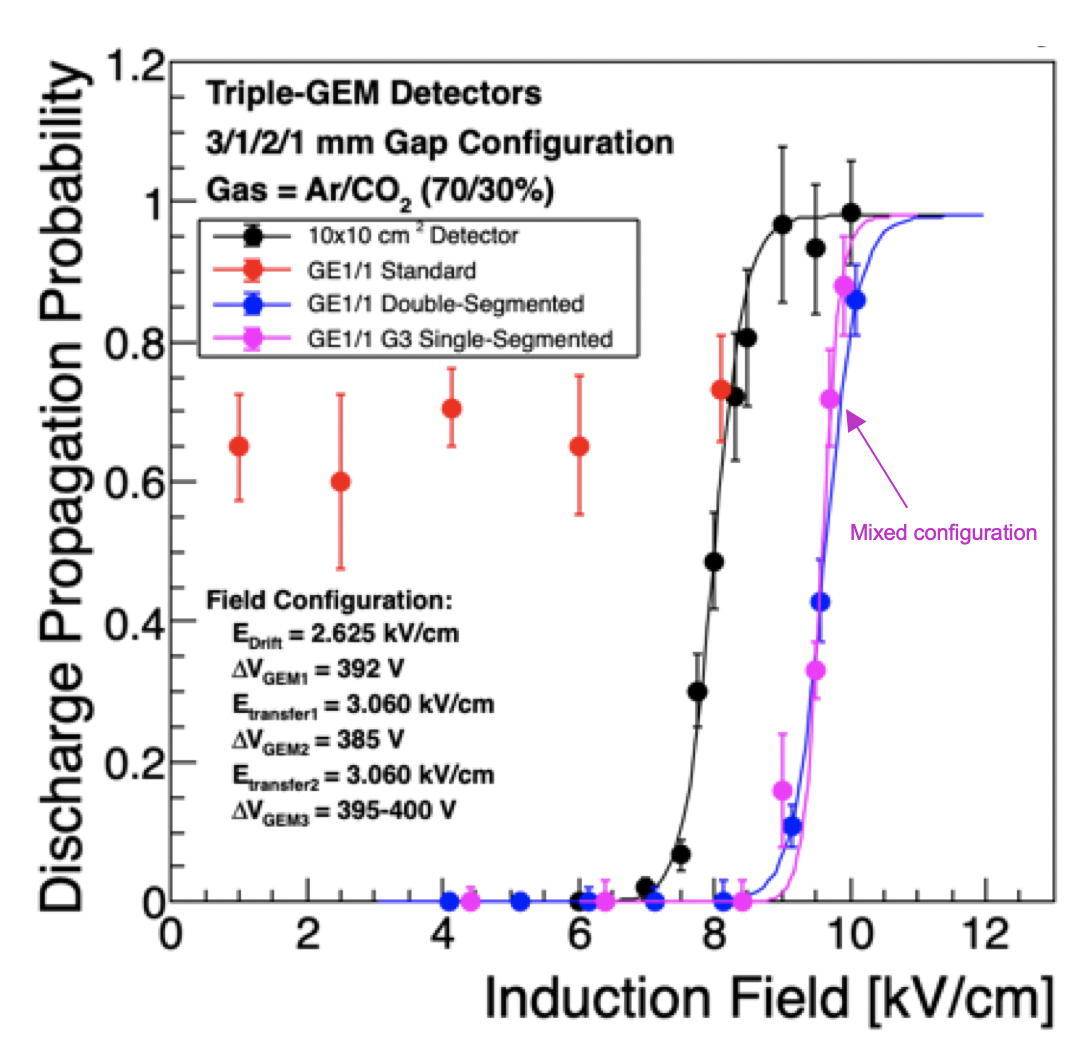}
    \caption{Discharge propagation probability as function of the induction field for different CMS GEM configurations \cite{Propa}.}
    \label{fig:MixedResults}
\end{figure}

A solution, elaborated in the framework of the CMS GE2/1 detector project, consists of combining the double-segmented design on the first two GEMs while maintaining the single-segmentation on last GEM facing the readout board. This configuration, shown on Fig.~\ref{fig:MixedDesign}, has a minimal impact on the detector design and production but it offers a strong protection against discharge propagation without introducing significant cross-talk side effect, as reported on Fig.~\ref{fig:MixedResults}.

%\subsubsection{New simulation tools for discharge studies}
% --> not sure if we want to go in this direction 

%\textcolor{blue}{note on new simulation tools for discharges ref/Resnati (finite element model) and ref/Gasik (MC model)}
%The critical charge density which can lead to the triggering of discharges was recently evaluated to 5 - 8 106 for Ar-CO2 mixtures and 7 - 9 106 for Ne-C02 mixtures 

Energetic discharges and their propagation are a major challenge for the next generation of GEM-based detector systems. The present solutions mostly rely on the establishment of smart designs and the implementation of complex production procedures with extra precautions at the levels of the construction, qualification and operation. Future applications will require the development of new propagation-free detector configurations but also the elaboration of stronger measures based on spark-free technologies in order to ensure detector stability, reliability and long-term performance.

\section{Alternative GEM-based designs}

\subsection{Resistive GEM detectors}

The development of resistive amplification structures has became of major interest in the MPGD community over the past few years \cite{RD51}. The access to new materials and the development of new manufacturing techniques in the PCB industry allows for the design and the production of new detector geometries with an intrinsic discharge protection while offering optimal detection performance.

The use of highly resistive materials, such as the Diamond-Like Carbon (DLC)\cite{dlc_ftm}, is particularly suitable to quench discharges or to reduce the energy stored in the detector electrodes. The increase of the overall resistance to a range of 10-100\SI{}{\mega\ohm}/$\square$ can drastically reduce effective rate capability, although high-rate layouts have been designed to allow the use of such technologies in environments up to 10-\SI{20 }{\mega\Hz\per\centi\m^2}.

On the other hand, the studies discussed in Section ~\ref{sec:discharge_studies} demonstrate that a de-coupling resistance of 30-100 k$\Omega$ between the hole and the main electrode is sufficient to prevent the formation of energetic discharges and their propagation inside the detector. One axe of development arising from this observation consists of minimizing the overall resistance by introducing resistive elements in a more surgical manner in order to locally protect the amplification channels. The recent experience on detector manufacturing with resitive materials sets the ground to new opportunities to design advanced resistive geometries to specifically protect strategic areas of the GEM structure. 

A promising solution, still in the very early stage its development, consists of de-coupling every single GEM hole from the main electrode by introducing resistive rings directly integrated to the GEM design. The resistance seen by GEM holes can be adjusted based on the choice of resistive material and the dimensions of the ring in order to reach an optimal protection with a minimal impact on the overall detector rate capability.

\begin{figure}[!h]
    \centering
    \includegraphics[width=0.80\textwidth]{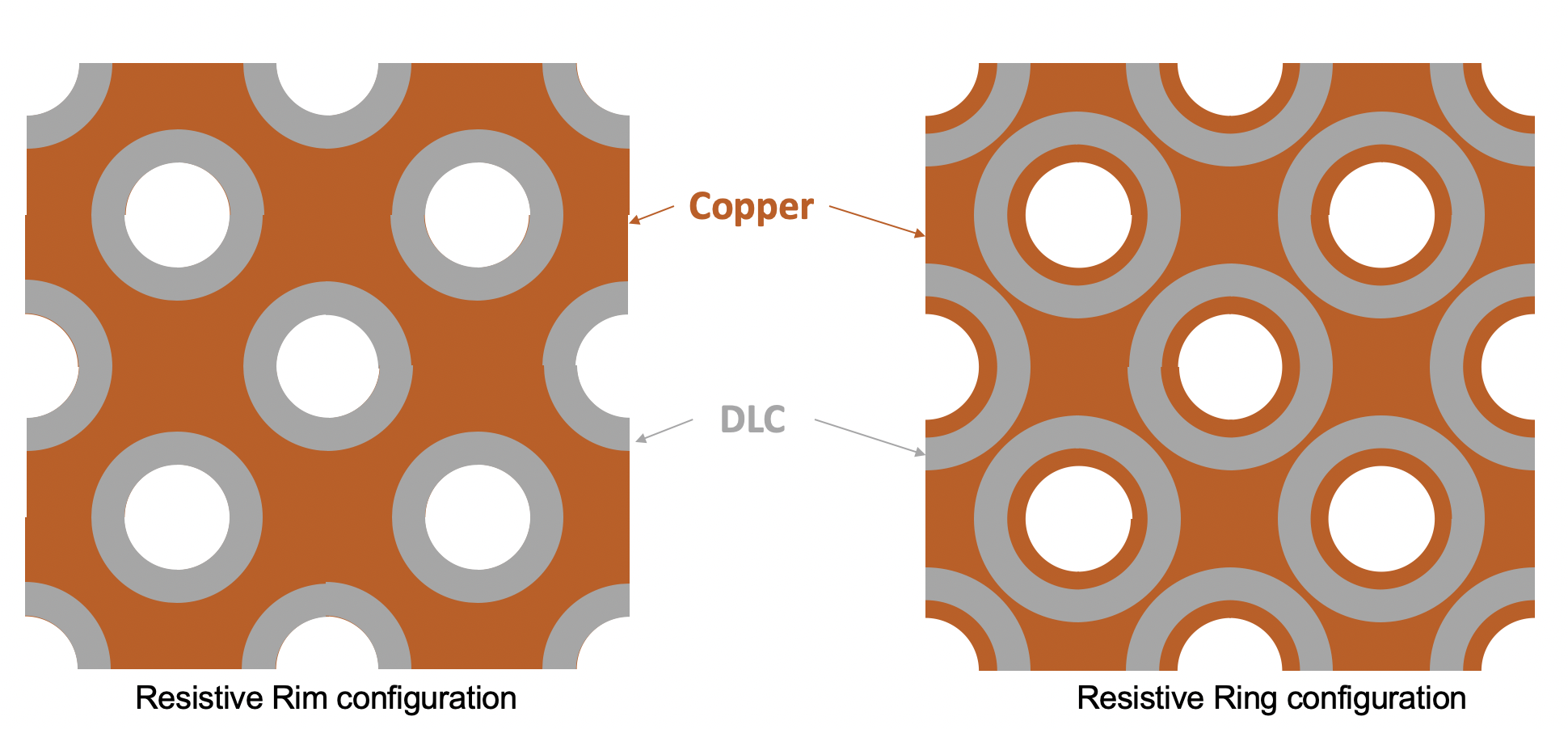}
    \caption{Overview of the discharge-protected GEM concept based on resistive rims (left) and resistive rings (right).}
    \label{fig:RRGEM}
\end{figure}

The schematics on Fig.~\ref{fig:RRGEM} shows two variants of this resistive GEM concept: the resistive rims and resistive rings designs.

With the resistive rings design, the discharges are not quenched but the amount of energy liberated during the process becomes insufficient to significantly damage the internal structure of the GEM. Compared to a standard 100 \SI{}{\centi\m^2} GEM foil, the de-coupled capacitance of a single hole can be reduced from 5-6 nF to 1-10 pF, reducing the discharge energy to a few µJ instead of mJ. The detector can therefore operate at unprecedented gas gain while being totally protected against deadly breakdown events.

A first prototype of $10\times10$ $cm^2$ resistive rim GEM was produced in 2021. Despite the poor quality of the base substrate available at that time, the foil is operational and already shows an outstanding resistance to discharges. The full characterization of this prototype is still an on-going activity. 

The Micro-Pattern Technology workshop at CERN is working on a new DLC manufacturing system \cite{Rui}, funded jointly by CERN and the Italian national institute for nuclear physics (INFN), which will allow the production of higher quality substrates with new possibilities in term of material, design choices and detector geometries. The DLC machine at CERN is expected to become active in the second half of 2022.

The main advantage of a local discharge protection is to operate each GEM at a significantly higher gain and still be able to operate in high rate environments. This result into simpler detector and electronics concepts, more compact and cost-effective while being fully protected against destructive events.

\subsection{Time resolution}

The time resolution of GEM detectors employed in most present-generation experiments is between \SI{2}{\nano\s} and \SI{10}{\nano\s}. This performance is sufficient for providing charged particle tracking at accelerators such as the High-Luminosity LHC, which at a luminosity of \SI{7.5e34}{\centi\m^{-2}\s^{-1}} is supposed to ultimately provide up to 200 pile-up collisions per event; at the higher luminosities expected at the Future Circular Collider and the Muon Collider, pile-up mitigation will take advantage by the possibility of a four-dimensional vertex reconstruction. MPGDs with time resolutions of the order of \SI{100}{\pico\s} will be a solution to provide fast timing over a large-area tracker or spectrometer.

Beside tracking, compact micro-pattern gaseous detectors with fast timing would have an application as cost-effective readout for hadronic of electromagnetic calorimeters. Outside high-energy physics, achieving fast timing with gaseous detectors is a topic of interest for medical diagnostic tools such as the positron emission tomography (PET), in which precise timing information could be used to improve the imaging resolution through time-of-flight techniques.

\begin{figure}
    \centering
    \includegraphics[width=0.75\textwidth]{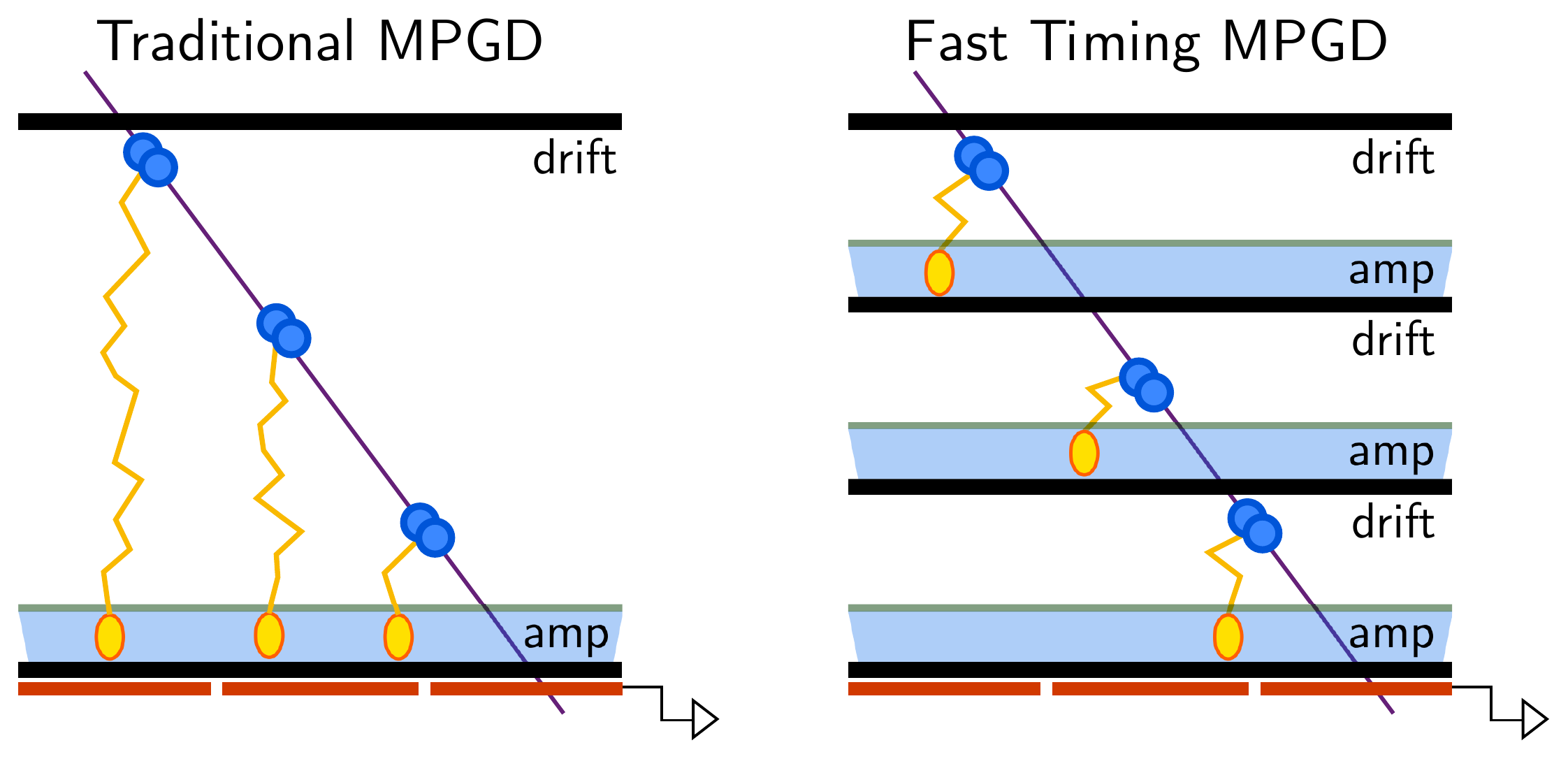}
    \caption{Geometry of a traditional MPGD compared with that of the FTM. The FTM active volume is divided in a stack of decoupled layers of fully resistive electrodes, each inducing a signal on a single external readout electrode.}
    \label{fig:mpgd_traditional_vs_ftm}
\end{figure}

The present time resolution limit is shared by all MPGDs divided in one drift gap -- where the primary ionization occurs -- and one or more amplification regions, as it is mainly determined by the fluctuations in the position of the first primary ionization cluster reaching the amplification region (Fig.~\ref{fig:mpgd_traditional_vs_ftm} left): the distance between the position where the primary ionization cluster is created and the beginning of the amplification region follows an exponential distribution with parameter $1/\lambda$, where $\lambda$ is the average number of primary ionization clusters created by a MIP per unit length. The signal arrival time $t=x/v_d$ (where $v_d$ is the average electron drift velocity in the gas) is then also an exponential variable:
\begin{equation}
    p(t) = \lambda v_d\,\text{e}^{-\lambda v_d t},
    \label{eq:time_resolution}
\end{equation}
and its sigma is $\sigma_t = 1/(\lambda v_d)$.

This equation can parameterize with reasonable approximation the time resolution of a GEM detector, which can be slightly lowered by different optimization procedures. One method consists of operating the detector at high drift field to have a pre-amplification between a factor 2 and 10 in the drift region, thus decreasing the time fluctuations on the earliest readout signal, at the cost of increasing the discharge probability in the amplification volume. A second possibility consists of employing gas mixtures (e.g. \ce{CF4}-based) with high electron drift velocity, reducing the drift time fluctuations due to the electron diffusion; however, most known mixtures with such properties are not compatible with the ongoing effort to limit the use of greenhouse and ozone-depleting gases. Finally, the drift time fluctuations can be lowered by a reduction in the thickness of the drift gap, with the drawback of decreasing the detector efficiency to charged particles, assuming the primary electrons are all created by gas ionization in the drift gap itself. A significant improvement in the time resolution of GEM detectors, then, would require a deeper change in the traditional detector structure.

An example of successful MPGD redesign aimed at improved time resolution is the PICOSEC-MicroMegas detector \cite{picosec}, which has demonstrated a time resolution of less than \SI{25}{\pico\second}. PICOSEC achieves fast timing by adding to the detector structure a Cherenkov radiator and moving the primary ionization from the drift gap to a photocathode, thus drastically reducing the fluctuations in the primary ionization position. The next section focuses instead on a detector design aimed at exploiting the possibilities of resistive materials for precise timing.

\subsubsection{Fast timing MPGD (FTM)}

The use of resistive materials in GEM detectors also opens up new possibility of achieving fast timing with MPGDs. Following a principle already successfully applied in the development of multi-gap RPCs, the gas volume of an MPGD can be divided in several independent layers, each instrumented with a drift and an amplification region (Fig.~\ref{fig:mpgd_traditional_vs_ftm} right); if all the electrodes are resistive, the signal can be read out by an external read-out plane. As the time of the event is defined by the first signal reaching an amplification layer\footnote{The arrival time of a cluster to the amplification region is exponentially distributed (Eq.~\ref{eq:time_resolution}) and the minimum of N identical exponential variables is still an exponential variable, with $\sigma_\text{N}=\sigma_t/N$.}, the fluctuations in the signal formation time are minimized by the competition of the different layers \cite{rui_ftm}. As the total gas volume remains unchanged, the efficiency of the entire stack remains identical to the one of a single-layer detector.

\begin{figure}
    \centering
    \begin{minipage}[b]{.5\textwidth}
        \includegraphics[width=.9\textwidth]{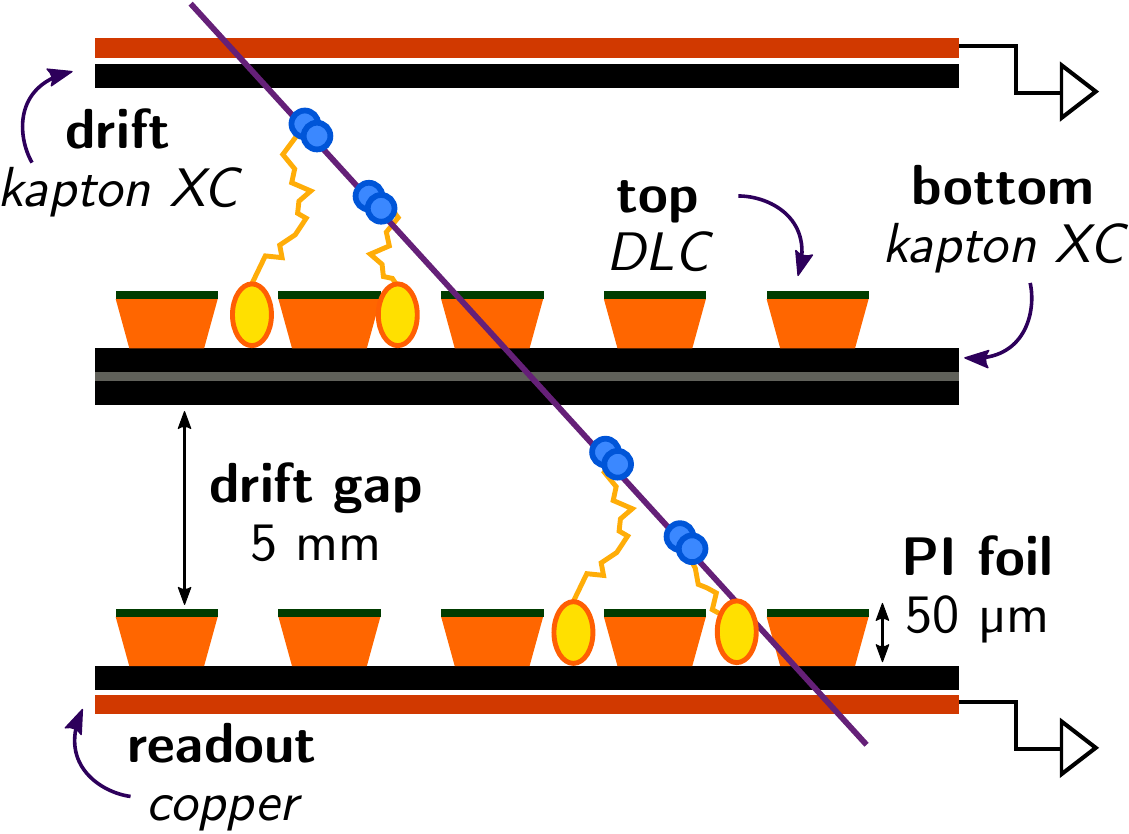}
    \end{minipage}
    \begin{minipage}[b]{.45\textwidth}
        \includegraphics[width=.9\textwidth]{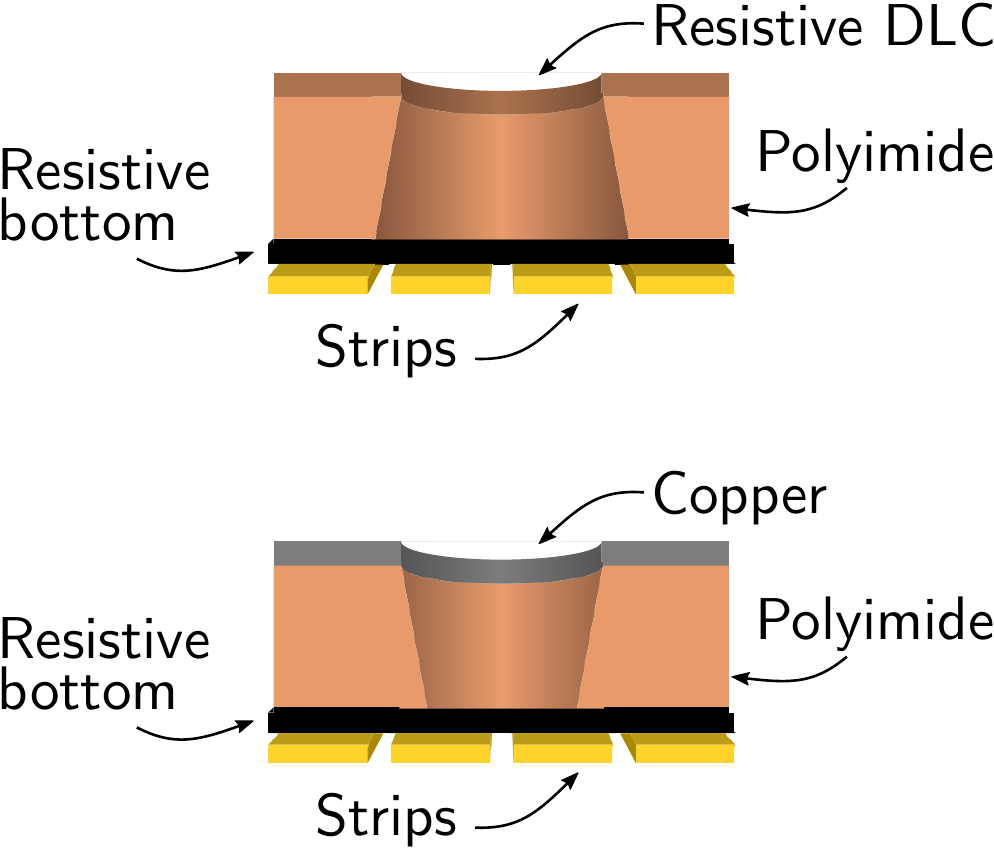}
    \end{minipage}
    \caption{On the left, structure of an FTM prototype with GEM foils instrumented with two layers. On the right, comparison between the FTM amplification foils and the foils used in the µ-RWELL detector, which differs for the copper electrode on top of the GEM foil as well as in the inverted hole shape.}
    \label{fig:ftm_2layers}
\end{figure}

The robust structure of a GEM detector makes it a candidate for the implementation of fast timing on an MPGD (Fig.~\ref{fig:ftm_2layers} left). Since the formulation of the FTM concept, results obtained with an FTM prototype in 2015 have proven the validity of its working principle, with the observation of a time resolution between 1.5 and \SI{2.5}{\nano\s} with muon and pion beams \cite{ilaria_ftm}. However, in addition to the improvements required for resistive GEM detectors, the development of a GEM-based fast timing MPGD (FTM) faces a sequence of unique challenges. The geometry of the amplification structure chosen for the FTM is inspired by the amplification foil of a µ-RWELL, but uses only resistive materials (DLC on the top and resistive polyimmide on the bottom, Fig.~\ref{fig:ftm_2layers} right).

\subsubsection{Production of resistive foils for the FTM}

A major lesson learnt from early and present studies on FTM prototypes is that it is not possible to decouple the physics performance of the FTM from the techniques used for the manufacturing of its main constituent, i.e. fully resistive GEM foils. Diamond-like carbon (DLC) has been the resistive material of choice in the FTM R\&D thanks to the possibility to create DLC layers of various thicknesses and resistivities.

The production of GEM foils with DLC anode can be divided in two major steps: the production of a flexible copper clad laminate (FCCL) by deposition of the DLC onto a polyimmide layer and the subsequent etching of the FCCL. Among the production techniques presently available for DLC-coated FCCL production, the most consolidated one is magnetron sputtering \cite{zhou_dlc}. A sputtering machine is a vacuum chamber filled with gas in plasma state; ions from the plasma extract from a target carbon atoms that are guided by the magnetic field, depositing to the polyimmide substrate. Magnetron sputtering has been the technique of choice for the foil manufacturing in past and present FTM prototypes, thanks to its reliablility for fast polyimmide coating on small to medium areas, but has to be improved for more stable DLC adhesion; a technique with similarly promising results, but not yet explored as extensively, is ion beam deposition. Conversely, for small-scale prototyping laser deposition has been observed to be a promising technique, with the possibility of tuning over a wide range of DLC resistivity \cite{dlc_ftm}.

\begin{figure}
    \centering
    \includegraphics[width=0.95\textwidth]{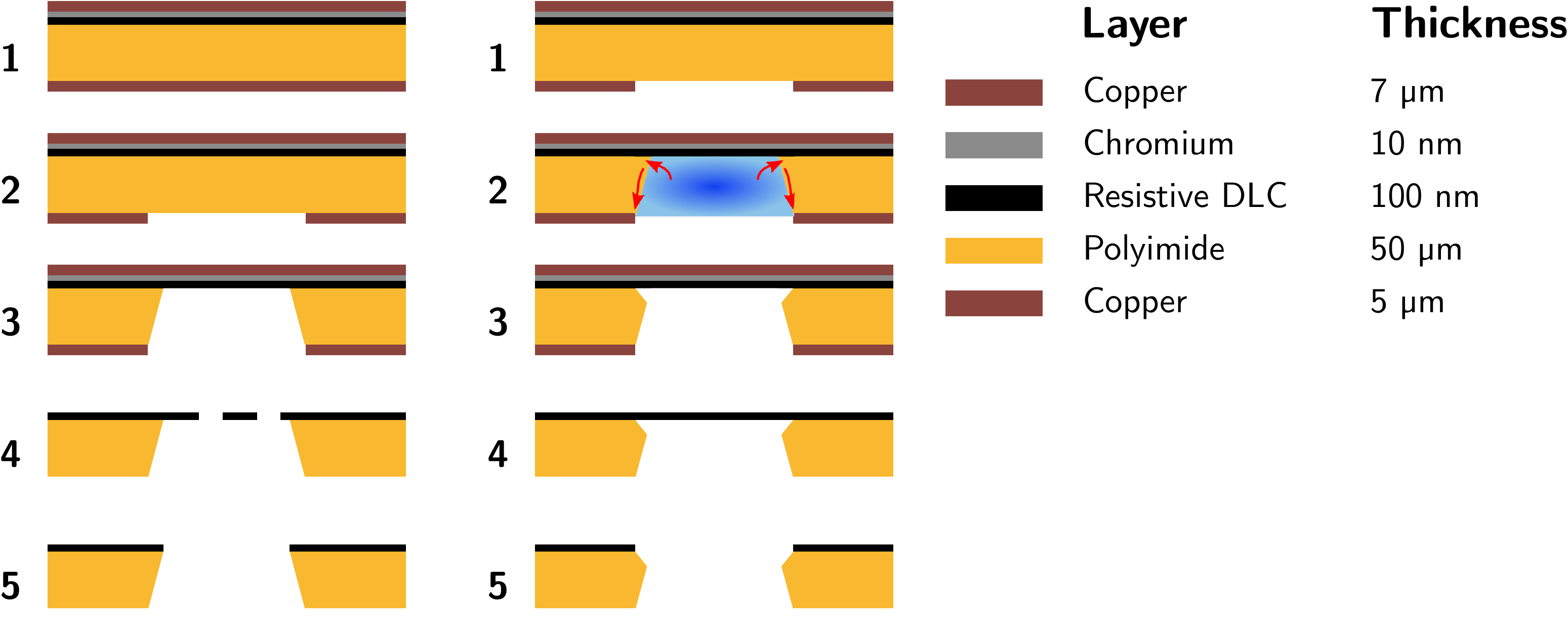}
    \caption{On the left, schematic steps of the production of resistive GEM foils from starting FCCL; on the right, production steps emphasizing the over-etching of the polyimmide resulting in hole walls irregularities.}
    \label{fig:ftm_foil_production}
\end{figure}

The second production step towards resistive foils, i.e. the etching of DLC-coated laminates, happens by masking and immersion in a chemical bath, with a process -- slightly different from the single-mask production of traditional GEM foils -- summarized in the following points (Fig.~\ref{fig:ftm_foil_production} left):
\begin{enumerate}
    \item The laminate obtained from the coating before undergoing the etching is made of the polyimide layer (typically of \SI{50}{\micro\m} covered on one side (top) by the DLC and a thin chromium layer to protect the DLC; additionally, both sides of the laminate are covered by a copper layer.
    \item The bottom copper layer is masked and then patterned by photolithography.
    \item The laminate is put in a chemical bath to allow the etching of the polyimmide.
    \item The chromium layer and copper residuals are removed and the DLC is left uncovered.
    \item The DLC left in the holes is removed mechanically (e.g. by sand blasting).
\end{enumerate}

\subsubsection{Performance of FTM prototypes}

\begin{figure}
    \centering
    \includegraphics[width=.6\textwidth]{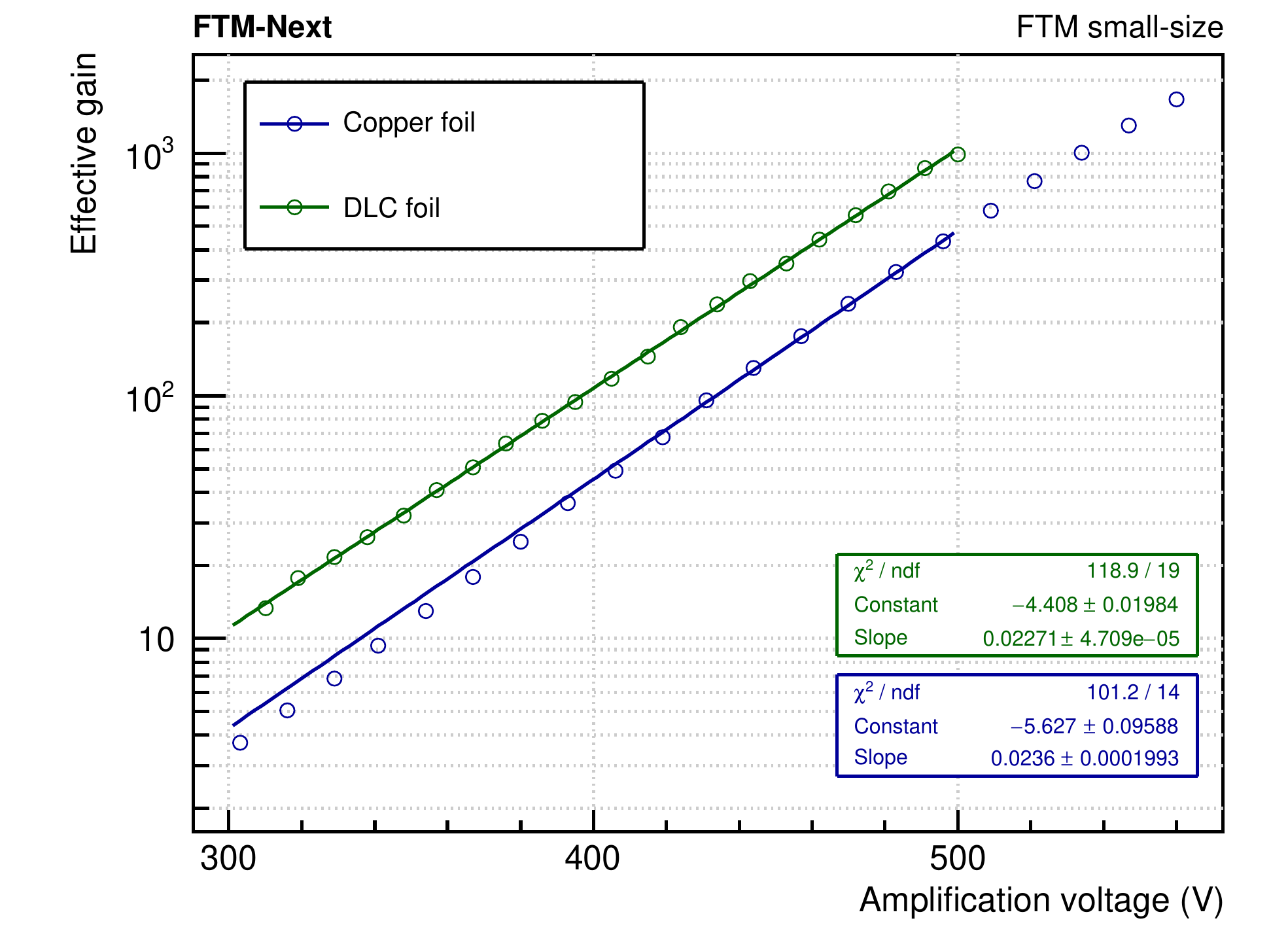}
    \includegraphics[width=.8\textwidth]{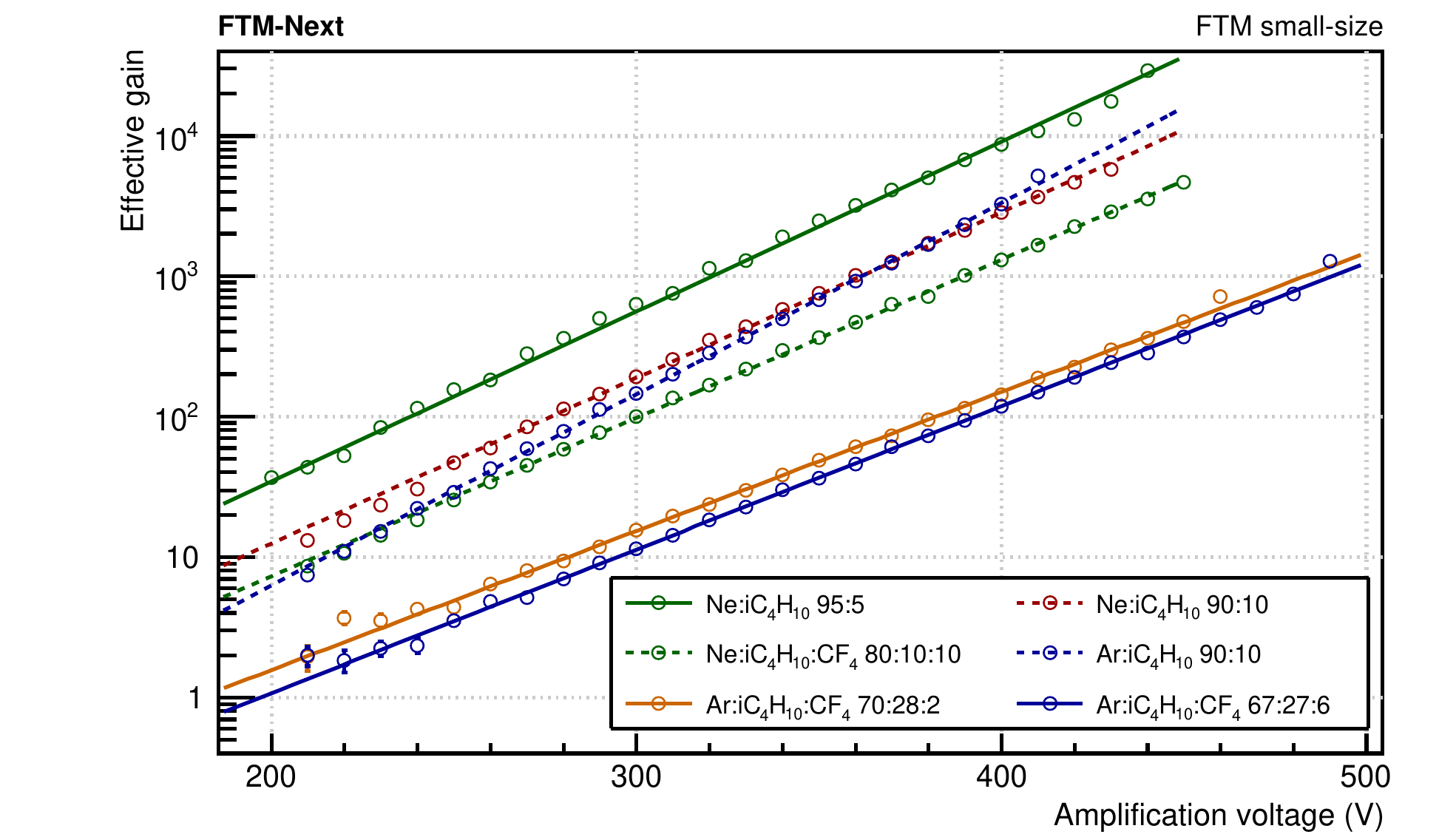}
    \caption{On top, comparison between the effective gain curves of conductive and resistive amplification foils measured in Ar:\ce{CO2} 70\%:30\%. At the bottom, comparison of effective gain curves obtained with resistive foil in \ce{iC4H10}- and \ce{CF4}-based mixtures. Images from \cite{pellecchia_ftm}.}
    \label{fig:ftm_gains}
\end{figure}

Several FTM prototypes have been designed, manufactured and tested with the goal of gaining insight on the technological innovations needed to achieve good physical performance \cite{christos_ftm}. A small-size FTM has been tested with different foils and several gas mixtures to compare the performance of merely resistive foils and copper-top foils. Gain measurements performed with a UV laser \cite{pellecchia_ftm} show that the maximum gain reachable with a resistive foil of DLC surface resistivity \SI{100}{\mega\ohm}/$\square$ is lower by at least a factor 2 than the one obtainable with a conductive foil (Fig.~\ref{fig:ftm_gains}) due to the higher instability (i.e. high probability of discharge) to high amplification fields, even though at equal amplification voltage the resistive foil has a higher gain than the conductive one as a consequence of the inverted hole shape. Comparison among different gas mixture show that it is still possible to obtain good performance with resistive foils by operating the FTM in isobuthane-based gases, with the highest gain of \SI{3e4} obtainable in \ce{Ne}:\ce{iC4H10} 95\%:5\%. Measurements in Ar:\ce{CO2} do not instead overcome gains of \SI{2e3}{}.

\begin{figure}
    \centering
    \includegraphics[width=0.99\textwidth]{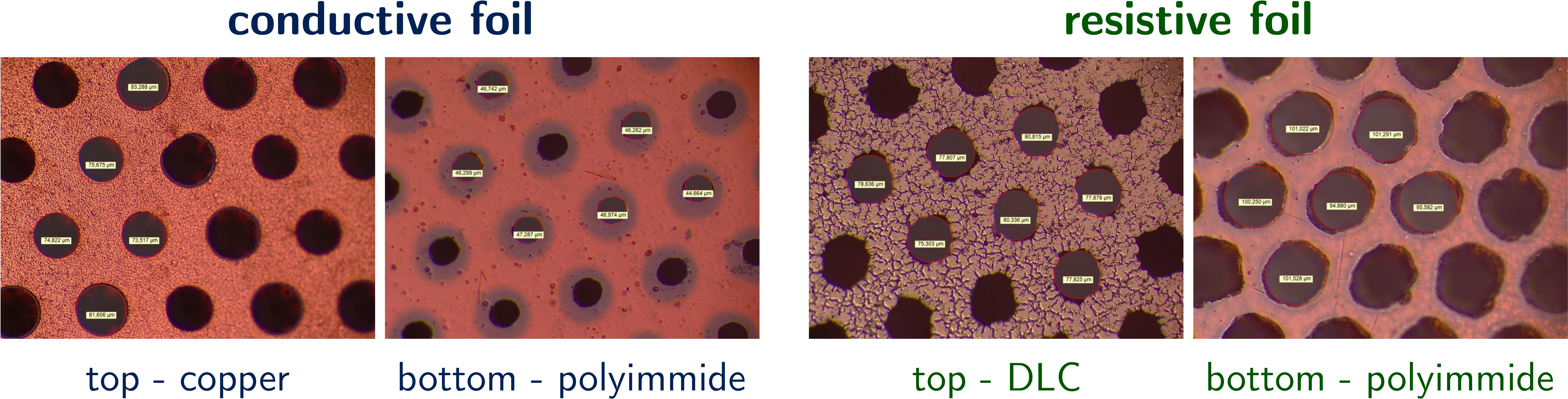}
    \caption{Conductive and resistive amplification foils observed at the microscope.}
    \label{fig:ftm_pics_microscope}
\end{figure}

The lower stability to discharges of resistive foils to high electric fields with respect to copper foils can be traced back to the higher risk of DLC delamination in the foil etching. Comparison of hole regularity and surface roughness between the two types of foils (Fig.~\ref{fig:ftm_pics_microscope}) show a worse surface regularity in the resistive foils, not only in the top (DLC) side but also in the bottom (polyimmide). One possible point of failure in the foil production process giving rise to this effect is the alminate immersion in the chemical bath (see Fig.~\ref{fig:ftm_foil_production} right): the etching solution entering in contact with the DLC can infiltrate through the DLC itself due to imperfect adhesion to the kapton, resuting in over-etching of the polyimmide walls.

One obvious solution to this issue would come from the possibility of producing DLC laminates with higher coating adhesion; in parallel to this goes the need for a reliable technique for the single-mask etching from the DLC side. One can conclude that at the present moment, the etching process of resistive-coated laminates is not fully optimized. A quicker development of these techniques will require communication with industrial partners and also better synergy between the main actors mastering the different steps involved in the foil manufacturing.
%Furthermore, the need for a good adhesion of the GEM foil polyimmide to the bottom layer and of the resistive layers to the readout electrode to maximize the capacitive coupling requires that the three layers be manufactured in a single piece.

\subsubsection{Progress for fast and sensitive readout electronics}

Beside the foil production quality, a second area of development of the FTM will involve the readout electronics, which requires particular attention with respect to other MPGDs due to the need to combine precise timing with high sensitivity. The time resolution of a front-end chip, often parameterized as
\[
    \sigma_t = \frac{\sigma_\text{noise}}{dV/dt} = \frac{t_\text{rise}}{S/N}, 
\]
is heavily influenced by the ASIC input noise and the chip shaping times. Commercially available electronics used for fast timing (such as the one developed for the RPCs) benefit of the good signal-to-noise ratio of their detectors, which can be safely operated at high gains (up to $10^8$). Therefore, their analog circuits have typically very short shaping times of the order of 1 ns or below and low sensitivity (below 5 mV/fC), with loose constraints on the input noise charge that can be as high as 4000 electrons.

On the other hand, all micro-pattern gaseous detectors share the same limitation in the total amount of charge produced by the avalanche, which cannot overcome a few fC due to their high sensitivity to discharges. This limits the gain of a GEM detector to about $10^{5}$ when it is operated to amplify primary charges of the order of a single electron and to 2-\SI{3e4} when the primary charge is of tens of electrons. Therefore, present-generation front-end ASICs used for MPGDs (such as the VMM and the VFAT chips) were developed to feature an excellent signal-to-noise ratio, thanks to their high sensitivity (between 10 and \SI{60}{\milli\V/\femto\coulomb}), a low equivalent-noise charge (below 1000 electrons) and a long signal integration time of tens of nanoseconds. The resulting time walk is higher than \SI{10}{\nano\second} for most MPGD front-end electronics. As a consequence, these solutions are typically not equipped with time-stamping outputs and are not suitable for precise timing applications.

One emblematic case of compromise and an important step in the direction of fast timing front-end for MPGDs is the case of the VFAT3 chip \cite{vfat3} designed for the CMS GEM upgrade, which introduces with respect to its previous version (the VFAT2 \cite{vfat2}) the possibility of a programmable long integration time up to 45 ns with the goal of improving the signal-to-noise ratio of the whole detector. To recover the timing loss, a constant fraction discriminator is built in the chip in sequence to the arming comparator; this addition allows to minimize the time walk and has demonstrated a good time performance of \SI{400}{\pico\second}, with the downside of an increasing complexity of the ASIC.

\begin{figure}
    \centering
    \includegraphics[width=0.6\textwidth]{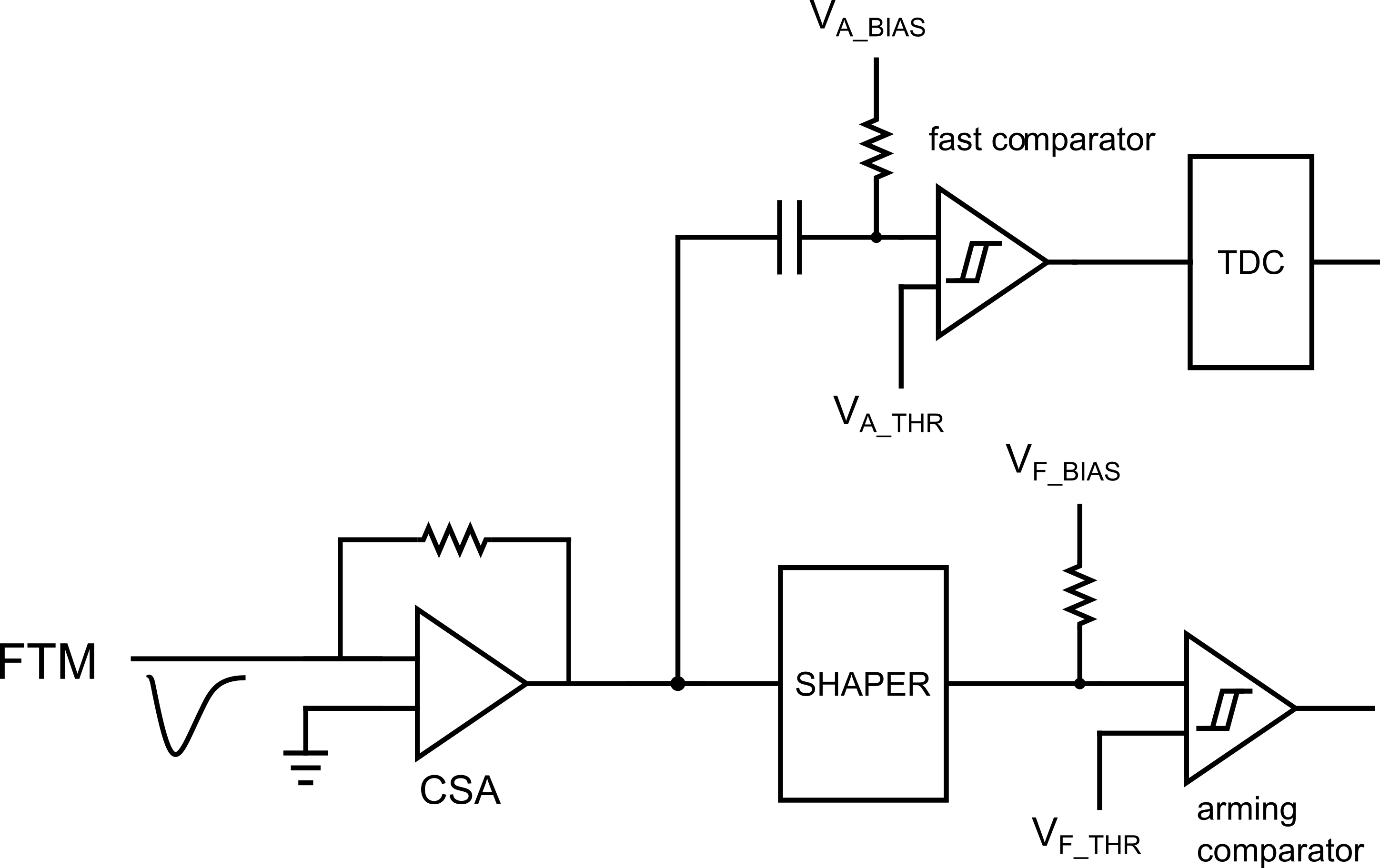}
    \caption{Analog circuit of a single FATIC channel \cite{fatic}. The fast branch is made by a discriminator and provides precise timing on the event, while the trigger branch is equipped with a shaper of long integration time followed by an arming comparator to validate the event.}
    \label{fig:fatic_analog}
\end{figure}

An ongoing effort towards the development of a front-end for the FTM is the FATIC (FAst Timing ASIC) \cite{fatic}. The FATIC is a 32-channel front-end chip designed for providing a good sensitivity and low time jitter with a layout that combines both a timing branch and a trigger branch. In the FATIC, the input signal from the detector is preamplified and split (Fig.~\ref{fig:fatic_analog}), feeding both a fast discriminator and a shaping circuit with a rise time of \SI{7}{\nano\s}. The shaper output (the "slow" signal) is then discriminated through an arming comparator which benefits of the improved signal-to-noise ratio and serves to validate the event; in case of event over threshold, the "fast" signal passes through a TDC that provides the event timing.

Simulations performed on the FATIC design (Tab.~\ref{tab:asic_comparison}) show promising results with an excellent sensitivity over \SI{50}{\milli\V/\femto\coulomb} and time jitter lower than \SI{50}{\pico\s}, suitable for an MPGD with an effective gain of the order of $10^4$ and a time resolution of a few hundreds of ps. Production and tests of the first FATIC prototypes are presently ongoing, with the first on-detector measurements to be performed with FTM and µ-RWELL detectors planned for the second half of 2022.

\begin{table}[t]
    \centering
    \begin{tabular}{ccccccc} \hline
        & {\bf Rise time} & {\bf Sensitivity} & {\bf ENC} & {\bf Time jitter} \\ \hline
        %& VFAT2 & VFAT3 & VMM & NINO & Cardarelli & FATIC \\ \hline
        VFAT2 \cite{vfat2} & 22 ns & 60 mV/fC & 1500 e- & 12 ns \\
        VFAT3 \cite{vfat3} & 15-45 ns & 48 mV/fC & 620 e- & 12 ns \\
        VMM3 \cite{vmm} & 25-200 ns & 16 mV/fC & 600 e- & / \\ \hline
        NINO \cite{nino} & 1 ns & / & 2000 e- & 25 ps \\
        Cardarelli \cite{cardarelli} & 300-600 ps & 2-3 mV/fC & 4000 e- & / \\ \hline
        FATIC \cite{fatic} & 7 ns & 50 mV/fC & 500 e- & 300 ps \\ \hline
    \end{tabular}
    \caption{Comparison of the simulated FATIC performance with different sensitive (VFAT, VMM) or fast (NINO, Cardarelli) front-end ASICs.}
    \label{tab:asic_comparison}
\end{table}

\section{Conclusion and perspectives}

As a a mature detector technology, GEM has become a reliable basis for application-specific advancements such as muon spectrometry over large areas in high-rate environments or fast timing for tracking and calorimetry. The time scales for each of these developments differ according to their complexity and the state-of-the art of their technological starting point.

The optimization of large-area GEM detectors for high rate capability and sustained operation in hostile environments is an ongoing effort supported by a large and expert community and benefits from already well-defined requirements -- such as those laid down by the expected operation in the CMS Phase-2 -- to be met on a relatively short time scale set by the High-Luminosity LHC schedule. Advancements requiring more substantial changes in the detector design, such as resistive-ring GEM foils, will be pushed from the increasing interest for resistive coatings in the MPGD community. Given the recent fast improvement in the quality of DLC-coated foils and the availability of new sputtering facilities such as the CERN DLC machine, the immediate next two years will turn out to be essential for GEM detectors with DLC rings to become competitive for specific applications.

Similar requirements as for resistive GEMs are shared by the FTM. On the other hand, in the FTM case the benefits drawn from the development and operational experience of all resistive MPGDs are merely a starting point for more specific areas of improvement, due to its peculiar complexity, despite its structural similarities to other detectors such as the µ-RWELL. Even after the estabilishment of a good time resolution on small prototypes -- as expected from the next two years of R\&D supported by the already available funding -- the manufacturing of high-quality fully resistive GEM foils with surfaces of hundreds of \SI{}{\centi\m\squared} will likely require a few more years of joint effort between foil manufacturing and etching specialists. Exploring different techniques of carbon deposition beside the most consolidated ones is an essential tool for achieving good control over the electrode resistivity -- which is related to the read-out transparency; collaborations between between research groups specialized in solid state physics and detector physics are already a reality and will continue to need support for the immediate next years. More into the future, applying the FTM technology to high-rate environments will require dedicated strategies to avoid the use of evacuation schemes with conductive elements.

Concerning the FTM readout electronics, the present performance expectations of the FATIC (with a time resolution of the order of \SI{100}{\pico\second}) will meet the timing requirements of the FTM and other MPGDs to be used for tracking and calorimetry, with the first on-detector results performed on the second generation of the chip (FATIC2) planned on a short time-scale for 2022 and early 2023; the second LHCb upgrade of 2035 is a first prospected application of the FATIC -- as µ-RWELL readout -- in a high-energy physics experiment.

%\section{RWELL for HEP experiments}

%\pubblock
\chapter{\centering $\mu$-RWELL for HEP experiments}
%\addcontentsline{toc}{chapter}{$\upmu$-RWELL for HEP experiments}
%\Title{$\upmu$-RWELL for HEP experiments}

\bigskip 

%\Author{I.~Balossino, G.~Cibinetto, R.~Farinelli, I.~Garzia, S.~Gramigna, M.~Melchiorri, G.~Mezzadri, M.~Scodeggio}
%\Address{INFN Ferrara - Via Saragat 1, 44122 Ferrara, Italy}

%\Author{V.~Cafaro, P.~Giacomelli}
%\Address{ INFN Bologna, Viale Berti Pichat, 6/2, 40127 Bologna, Italy}

%\Author{G.~Bencivenni, M.~Bertani, E.~De~Lucia, D.~Domenici, G.~Felici, M.~Gatta, M.~Giovannetti, G.~Morello, G.~Papalino, M.~Poli~Lener}
%\Address{Laboratori Nazionali di Frascati - INFN, Via Enrico Fermi 54, 00044 Frascati (RM), Italy}

%\Author{L.~Lavezzi}
%\Address{INFN Torino -  Via Pietro Giuria, 1, 10125 Torino (Italy)}

\medskip

% \begin{Abstract}
%\noindent The $\upmu$-RWELL (micro Resistive WELL) is a novel resistive  Micro Pattern Gas Detector (MPGD); it takes some of the best characteristics of existing MPGDs, like GEMs and MicroMegas, while simplifying the detector construction. It also improves significantly the spark protection by incorporating in the design a resistive layer realized with a Diamond-Like-Carbon (DLC). The $\mu$-RWELL is in fact  composed of two elements: a cathode, a simple FR4 PCB with a thin copper layer on one side, and the $\upmu$-RWELL PCB, that encompasses the amplification stage and the readout. The performance is excellent: gain exceeding 10$^4$, efficiency $>$97\%, time resolution $<$6 ns, spatial resolution $<$100 $\mu$m. The rate capability is up to few tens of kHz/cm$^2$ for the simpler low rate design, and above than 10 MHz/cm$^2$ for the high rate one which have a more effective current evacuation.
%The $\upmu$-RWELL technology therefore provides a cost-effective and high-performance particle tracker with many potential applications. An overview of the status of the art of the technology and ongoing activities will be presented in this section, together with a brief review of some applications currently under study.
%\end{Abstract}

%\snowmass

%\snowmass

\def\thefootnote{\fnsymbol{footnote}}
\setcounter{footnote}{0}
Micro-pattern gaseous detectors (MPGD, \cite{MPGD}), based on modern photolithographic technology, allow operation at very high background particle flux with high efficiency and spatial resolution. These features determines the main applications of these detectors in particle physics experiments as precise tracking in high radiation environment as well as muon identifier in general purpose detectors (LHC and future generation e-h colliders, FCC-ee/hh, CepC). In addition, the reduced impact in terms of material budget and the flexibility of the base material makes these devices suitable for the development of very light, full cylindrical fine tracking inner trackers at high luminosity tau-charm factories (STCF in Russia and SCTF in China). Among the most prominent MPGD technologies, the Gas Electron Multiplier (GEM, \cite{GEM}) and MicroMegas (MM, \cite{MM}) have been successfully operated in many different experiments, such as Compass \cite{Compass}, LHCb \cite{LHCb}, TOTEM \cite{Totem}, KLOE-2 \cite{Kloe2}, and are being built for the upgrades of ATLAS \cite{ATLAS-MM} and CMS \cite{CMS-GEM} at LHC, and BESIII \cite{CGEM-BES3} at IHEP.
The secret of the success of MPGDs, compared to classic gaseous detectors, lies in the sub-millimeter distances between the anodic and cathode electrodes. This feature, reducing the collection times of the ions in the gas, allows the operation of these devices in environments with very high radiation fluxes. At the same time, due to the fine structure MPGDs generally suffer from spark occurrence that can eventually damage the detector, as well as the readout electronics.
One of the most efficient solutions to this problem is provided by the introduction of thin resistive film deposited between the amplifying stage and the PCB readout of the detector. In this layout the amplifying mesh (or top electrode) is kept at high voltage while the resistive film is generally grounded and capacitively coupled with the underlying readout plane. The principle of operation is the same of the resistive electrode used in Resistive Plate Counters (RPC \cite{Pestov, santo, bencivenni}): the streamer, discharging a limited area around its location, is quenched and the transition to spark is strongly suppressed giving the possibility to achieve large gains.\\
The $\upmu$-RWELL, a recently introduced resistive-MPGD \cite{uRwell}, inherits the best characteristics of the GEM and MM detectors, while further simplifying the manufacturing process, thus enabling for the first time in the MPGD history a complete technology transfer to standard PCB industry. The $\upmu$-RWELL is the baseline option for the phase-2 upgrade of the innermost regions of the Muon system of the LHCb experiment. The technology, among the candidates for large muon detection systems at the FCC-ee and CEPC future large circular leptonic colliders, has been recently considered for the upgrade of the muon spectrometer of the CLAS12 experiment at Jlab.

\begin{figure}
\begin{center}
\includegraphics[width=0.70\hsize]{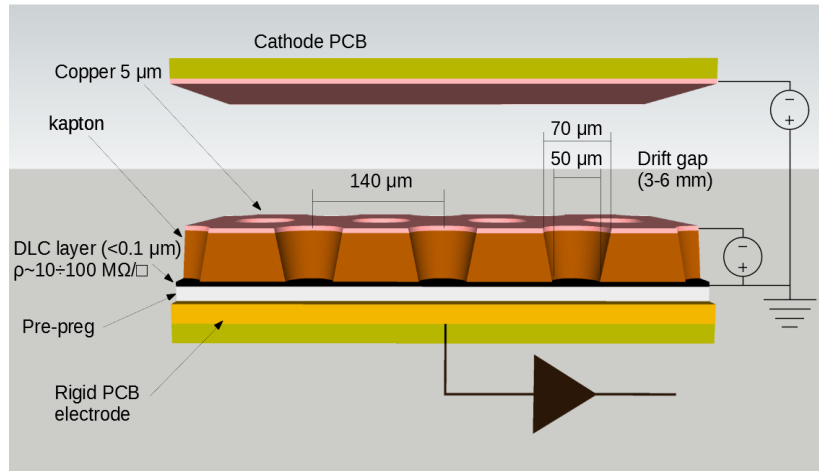}
\end{center}
\caption{Basic layout of a $\upmu$-RWELL.}
\label{fig:RWELL-cross-section}
\end{figure}

\begin{figure}
\begin{center}
\includegraphics[width=0.70\hsize]{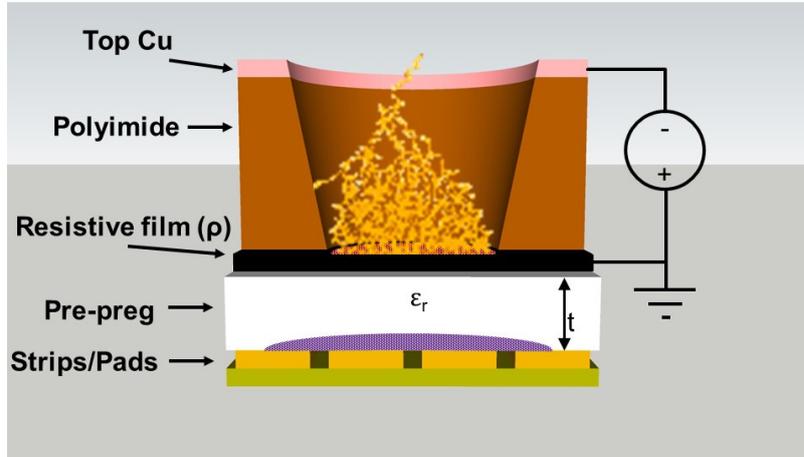}
\end{center}
\caption{Principle of operation of the $\upmu$-RWELL.}
\label{fig:amplification-well}
\end{figure}

As shown in Fig.\ref{fig:RWELL-cross-section} the $\upmu$-RWELL detector is composed of two PCBs: a mono-layer PCB acting as the cathode, defining the gas detector gap,  and a $\upmu$-RWELL-PCB that couples in a unique structure the electron amplification stage with the readout board.
The amplification stage, based on a 50 $\mu$m thick polyimide foil, copper clad on the top side and sputtered with Diamond Like Carbon (DLC \cite{DLC}) on the opposite (bottom) side, is coupled to a standard PCB readout board, through a 50 $\mu$m thick pre-preg foil. The thickness  of the DLC layer (typically in the range 10-100 nm) is adjusted according to the desired surface resistivity value (10-100 M$\Omega$/square) in order to provide discharge suppression as well as current evacuation.
A chemical etching process of the polyimide foil is performed on the top surface of the overall structure in order to create a pattern of micro-well, that represents the basic amplification element of the detector. The well has a truncated cone shape with a 70 $\mu$m (50 $\mu$m) top (bottom) diameter and 140 $\mu$m pitch). The high voltage applied between the copper and the resistive DLC layers produces the required electric field within the wells that is necessary to develop charge amplification, Fig.\ref{fig:amplification-well}. The signal is capacitively collected on the strips/pads of the readout board.
The introduction of the resistive layer allows to achieve large gains ($\geq$10$^4$) with a single amplification stage, while partially reducing the capability to stand high particle fluxes. The simplest resistive layout, designed for low-rate applications, is based on a single-resistive layer with edge grounding. At high particle fluxes this layout suffers of a non-uniform response. In order to get rid of such a limitation different current evacuation geometries have been designed. Several high rate layouts have been developed which allow the detector operation up to mip rates of the order of 10-20 MHz/cm$^{2}$.\\
The overall performance of the detector is excellent: gain exceeding 10$^4$, efficiency $>$97\%, time resolution $<$6 ns, spatial resolution $<$100 $\mu$m.
In the following sections the high rate layouts developed in recent years will be described together with their performances. The typical tracking performance in micro-TPC mode obtainable with this detector will be then discussed, together with the description of a peculiar low-mass cylindrical tracker layout as inner tracker for low momentum e$^{+}$e$^{-}$ colliders (STCF in Russia, SCTF in China, EIC in USA).
In addition to these applications it should be stressed that the basic version of the $\upmu$-RWELL, characterized by high reliability and constructive simplicity, can be exploited as active device in digital calorimetry, while for non-HEP applications it is proposed as gamma and neutron detection (with suitable $^{10}$B converters \cite{urania}) in homeland security for radiation portal monitor or radiation waste monitoring.

\section{High Rate layouts for muon detection}

\begin{figure}
\begin{center}
\includegraphics[width=0.70\hsize]{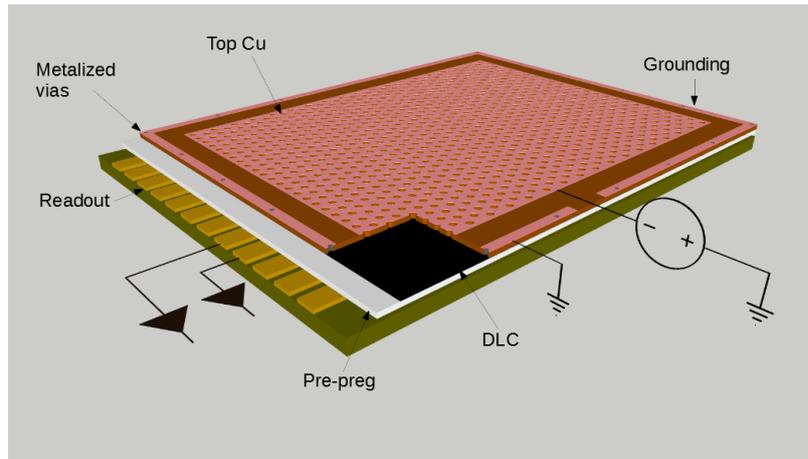}
\end{center}
\caption{Sketch of the Single-Resistive layout.}
\label{fig:resistive-stage1}
\end{figure}

The simplest current evacuation layout of the $\upmu$-RWELL is based on a single resistive layer with a grounded conductive line all around the active area, Fig.\ref{fig:resistive-stage1}, (Single Resisitive layout - SRL).
For large area SRL devices the path of the current to ground could therefore be large and strongly dependent on the incidence point of the particle. \\
%This limitation can be overcome creating a high density grounding network directly on the DLC layer.
% 
In order to cope with this effect the solution is to reduce as much as possible the average path towards the ground connection, introducing a high density grounding network on the resistive stage.\\
Several high rate (HR) layouts have been developed \cite{HR-layouts}:
\begin{itemize}
    \item the Double-Resistive Layer (DRL)
    \item the Silver-Grid (SG)
    \item the Patterning-Etching-Plating (PEP)
\end{itemize}

\begin{figure}
\begin{center}
\includegraphics[width=0.70\hsize]{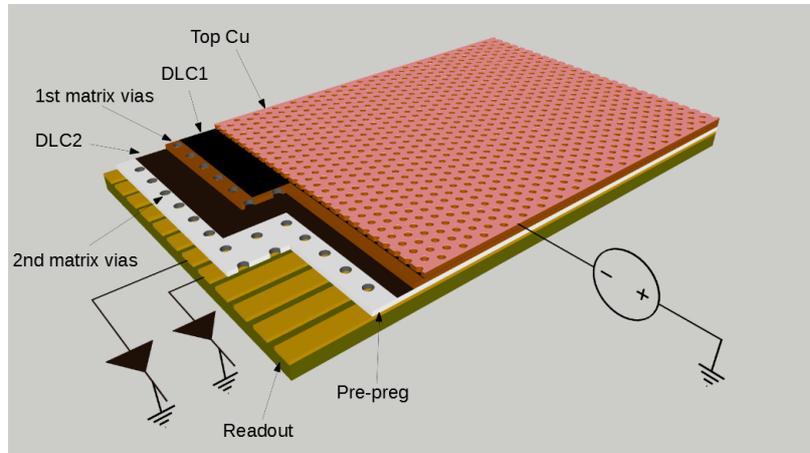}
\end{center}
\caption{Sketch of the Double-Resistive layout.}
\label{fig:resistive-stage2}
\end{figure}

\begin{figure}
\begin{center}
\includegraphics[width=0.70\hsize]{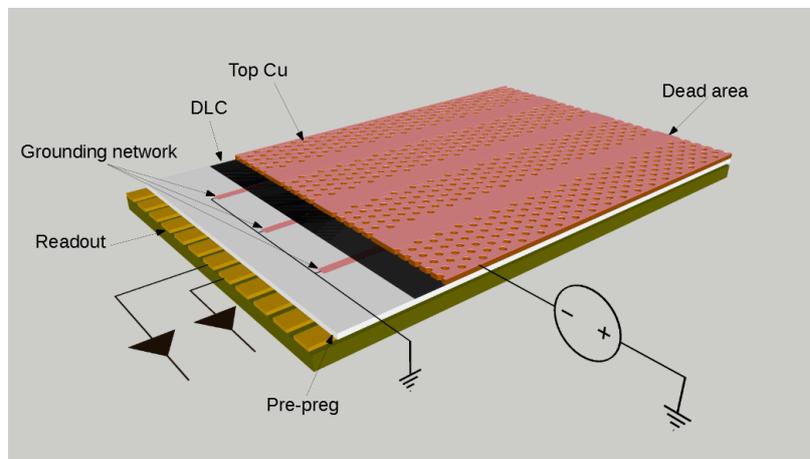}
\end{center}
\caption{Sketch of the Silver-Grid layout.}
\label{fig:SG-sketch}
\end{figure}

\begin{figure}
\begin{center}
\includegraphics[width=0.70\hsize]{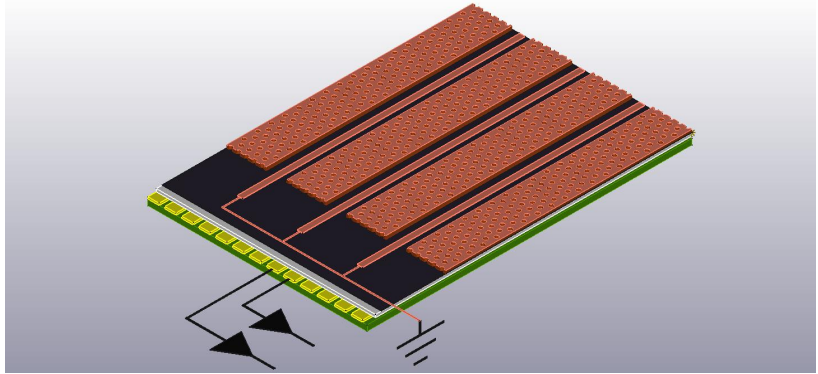}
\end{center}
\caption{Sketch of the Patterning-Etching-Plating layout.}
\label{fig:PEP-sketch}
\end{figure}

All these layouts are characterized by a dense matrix of conductive strip-lines or dots patterned on the DLC or the amplification stage. The validation of each layout is performed in terms of rate capability, safe operation, efficiency and simplified manufacturing.

In the DRL layout (Fig.~\ref{fig:resistive-stage2}) two through-vias matrices (density 1/cm$^{2}$) in cascade connect the DLC films to ground. The DRL shows very good performance with no dead zone in the amplification stage, but it is realized with a quite complex manufacturing procedures.
In the SG layout, Fig.~\ref{fig:SG-sketch}, a copper grid (1 cm pitch) patterned on the unique DLC film, acts as the grounding system. The SG layout, much simpler than the DRL because it is based on a single DLC layer, does not require complex production steps (i.e. double matrix of vias). In order to avoid instability a very accurate alignment between the tiny dead zone in the (top) amplification stage and the underlying (bottom) Cu grid is required  \footnote{In the following we refer to the SG detector with the acronymis SG2++ due to the optimization carried out for the geometrical parameters of its layout.}. 

%The SG and DRL performance are very similar and will be reported and discussed in sec.~\ref{HR-performance}, while in sec. preliminary results of the PEP layout are shown. 

The PEP layout (Fig.~\ref{fig:PEP-sketch}), a new single DLC layer HR scheme, has been recently introduced as a synthesis of the previously developed HR layouts.
%
%The grounding vias are created by patterning and etching the polyimide base material from the top; the grounding is done plating the Cu top layer and DLC stage without requiring the deposition of the Cu grids on the DLC, such as in the SG one, and simplify the manufacturing.
%
The grounding of the DLC stage is done through a micro-strips grid patterned on the top side of the polyimide foil and connected to the resistive film through metalized vias.  
For both SG and PEP layouts, the presence of a grounding grid close to the amplification stage, can induce discharges on the detectors. This effect, depending on the DLC resistivity, requires the introduction of a small dead zone in the amplification stage above (for SG) or close (for PEP) the grid lines.

In table \ref{tab:det-caracteristiche} we report the characteristics of the HR-layouts described in this paper. For the PEP layout, the a tuning of the geometrical parameters could be done to achieve an acceptance above 90\% (PEP++), such as already done with SG layout. 
%For completeness also the parameters of the SRL have been included.
%

\begin{table}\footnotesize
\begin{center}
\begin{tabular}{|l|c|c|c|c|c|c|} 
\hline
\textbf{Layout} & $\rho$ (M$\Omega$/sq.) &  Ground-pitch & Ground-type & Dead-zone & DOCA & Geometric  \\
\textbf{} & beam facility & (mm) & & (mm) & (mm) & acceptance (\%) \\
\hline
%\textbf{SG1} &  70 & 6 & 2 & 0.85 & 66 & 134 \\
%\hline
%\textbf{SG2} &  65 & 12 & 1.2 & 0.45 & 90 & 209 \\
%\hline
               & 64 - PSI   &     &      &     &      &    \\
\textbf{SG2++} &            & 12 & grid  & 0.6 & 0.25 & 95 \\
               & 10 - X-ray &     &      &     &      &    \\
\hline
             & 54  - PSI    &     &      &     &      &     \\
\textbf{DRL} &              & 6   & dot  & 0   & 7    & 100 \\
             & 65 - X-ray   &     &      &     &      &     \\
\hline
\textbf{PEP} & 10 - X-ray   & 6   &      & 1   & 0.475& 66 \\
             &              &     & grid &     &      &     \\
%\hline
\textbf{PEP++}&             & 12  &      & 0.85& 0.400& 93 \\
\hline
%\textbf{SRL} &  70 & 100 & 0 & 5.5 & 100\\
%\hline

\end{tabular}
\caption{Resistive and geometrical parameters of the HR layouts. In all the $\upmu$-RWELL layouts the DOCA can be defined as the minimum distance between a grounding line and the closest amplification well.}
\label{tab:det-caracteristiche}
\end{center}
\end{table}

The validation of the different layouts includes: high intensity local irradiation with the X-ray facility at LNF as well as pion/muon beams at PSI/CERN; full irradiation for ageing studies at the Calliope gamma facility at the ENEA Casaccia.
\subsection{Performance of the HR-layouts with pion beam}
\label{HR-performance}
The performance of the DRL and SG layouts have been measured with a high intensity pion beam at the $\uppi$M1 of PSI.
The experimental set-up used in the beam tests is composed of:
\begin{itemize}
\item two couple of plastic scintillators (up-stream/down-stream), providing the DAQ trigger
\item two external triple-GEM trackers equipped with 650 $\upmu$m pitch X-Y strip read-out with analog APV25 front-end electronics \cite{apv}, defining the particle beam with a spatial accuracy of the order of 100 $\upmu$m 
\item six $\upmu$-RWELL detectors based on different resistive layouts, equipped with 0.6$\times$0.8 cm$^2$ pads and read-out with APV25 and current monitored
\end{itemize}
The gaseous detectors have been operated with Ar/CO$_2$/CF$_4$ (45/15/40) gas mixture. The relevant physical quantities are reported in this paper as a function of the detector gas gain to take into account small manufacturing differences in the amplification stage of the $\upmu$-RWELL prototypes (i.e. well diameter and shape).
\subsubsection{Efficiency studies}
\begin{figure}
\begin{center}
\includegraphics[width=0.70\hsize]{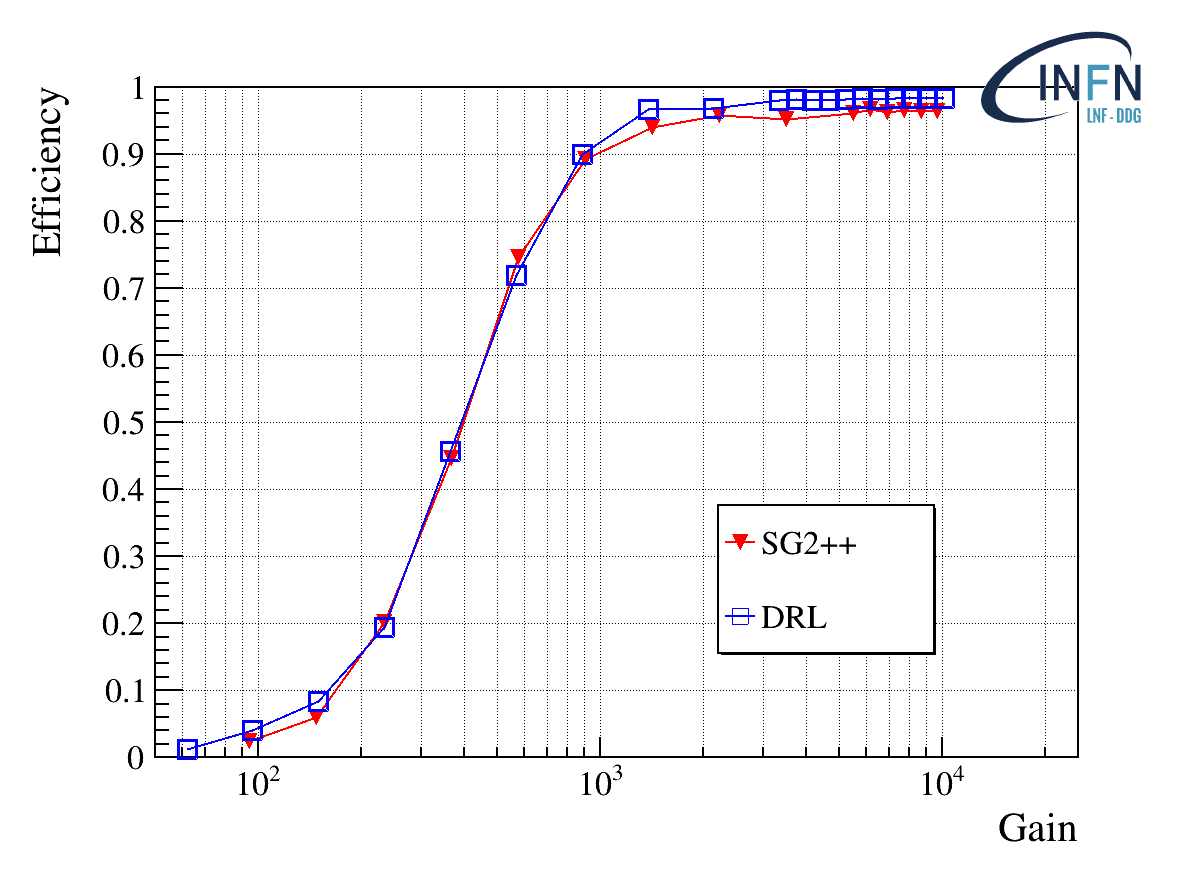}
\end{center}
\caption{Efficiency as a function of the gas gain for the DRL and SG layouts. The measurement has been performed equipping the detectors with APV25 front-end electronics.}
\label{fig:global-efficiency-HR}
\end{figure}

\noindent In Fig.~\ref{fig:global-efficiency-HR} the efficiency of the SG and DRL layouts is reported as a function of the detectors gain. The measurement has been performed with a flux of $\sim$300 kHz/cm$^2$ $\pi^{-}$ (350 MeV/c) and an average beam spot of 5$\times$5 cm$^2$ (FWHM$^2$). The efficiency has been evaluated considering a fiducial area of 5$\times$5 pads around the expected hit. At a gain of 5000 the DRL shows an efficiency of 98$\%$, while the SG tends to about 97$\%$, larger than its geometrical acceptance.

\subsubsection{Rate capability measurement}
\begin{figure}
\begin{center}
\includegraphics[width=0.70\hsize]{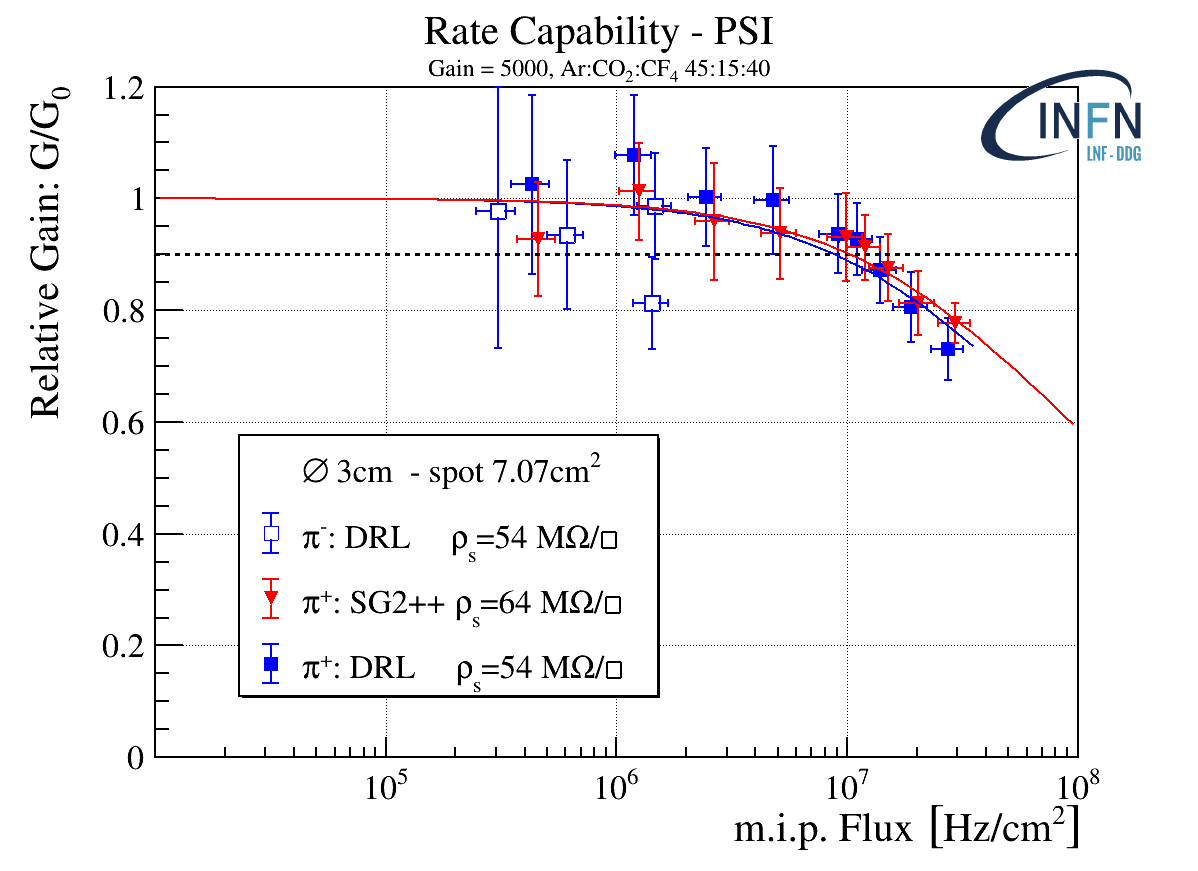}
\end{center}
\caption{Normalized gas gain for the HR layouts as a function of the pion flux. The function used to fit the points is the one derived in \cite{uRwell}.}
\label{fig:rate-capability}
\end{figure}

The rate capability of the DRL and SG layouts has been measured at the PSI $\uppi$M1 facility that provides a quasi-continuous high-intensity secondary beam with a fluence of $\sim$10$^7$ $\pi^{-}$$/s$  and $\sim$10$^8$ $\pi^{+}$$/s$, for a momentum ranging between 270$\div$350 MeV$/c$.
%This measurement can be considered reliable because 
The dimension of the average beam spot, $\oslash$ $\sim$ 3 cm - 7 cm$^2$, has been tuned in order to be constant in the two orders of magnitude of the flux scan and larger than the basic grounding cells of the HR prototypes.
The result of this study is reported in Fig.~\ref{fig:rate-capability}.
The low rate measurements ($\leq$1 MHz/cm$^2$) have been performed with the $\pi^{-}$ beam, while the high intensity have been obtained with the $\pi^{+}$ beam. The detectors have been operated at a gain of about 5000.
The particle rate has been estimated with the current drawn by the GEM, that shows a linear behaviour up to several tens of MHz/cm$^2$ \cite{GEM-ELBA-2003}.
The beam spot has been evaluated with a 2-D gaussian fit of the hits reconstructed on the X-Y plane for each detector.
The gain drop observed at high particle fluxes is correlated with the ohmic behaviour of the detectors due to the presence of the DLC film. The larger the radiation rate, the higher is the current drawn through the resistive layer and, as a consequence, the larger the drop of the amplifying voltage. 
%The different behaviour of the HR-layouts depends on their resistivity and current evacuation scheme. 
The two HR layouts tested at PSI stand particle fluxes up to 10 MHz/cm$^2$ with a gain drop of 10\%, still corresponding to a full detection efficiency.

\subsection{Rate capability measurement with X-ray}
\begin{figure}
\begin{center}
\includegraphics[width=0.70\hsize]{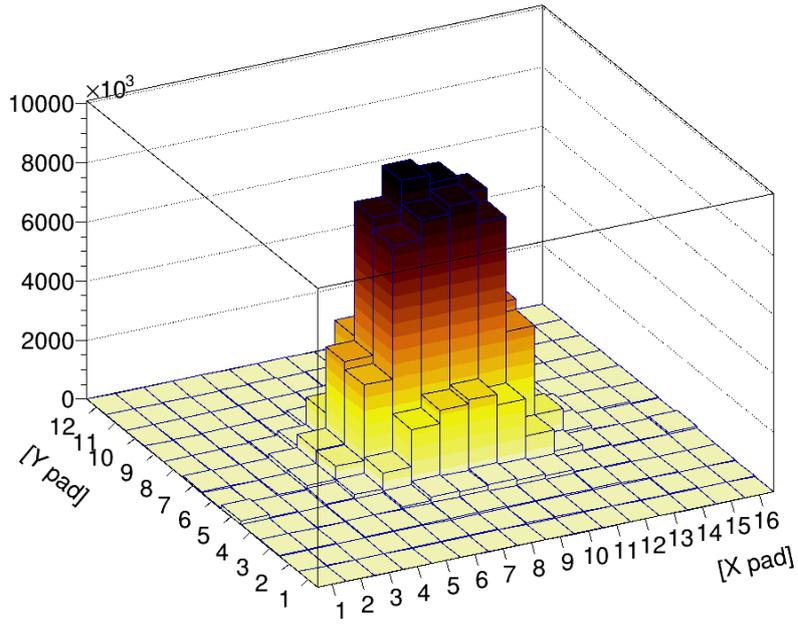}
\end{center}
\caption{X-ray irradiation spot as measured on the detectors equipped with APV25 front-end electronics.}
\label{fig:x_ray}
\end{figure}

\begin{figure}
\begin{center}
\includegraphics[width=0.70\hsize]{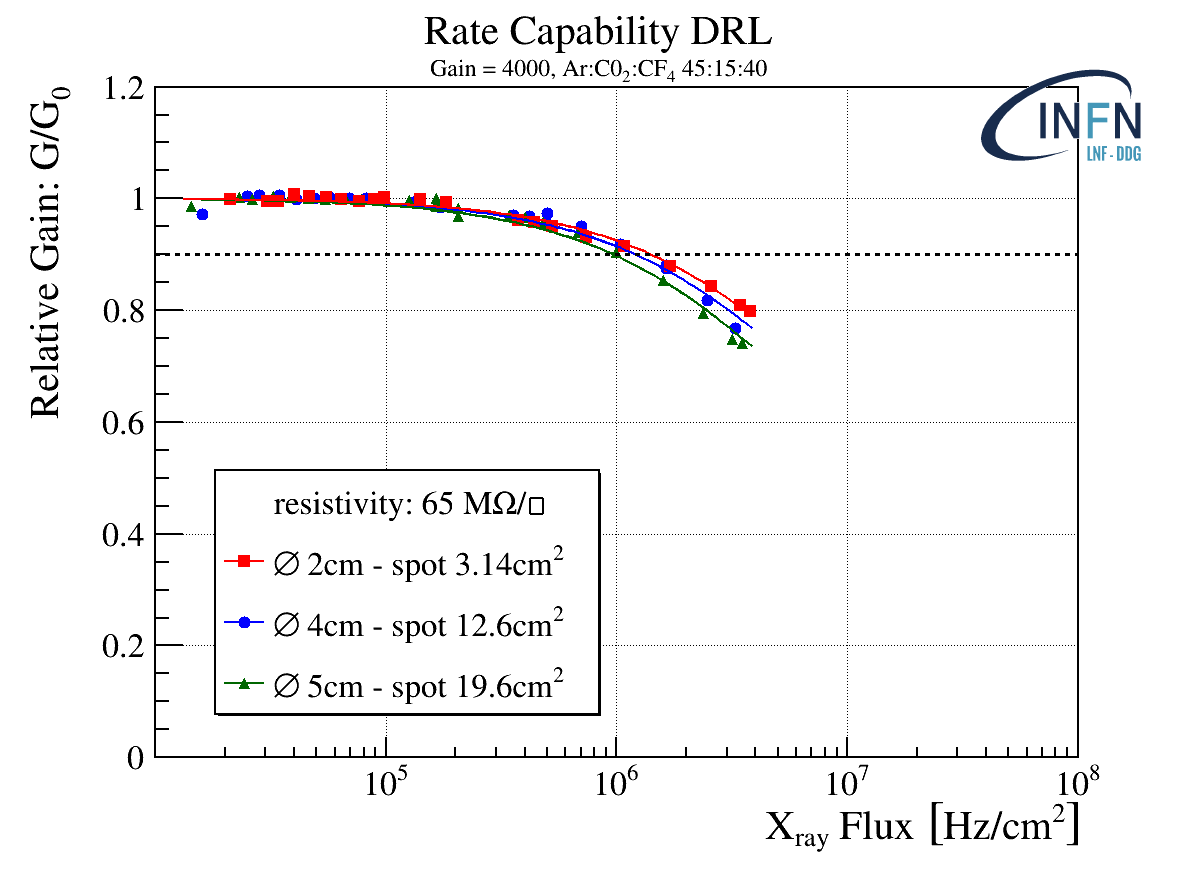}
\end{center}
\caption{Normalized gain for the DRL as a function of the X-ray flux and different beam spot at a gas gain of 4000.}
\label{fig:DRL}
\end{figure}

\begin{figure}
\begin{center}
\includegraphics[width=0.70\hsize]{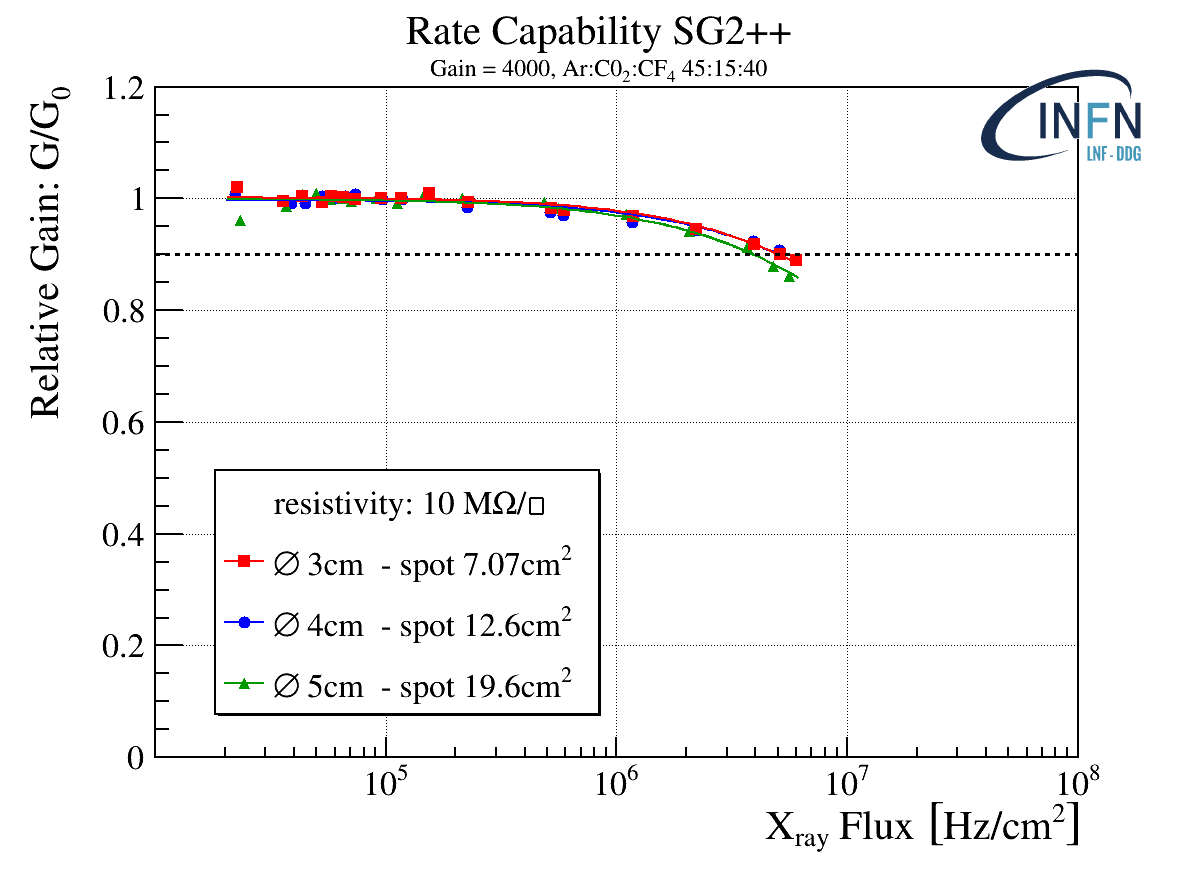}
\end{center}
\caption{Normalized gain for the SG as a function of the X-ray flux and different beam spot at a gas gain of 4000.}
\label{fig:SG}
\end{figure}

\begin{figure}
\begin{center}
\includegraphics[width=0.70\hsize]{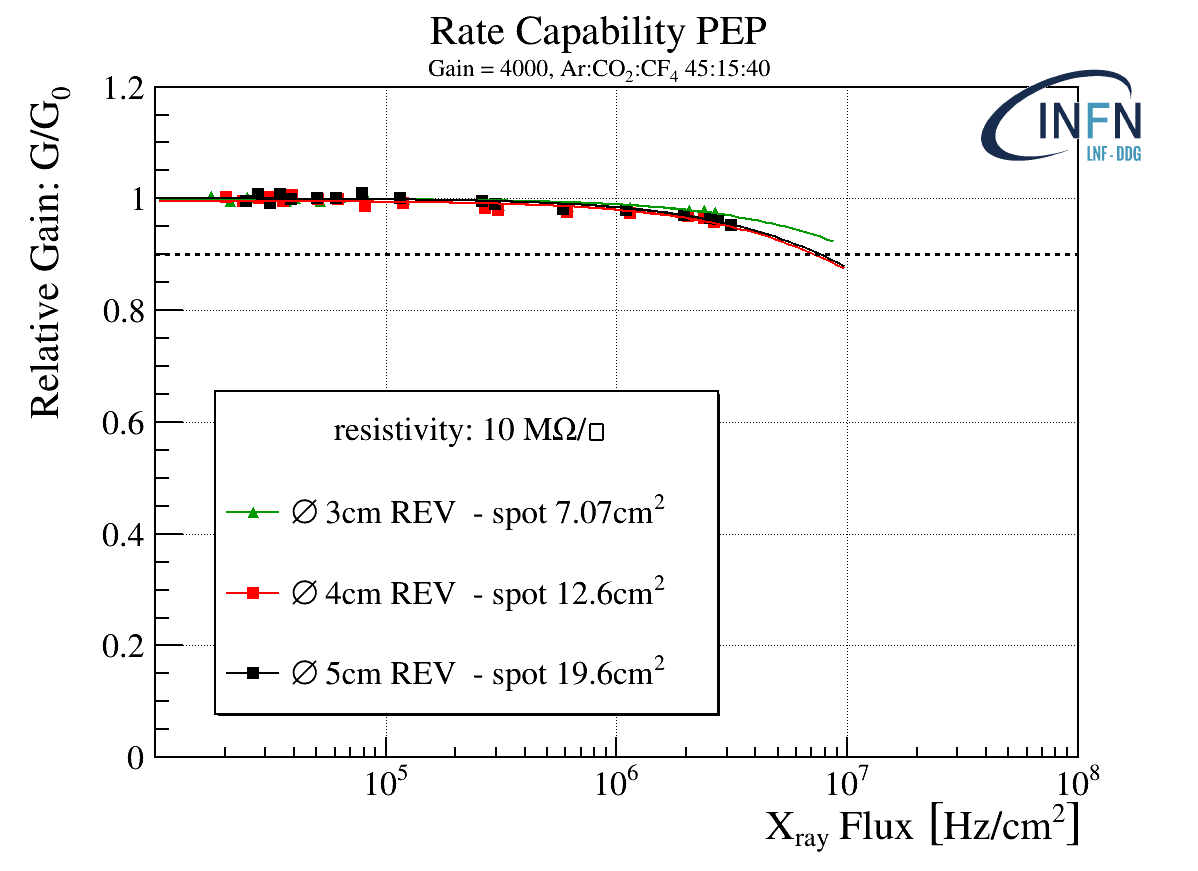}
\end{center}
\caption{Normalized gain for the PEP as a function of the X-ray flux and different beam spot at a gas gain of 4000.}
\label{fig:PEP}
\end{figure}

The rate capability of the HR layouts has been also performed with a high intensity 5.9 keV X-ray tube. Several 2 mm thick lead collimators with different hole diameters, from 1 to 5 cm, have been used to define an almost uniform X-ray spot (Fig.~\ref{fig:x_ray}).
%
%The gain drop effect due the particle flux on the HR layouts as a function of the beam spot has been investigated by a high intensity 5.9 keV X-ray tube. Several 2 mm thick lead collimators with different hole diameters, from 1 to 5 cm, have been used to define the X-ray spot (Fig.~\ref{fig:x_ray}).
%shows the irradiation measured with APV25 is nearly uniform and compatible the Pb hole dimension.
%
The normalized gas gain curves of the different HR layouts as a function the X-ray flux for several X-ray spots are shown in fig.~\ref{fig:DRL}, ~\ref{fig:SG}, ~\ref{fig:PEP}. The  measurements have been performed at a gas gain of 4000. Since the irradiated area is larger than the basic current evacuation cell defined by the density grounding network, the gain drop of the HR layouts as expected is almost independent on the spot size. 
The measured rate capability (with X-ray) of the HR layouts are:
\begin{itemize}
    \item $\sim$ 1 MHz/cm$^2$ for the DRL 
    \item $\sim$ 4 MHz/cm$^2$ for the SG
    \item $\sim$ 10 MHz/cm$^2$ for the PEP
\end{itemize}
In order to compare this result with the one obtained at PSI \cite{HR-layouts}, the rate capability with X-rays must be multiplied by a factor of three to take into account the different primary ionization between X-ray and m.i.p..
The different rate capability of the layouts is mainly due to the different DLC surface resistivity the detectors. 
%

%%%%%%%%%%%%%%%%%%%%%%%%%%%%%%%%%%%%%%%%%%

\section{Muon tracking at FCC-ee and CepC}

The $\upmu$-RWELL inherits the best characteristics of the GEM and MicroMegas
detectors, while further simplifying the manufacturing process. A technology transfer with a few
industries is already in place and this should allow to be able to mass manufacture this
detector in the near future at a very competitive cost of about 1000 euros/m$^2$. 
This detector is therefore an ideal
candidate to be the technology of choice for building future large muon detection systems~\cite{Giacomelli:muon-detection-FCCee,Giacomelli:mpgd-muon-detectors}. The $\mu$RWELL technology is in fact envisaged to realise the muon detection system and the preshower of the IDEA detector concept~\cite{IDEA:test-beam} that is proposed for the FCC-ee~\cite{Abada:2019lih} and CEPC~\cite{cepc:cdr-2018} future large circular leptonic colliders.

\begin{figure}
\begin{center}
\includegraphics[width=0.50\hsize]{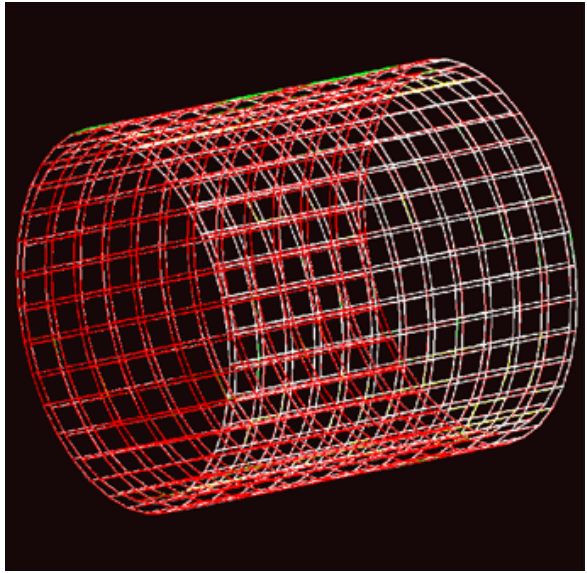}
\end{center}
\caption{Schematic drawing of one of the three barrel stations of the IDEA muon detector. The modular structure of the station, based on a large mosaic of same size $\upmu$-RWELL detectors is clearly visible.}
\label{fig:IDEA-muon}
\end{figure}

Both the preshower and the muon detector would follow the IDEA geometry with a central cylindrical barrel region closed at the two extremities by two endcaps to ensure hermeticity. The preshower detector would consist of a single layer of $\upmu$-RWELL detectors in both the barrel and the endcap regions.  The preshower will have a modular design and will be made of a mosaic of $\upmu$-RWELL detectors with two layers of strip readout placed perpendicularly to each other. In order to achieve a good position resolution, of the order of 60 $\mu m$, a fine strip pitch of 400 $\mu m$ is envisaged.
The muon detection system will instead consist of three muon stations in the barrel region, at increasing radial distance from the interaction point, housed within the iron yoke that closes the solenoidal magnetic field. Each station will consist of a large mosaic of $\upmu$-RWELL detectors.
In order to profit of the industrial production capabilities of this technology a modular design has been adopted for both the preshower and the muon detection system. The basic $\upmu$-RWELL "tile" will have an active area of 50x50~cm$^2$, and will have a design for the preshower and the muon system almost identical, the main difference being the pitch between the readout strips, that will be finer for the preshower to obtain the best possible position resolution. For the muon detector a lower position resolution is perfectly adequate and this reduces the number of readout channels.
The two layers of strips will both have a strip pitch of, respectively, 400 $\mu m$ and 1 mm, for the preshower and the muon system. This translates into a total of 500$\times$2 strips and consequently 1000 readout channels per tile for the muon detectors. 
The detector dimensions, the strip pitch and width are a compromise between the largest $\upmu$-RWELL detector that can be industrially mass produced while maintaining a not too high input capacitance to the readout electronics.
In table~\ref{tab:barrel-muon} are listed the dimensions, number of basic $\upmu$-RWELL tiles and readout channels of the 3 muon stations. The two endcaps are made of 3 disks, at increasing distance from the interaction point in the direction along the beam line equipped with similar $\upmu$-RWELL tiles. 
\begin{table}[]
\centering
    \begin{tabular}{|c|c|c|c|c|c|c|c|}
     \hline 
     Station & Radius  & Length  & Strip pitch & Strip length & Area     & N. of tiles & Channels    \\
             &  (m)    &  (m)    & (mm)        &  (mm)        & (m$^2$)  &       &       \\
     \hline 
        1    &  4.52   &  9.0    & 1           & 500          &  260     &  1040 & 1040000\\
        2    &  4.88   &  9.0    & 1           & 500          &  280     &  1120 & 1120000\\
        3    &  5.24   &  10.52  & 1           & 500          &  350     &  1400 & 1400000\\
     \hline 
   \end{tabular}
   \caption{Dimensions of the 3 IDEA barrel muon stations, together with the number of individual detector tiles and the number of readout channels.}
\label{tab:barrel-muon} 
\end{table}
In total, between the barrel and endcap muon stations, there would be about 5800 $\upmu$-RWELL tiles with a total of roughly 6 million readout channels. A schematic drawing of a barrel station of the muon detector is shown in Fig. ~\ref{fig:IDEA-muon}. Each $\upmu$-RWELL would be able to identify muon hits with 98-99\% efficiency and measure the coordinate perpendicular to the strip direction with a precision of about 200 $\mu$m. 
Such a detector would be able to provide 3 three-dimensional space points (the third coordinate coming from the known radial position of the $\upmu$-RWell tile) and from these reconstruct the tracks crossing the muon stations. These muon tracks could then be used to complement the tracks reconstructed in the central tracker providing the best momentum measurement of muons. The muon detector could also reconstruct charged particle tracks coming from the decays of hypothesized long-lived particles (LLP) that would produce a secondary vertex outside ($\geq$ 2.5 m from the primary interaction point) of the central tracker. This could significantly enhance the detector capabilities to study these interesting signatures of possible new physics beyond the Standard Model. 

Since the $\upmu$-RWELL technology has not yet been used to realise a full detector system a vigorous R\&D program to study integration issues will be carried out in the coming years. The detailed layout of the muon detector, together with all its services, will have to be accurately developed and optimised. The optimal characteristics of the basic $\upmu$-RWELL tile, like gas gap, DLC resistivity, strip pitch, and gas amplification will be finalised to match the requirements of the IDEA muon detection system. Another important aspect of the R\&D program will be the design and development of a dedicated front-end electronics based on a custom-made ASIC for the muon detector readout.  

\section{Low mass cylindrical Inner Trackers}

\begin{figure}
	\centering
	\begin{minipage}[t]{\textwidth}
		\centering
		\vspace{0pt}
		\includegraphics[width=.6\textwidth]{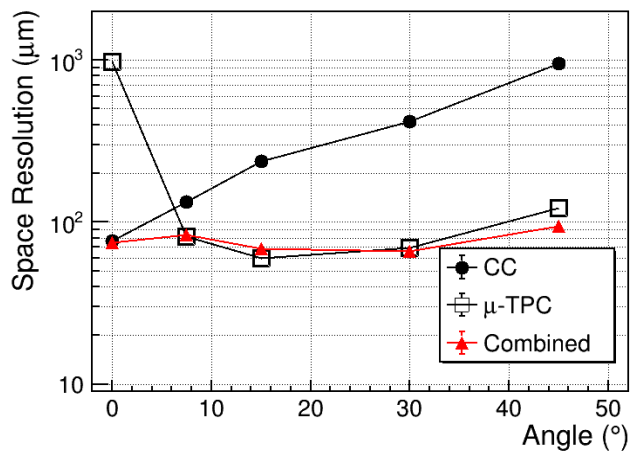}
		\caption{Comparison of the two algorithms with their combined reconstruction, at a drift field E$_{D}$=1 kV/cm.}
		\label{1utpc}
	\end{minipage}
\end{figure}

\begin{figure}
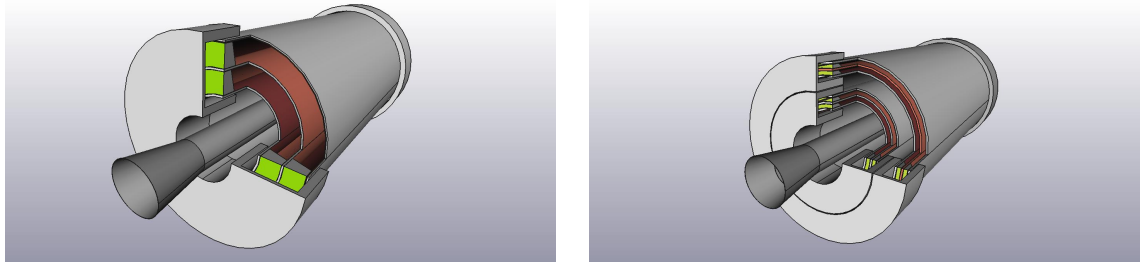

    \centering
    \begin{subfigure}
    {
    \includegraphics[scale=0.24]{Figures/CREMLINplus_single_large_gap.jpeg}
    }   
    \end{subfigure}
    \hspace{-0.2cm}
    \begin{subfigure}
    {
    \includegraphics[scale=0.24]{Figures/CREMLINplus_double_small_gap.jpeg}
    }
    \end{subfigure}
    \caption{Inner Tracker ideas: double large gas gap (top) and pair of double thinner gas gap (bottom).}
    \label{fig:ITDraw}
\end{figure}

The main feature of the $\upmu$-RWELL technology is to have the amplification stage and the readout board embedded, through the thin resistive layer, in one single element. The possibility to realize this element with flexible substrates (namely polyimide) makes the technology suitable for non-planar geometry.\\
Exploiting this feature a C-RWELL has been designed and it is under development, in the framework of the EU project CREMLINplus, as a low-mass inner tracker at the Super Charm-Tau Factory (SCTF) scheduled to be realized in Sarov (RU). 
In addition being designed as an “openable” and “modular” detector it will be a highly reliable and performing IT, while the spark suppression mechanism, intrinsic to the µ-RWELL technology, makes the operation of this detector more safe with respect to previous generation of cylindrical MPGD based devices, developed ten years ago in the framework of KLOE2 Collaboration \cite{Kloe2}, and successively by BESIII CGEM group \cite{CGEM-BES3}.
Two ideas are under study, both based on a common double-faced cathode layout. In one case (Fig. \ref{fig:ITDraw} (left)) two large radial gaps for a 10cm global sampling gas along the radial direction has been considered, while in the second case (Fig. \ref{fig:ITDraw} (right)) four thinner gaps  for a 4cm global sampling gas are foreseen.
Depending on the material choice the former layout could be realized with a global material budget in the range 0.75$\div$0.86\% X$_{0}$, while the latter layout in the range 1.46$\div$1.72\% X$_{0}$.\\
For both layouts, the cylindrical $\upmu$-RWELL\_PCB is divided in three "roof tiles" detectors that thanks to the  the possibility to open (and re-close) the cylindrical support, are removable  in order to be replaced in case of malfunctioning.\\ 
A 1 cm large single drift-gap prototype, composed of a coaxial cylindrical anode and cathode structures, has been designed and realized in by the  LOSON S.r.l., a company with a remarkable expertise in composite elements. The substrate for the electrodes is Millifoam.
The dimensions and all relevant numbers of the C-RWELL prototype are summarized in table~\ref{tab:C-RWELL-proto}.\\
The reconstruction of the particle track traversing the gas sensitive gap of the prototype will be based on a combination of two algorithms: Charge Centroid (CC) and $\upmu$-TPC (\cite{uTPC-RWELL}).
Tests performed with planar prototypes shows that an almost uniform space resolution below 100~$\upmu$m over a wide angular range of track incidence (0$\div$45$^{\circ}$) is obtained. As shown in  Fig.~\ref{1utpc}, at low drift fields the measured space resolution improves reaching values down to 65~$\upmu$m.
\begin{table}[]
\centering
    \begin{tabular}{|c|c|c|c|c|c|c|}
     \hline 
     anode dia. & cathode dia  & drift gap  & active length &  HV chs & r/out chs & strip pitch \\
         (mm)   &  (mm)        &    (mm)    &   (mm)        &           &              &    (mm)   \\
     \hline 
        168.5   &  188.5       &    10      &    600        &     12    &     768      &   0.680    \\
     \hline 
   \end{tabular}
   \caption{Dimensions and relevant numbers of the C-RWELL prototype}
\label{tab:C-RWELL-proto}
\end{table}
\subsection{The cathode}
The cathode is the  outermost electrode of this prototype, it has been stratified starting from a 50+5$\mu$m thick kapton-copper foil (Fig. \ref{fig:catLoson}). The stratification has been continued with a 100$\mu$m thick skin of fiberglass (fig. \ref{fig:fgLoson}), a layer of a 3 mm thick Millifoam  (Fig. \ref{fig:mfLoson}), a second fiberglass skin and a copper layer operating as Faraday cage. The flanges at the two edges of the cathode are made of Peek (standing for polyether ether ketone). 
\subsection{The anode}
The cylindrical anode is composed of three roof-tiles, each then covering 120$^\circ$ (Fig. \ref{fig:3-tiles}). Each flexible $\upmu$-RWELL\_PCB, designed at LNF and built at the CERN-PH-DT Workshop (Fig. \ref{fig:detector-tiles}), is equipped with axial strips parallel to the axis of the cylinder. The roof-tile support for the $\upmu$-RWELL\_PCB is done with a 3 mm thick Millifoam\textregistered (Fig. \ref{fig:mf_anode}): this material must be handled very carefully, so it has been quite challenging to obtain a certified procedure to realize such light roof tile at the same time robust enough to keep the $\upmu$-RWELL\_PCB  at a given shape with a very tight mechanical tolerance ($\pm$100$\mu$m) on the radial direction.\\ 
The anode flanges, realized in Peek, have been designed with proper windows to host the boards for the HV distribution (Fig. \ref{fig:hvb}) and the interface boards for the front-end electronics (Fig. \ref{fig:feeb}). The presence of these openings required a dedicated test for the gas tightness of the flange-boards system. The boards are actually glued on the inner surface of the flange with Araldite\textregistered 2011. The gas tightness test has been performed with a dedicated tool Fig. \ref{fig:gastest}, where the system has been flushed with nitrogen, comparing the entering and the exiting flow; a second test has been done setting a 20 mbar over-pressure (condition much worse than the operating condition, that foresee $<$5 mbar gas over-pressure) inside the flange and monitoring the pressure drop as a function of the time. The over-pressure decreased of about 1 mbar after 2 hours, when the test was stopped and considered successful.\\
The construction of the prototype has been completed in December 2021 and the final assembly of the whole detector, including the flexible detector tiles, will be performed in the next months.

\begin{figure}
    \centering
    \begin{minipage}[b]{0.32\textwidth}
    \includegraphics[width=\textwidth]{Figures/laminazione_catodo.jpg}
    \caption{Cathode lamination.}
    \label{fig:catLoson}
    \vspace{-0.36cm}
    \end{minipage}
    \begin{minipage}[b]{0.67\textwidth}
    \includegraphics[width=\textwidth]{Figures/fiberglass1.jpg}
    \caption{Fiberglass deposition on cathode.}
    \label{fig:fgLoson}
    \end{minipage}
\end{figure}
\begin{figure}    
    \centering
    \includegraphics[scale=0.6]{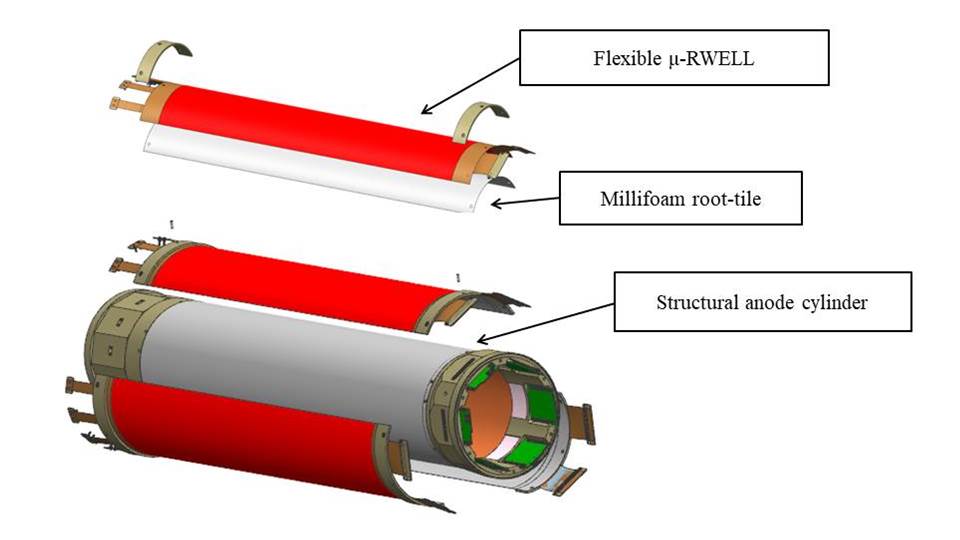}
    \caption{Sketch of the assembly of the three roof-tile detectors on the anode cylinder. The anode cylinder is made of composite material: FR4 - Millifoam - FR4 sandwich plus an additional finely machined layer of Millifoam.}
    \label{fig:3-tiles}
\end{figure}
\begin{figure}
    \centering
    \includegraphics[scale=0.05]{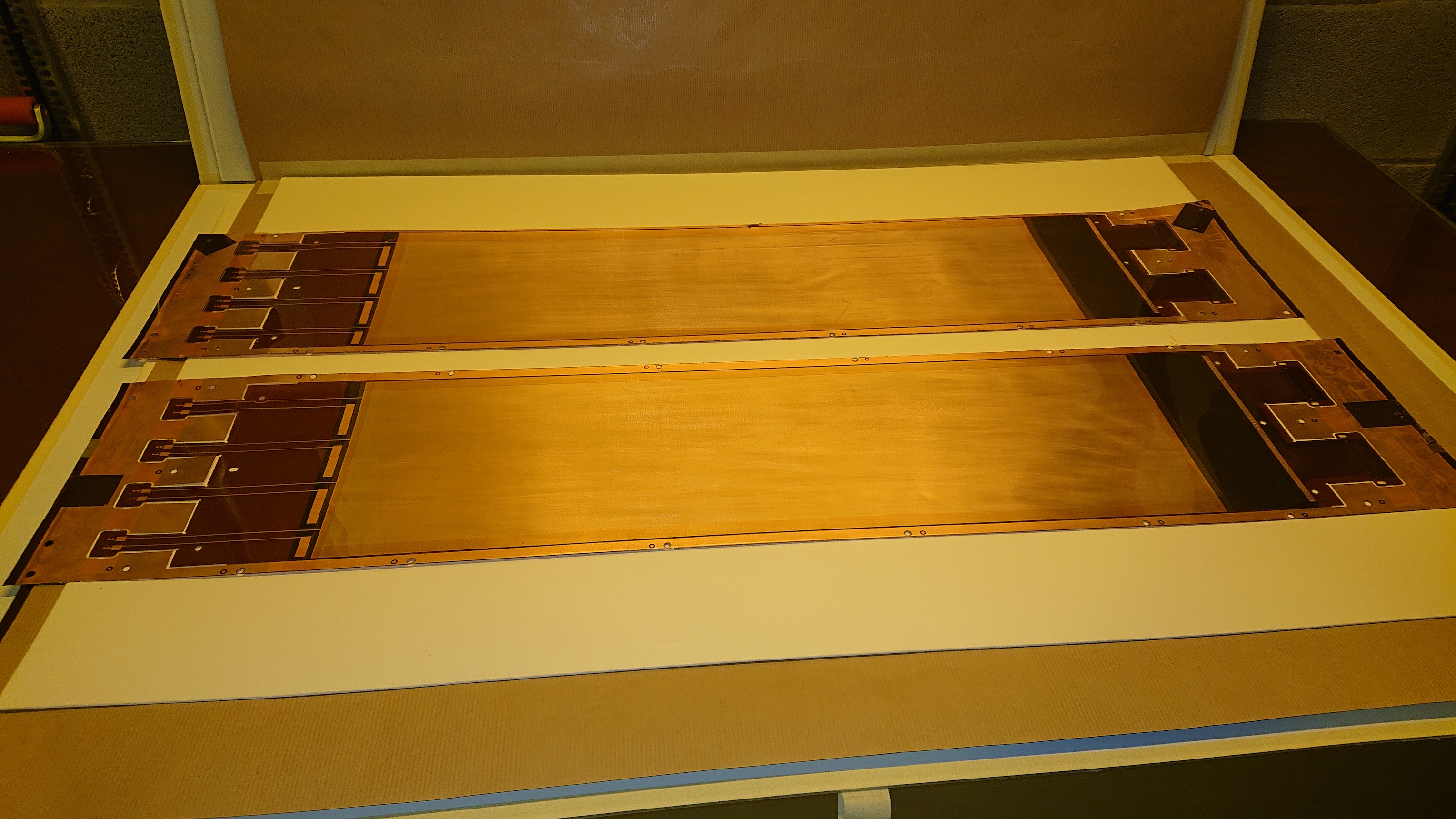}
    \caption{Flexible $\mu$-RWELL\_PCB detector tiles manufactured at the CERN-PH-DT Workshop
.}
    \label{fig:detector-tiles}
\end{figure}
\begin{figure}    
    \centering
    \includegraphics[scale=0.8]{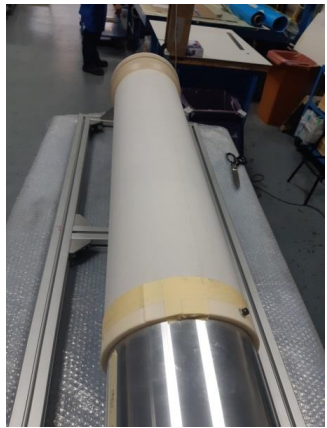}
    \caption{Millifoam layer glued on cathode.}
    \label{fig:mfLoson}
\end{figure}
\begin{figure}
    \centering
    \includegraphics[scale=0.1]{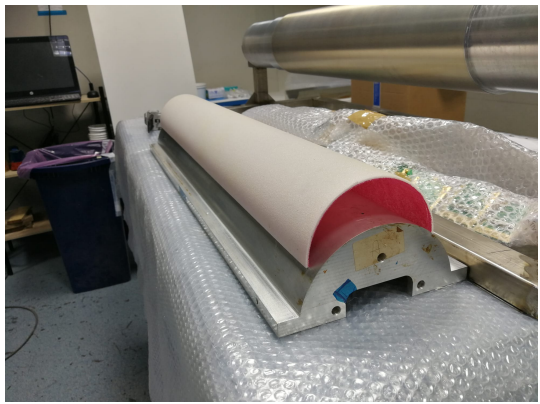}
    \caption{Coupling test done to validate the Millifoam support for the flexible $\upmu$-RWELL\_PCB.}
    \label{fig:mf_anode}
\end{figure}
\begin{figure}
    \centering
    \begin{minipage}[b]{0.54\textwidth}
    \includegraphics[width=\textwidth]{Figures/hvboard.jpg}
    \caption{A HV distribution board glued on the flange.}
    \label{fig:hvb}
    \vspace{0.4cm}
    \end{minipage}
    \hspace{0.5cm}
    \begin{minipage}[b]{0.3\textwidth}
    \includegraphics[width=\textwidth]{Figures/feeboard.jpg}
    \caption{FEE intermediate board glued on the dedicated flange.}
    \label{fig:feeb}
    \end{minipage}
\end{figure}

\begin{figure}
    \centering
    \begin{subfigure}
    {
    \includegraphics[scale=0.6]{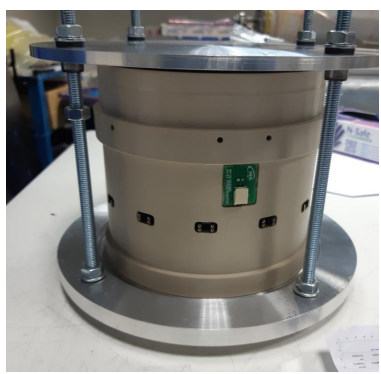}
    }   
    \end{subfigure}
    \begin{subfigure}
    {
    \includegraphics[scale=0.6]{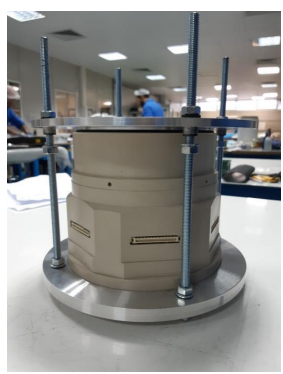}
    }
    \end{subfigure}
    \caption{Setup for the gas tightness tests of the HV flange (left) and FEE flange (right).}
    \label{fig:gastest}
\end{figure}

\begin{figure}
    \centering
    \begin{subfigure}
    {
    \includegraphics[scale=0.04]{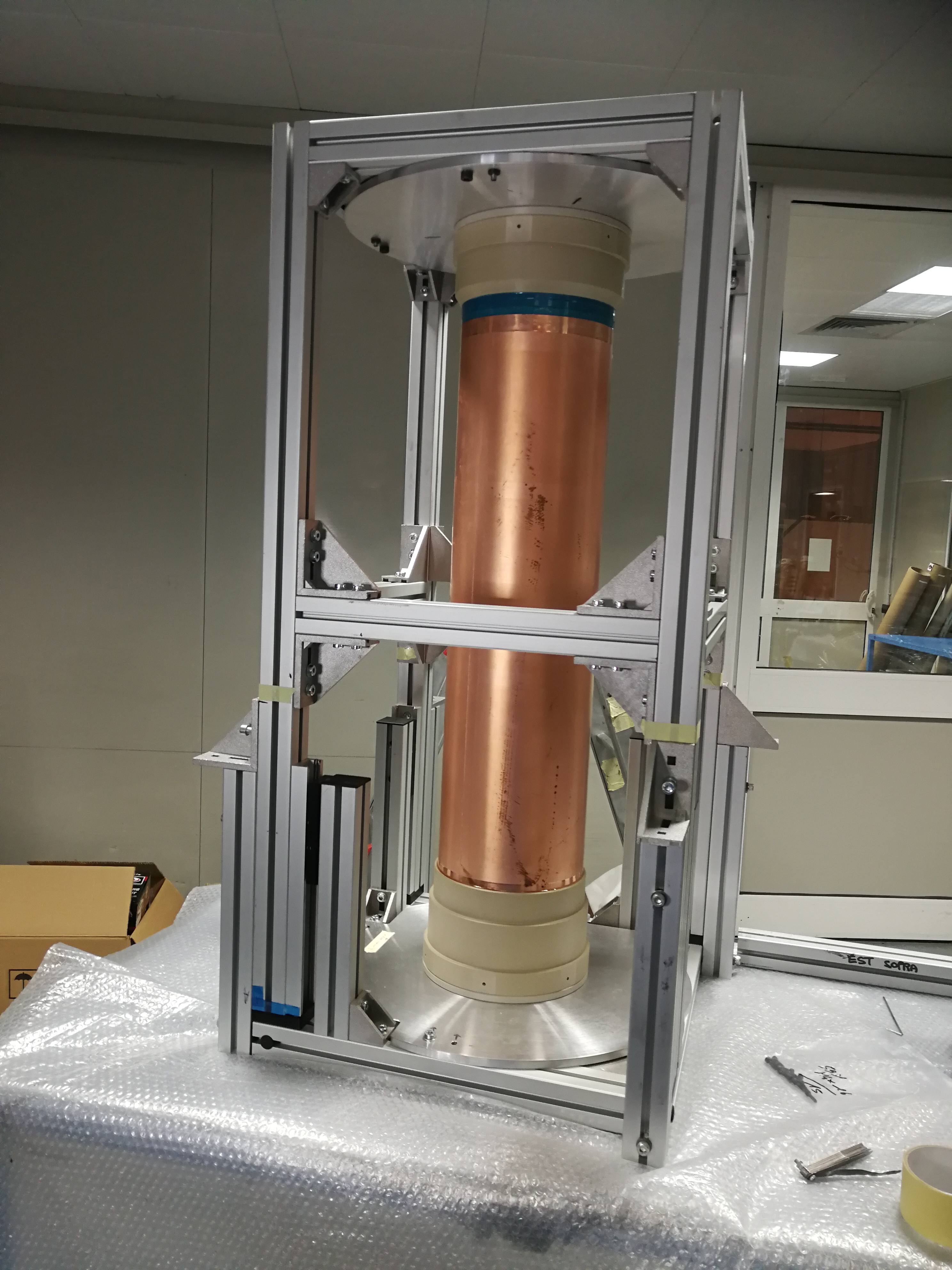}
    }
    \end{subfigure}
    \begin{subfigure}
    {
    \includegraphics[scale=0.05]{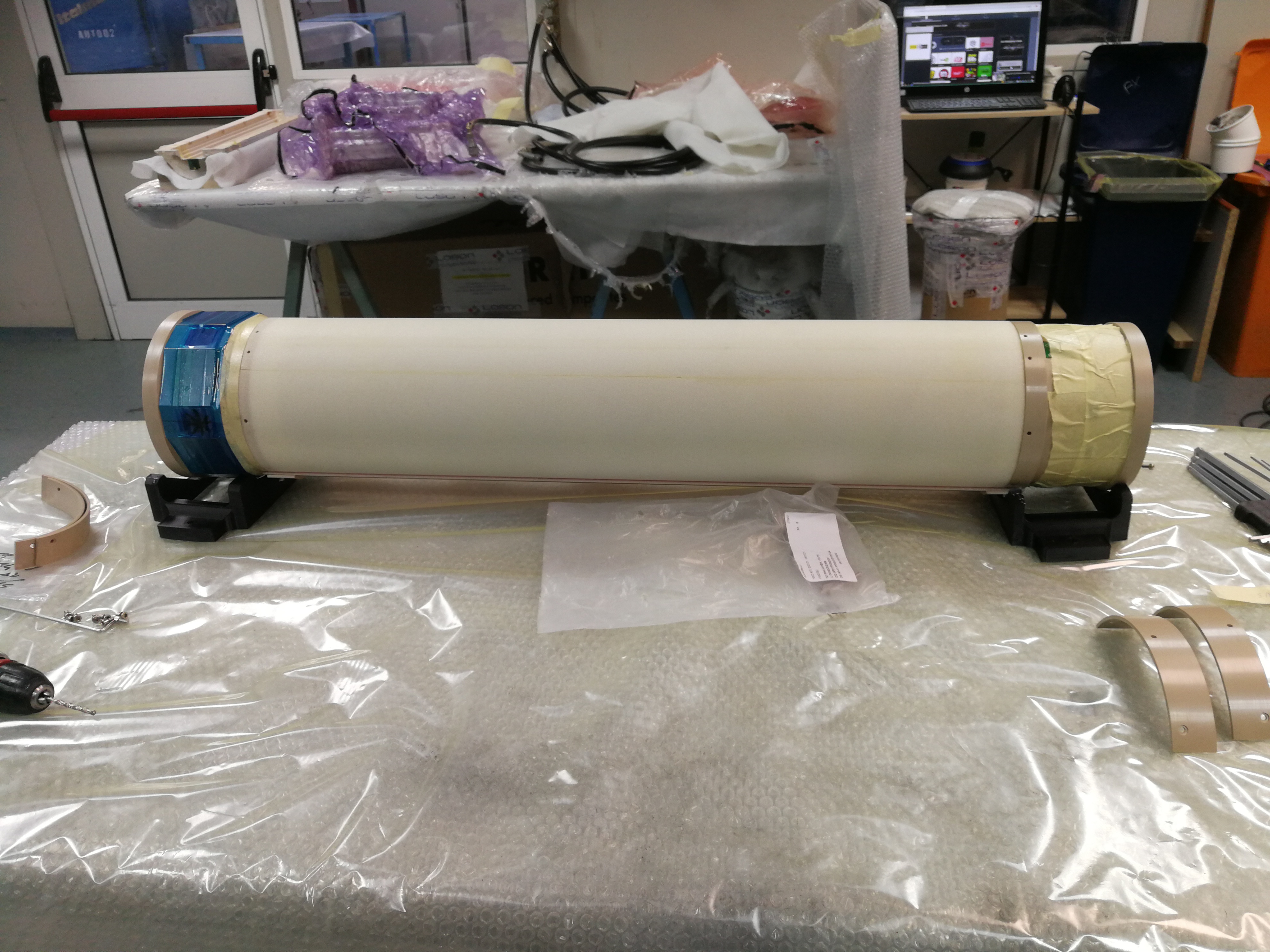}
    }
    \end{subfigure}
    \begin{subfigure}
    {
    \includegraphics[scale=0.04]{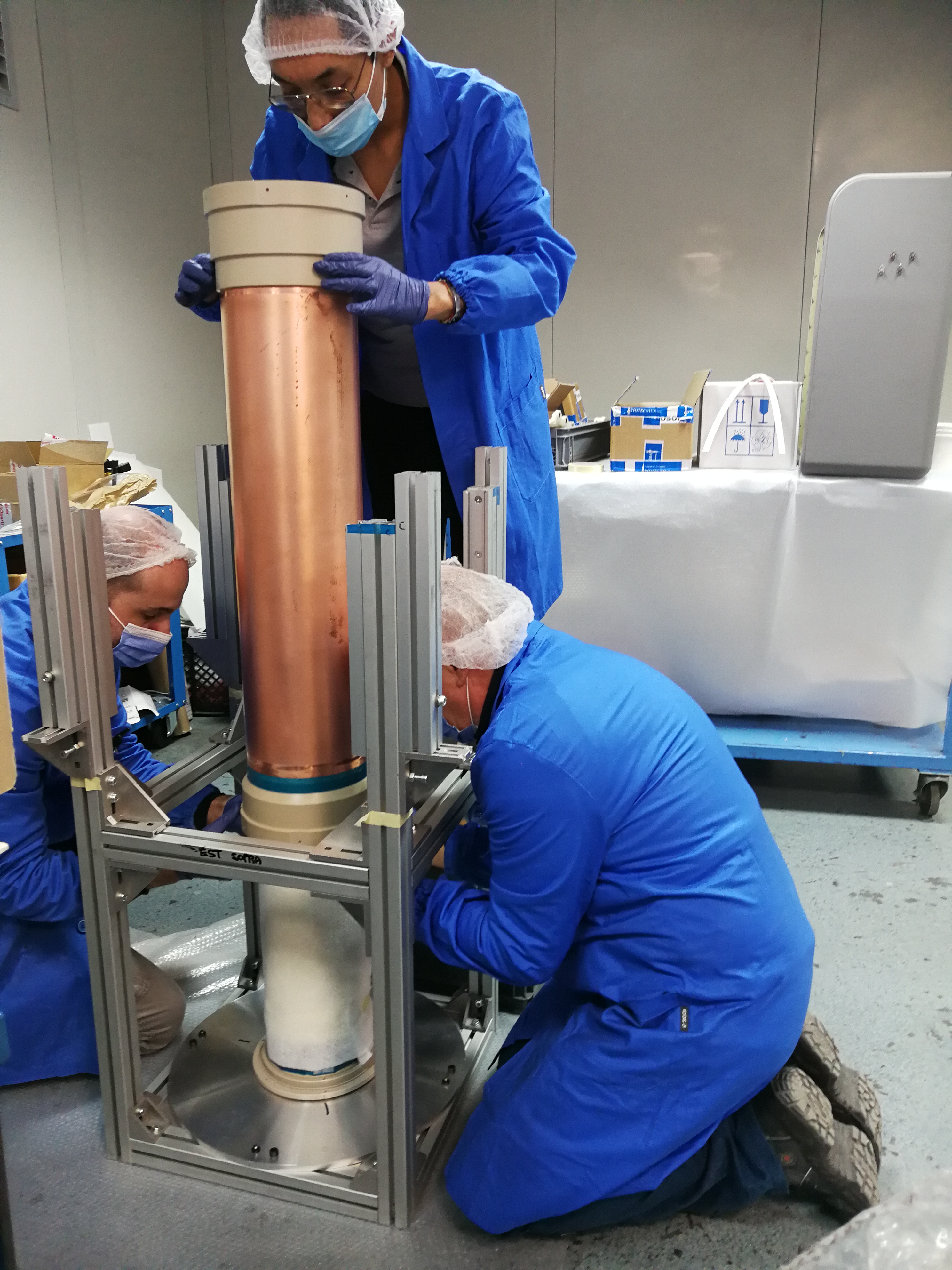}
    }
    \end{subfigure}
    \caption{Assembly of the prototype. From top left: the cathode completed mounted on the assembly machine; the rectified millifoam support for the $\upmu$-RWELL tiles; the manual insertion of the cathode on the cylindrical anode.}
    \label{dec2021}
\end{figure}

%%%%%%%%%%%%%%%%%%%%%%%%%%%%%%%%%%%%%%%%%%

%
%\begin{figure}
%\begin{center}
%\includegraphics[width=0.70\hsize]{Figures/}
%\end{center}
%\caption{}
%\label{fig:}
%\end{figure}
%

%%%%%%%%%%%%%%%%%%%%%%%%%%%%%%%%%%%%%%%%%%

%  If you would like to use BibTEX for the bibliography, please feel free to do so.  It is not required.

%  To use BibTeX,

%    1.  uncomment the following two lines, 
%    2.  comment out everything below from  \begin{thebibliography}{99}   to \end{thebibliography).
%    3.  create the file  myreferences.bib, and process this file in the usual way

%\bibliographystyle{JHEP}
%\bibliography{myreferences}  % file myreferences.bib

%%%%%%%%%%%%%%%%%%%%%%%%%%%%%%%%%%%%%%%%%
\newpage

%\section{Gas systems for particle detectors}

%\pubblock

%\Title{Gas systems for particle detectors}
\chapter{\centering Gas systems for particle detectors}
%\addcontentsline{toc}{chapter}{Gas systems for particle detectors}
\bigskip 

%\Author{R. Guida, B. Mandelli, M. Corbetta, G. Rigoletti}

\medskip

\medskip

%\snowmass

\def\thefootnote{\fnsymbol{footnote}}
\setcounter{footnote}{0}

\section{Introduction}
Gases are used as active medium in different research fields. Ionization processes are used in many types of particle GDs like proportional counters, MPGDS, resistive plate chambers (RPC). Furthermore, gases are also used as light radiator (Cherenkov radiation) in RICH detectors. Specific conditions are achieved with gases in reaction vessel (like it is the case of the CLOUD experiment which aims in re-creating process that happens in the atmosphere) or for the production of primary particles for all the accelerators. In some cases, the use of expensive and/or greenhouse gases (GHGs) cannot be avoided because of specific physics requirements. This is particularly important for the gas systems of the LHC experiments. Indeed, the gas volume involved can be as high as several hundred cubic meters. In addition, many different types of expensive and/or GHGs are used (i.e. Freon like C$_2$H$_2$F$_4$, C$_4$F${10}$, C$_3$F$_8$, CF$_4$, ... or expensive like Xenon, ...). Reducing the use of GHGs is nowadays a worldwide objective to which the scientific community wants to contribute. In this context and with the idea of preparing the very long-term operation, CERN has elaborated a strategy based on several action lines, based on the experience of the LHC experiments.

\section{Strategies for reducing the gas consumption}
The LHC gas systems extend from the surface building where the primary gas supply point is located to the service balcony on the experiment following a route few hundred meters long. The basic function of the gas system is to mix the different gas components in the appropriate proportion and to distribute the mixture to the individual chambers.
\\\\
At the LHC about 30 gas systems are delivering the proper gas mixture to the corresponding detectors. In few numbers the gas systems for the LHC experiments consist of about 300 standard racks, 70 PLCs and kms of pipe. In order to facilitate the construction and, later on, the maintenance, the gas systems were designed starting from functional modules with similar functionalities (i.e. elementary building blocks). Functional modules are, for example: mixer, pre-distribution, distribution, circulation pump, purifier, humidifier, membrane, liquefier, gas analysis, etc. For example, the mixer module is basically identical between every system, but it can be configured in order to satisfy the specific needs of each detector. This module-oriented design is reflected by the implementation: each system has a control rack where the PLC and all the other crates corresponding to all functional modules are located. The control software for the gas system runs in the PLC, while the crates collect all the I/O information from the corresponding modules and, finally, they are connected to the PLC through Profibus. This approach was also facilitating the installation work and the commissioning especially when all modules of a particular system were not ready for installation at the same time.
\\\\
Reliability, automation and stability are keywords for the CERN LHC gas systems. Only thanks to a reliable and fully automated systems the large gas systems infrastructure can be operated efficiently by a relatively small team. Stability is fundamental to ensure good detector performance (stable mixture composition, detector pressure, flows, …). Several approaches with different levels of complexity can be adopted to control the gas consumption. In particular four categories have been identified and they will be described in the following \cite{strategies}.

\subsection{Open mode gas system} 
\begin{figure}[h!]
	\centering
	\includegraphics[width=12cm]{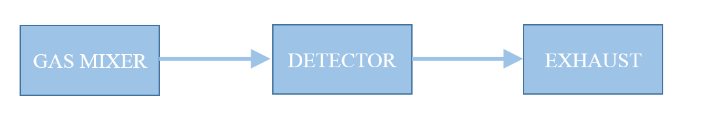}
	\caption{Schematic view of the simplest possible gas system: in open mode operation the gas mixture is vented after being used in the particle detector.}
	\label{gas1}
\end{figure}

In a very basic gas system, the gas mixture is prepared and it is sent to the detector. After passed through the detector, the gas mixture is vented to atmosphere (Figure \ref{gas1}). The big advantage of this approach consists in its simplicity. There is no particular need for a gas mixture monitoring after the detector. Unfortunately, for large detector systems it is no longer applicable since the operational cost and the consumption can become easily very high as well as the greenhouse emission to atmosphere in case GHGs are used. 
    
\subsection{Gas recirculation system}
\begin{figure}[h!]
	\centering
	\includegraphics[width=12cm]{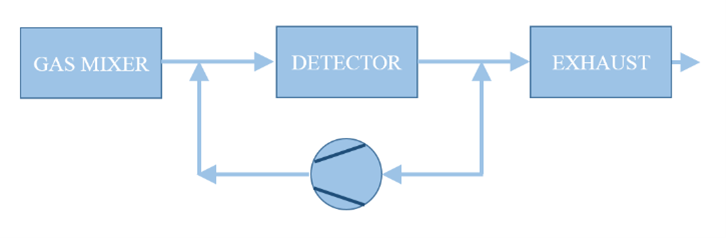}
	\caption{Schematic view of the gas recirculation system. After being used in the particle detector, the gas mixture is passing through dedicated purification modules and it is then in large part re-used. The maximum recirculation fraction is fixed by detector requirements or filtering capacity for specific impurities.}
	\label{gas2}
\end{figure}

In order to reduce operational costs and emissions, most of the LHC gas systems were already designed to operate in re-circulation mode. In this layout, after being used, the gas mixture passes through specific gas purification units. A small fraction of fresh gas is added before resending the mixture to the detector system. The maximum recirculation rate is fixed by detector leak or need of controlling impurities that cannot be filtered. Figure \ref{gas2} shows a schematic view of a gas recirculation system.
\\
The advantage of the gas recirculation system is that the gas consumption can be drastically reduced. In some systems the gas recirculation fraction is well above the 99\% level.
\\\\
The disadvantages of the gas recirculation systems are related to their complexity. Detector pressure regulation and flow are in general much more complex with respect to an open mode basic gas system.
Moreover, particular attention must be put during construction and the selection of components for the detector and related services. Indeed, the presence of standard gas leak or diffusion leaks can limit or compromise the operation in gas recirculation by requiring to reduce the recirculation fraction. While for the presence of standard gas leaks it is clear why more gas needs to be injected, for diffusion leaks trough the detector components the need is coming from the accumulation of large amount of N$_2$ in the gas stream that cannot be removed with online purification systems.   
\\\\
In addition, when operating with gas re-circulation, a constant monitoring of the mixture composition and of the presence of impurities is mandatory. Gas mixture is indeed the primary element influencing GD performance, as its quality and stability are fundamental for good and safe long-term operation. Common LHC gas systems impurities are O$_2$ and H$_2$O, removed before gas re-injection thanks to dedicated purifier modules, developed to maintain a good gas mixture quality in gas recirculation systems. The module contains two 24 liters cartridges which can be filled with the suitable purifier agent: in general, molecular sieves are used for water removal, metallic catalysts for Oxygen absorption or other specific materials. During normal operation the gas mixture is passing through one column, while the other is in regeneration or it has just completed the regeneration cycle and it is ready to be used. The purifier cycle is completely automated. When operating detectors with Fluorinated gases, also Fluoride impurities are created in the gas system and could potentially accumulate due to gas recirculation. Such impurities come from the breaking of the Fluorinated gas mixture components under the combined action of electric field, charge multiplication and high radiation background. Though purifiers modules were designed for trapping specifically O$_2$ and H$_2$O, it was found that they can efficiently also trap Fluoride impurities, reducing their accumulation in the recirculating system and the possible damage they could do  to GD operation \cite{fluoride}. Examples of effects of the presence of gas impurities in a Triple-GEM detectors are reported in the figures below, showing how significant is their impact and therefore how important it is to take actions for their removal.
\begin{figure}[h!]
	\centering
	\includegraphics[width=5.6cm]{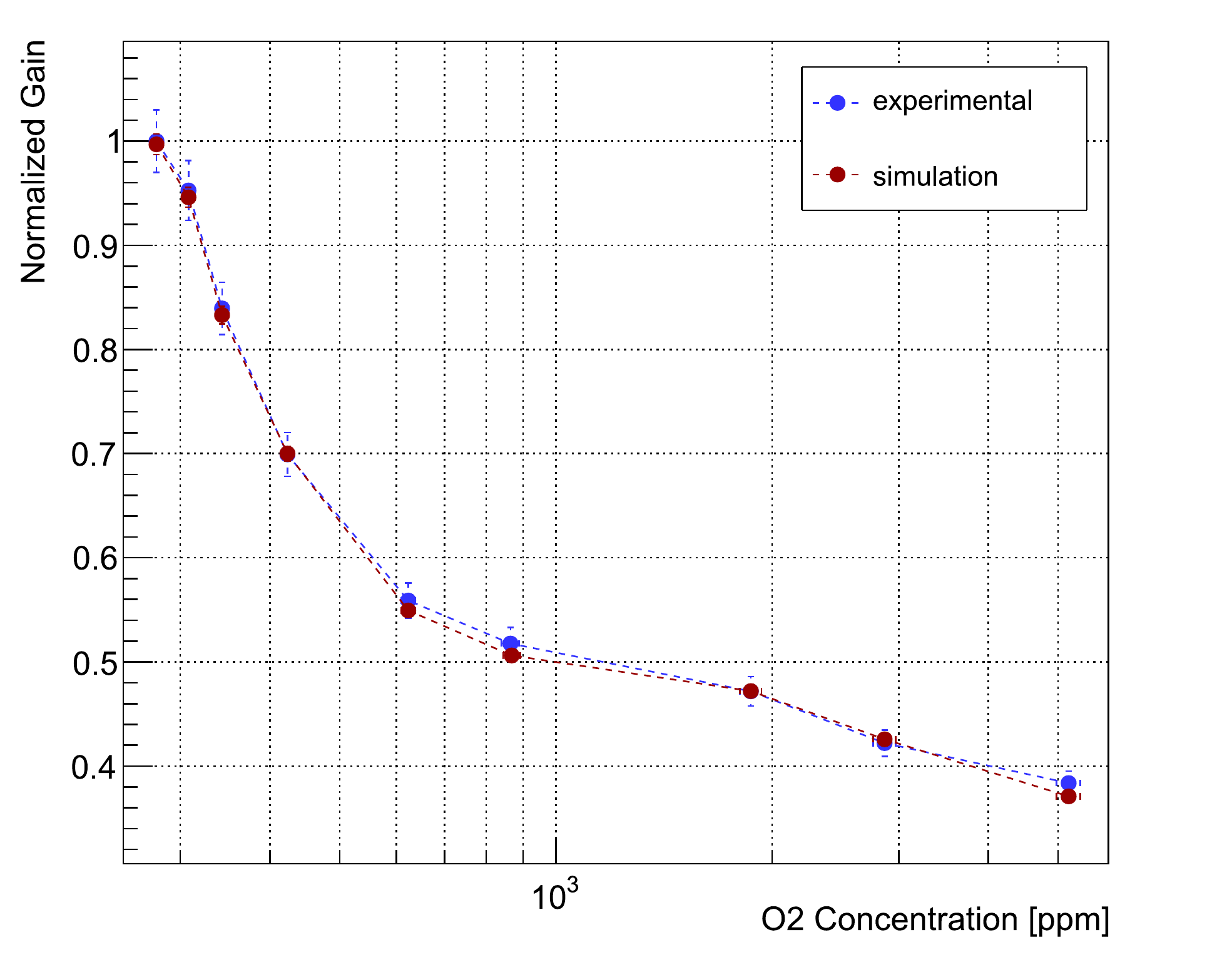}
	\includegraphics[width=7cm]{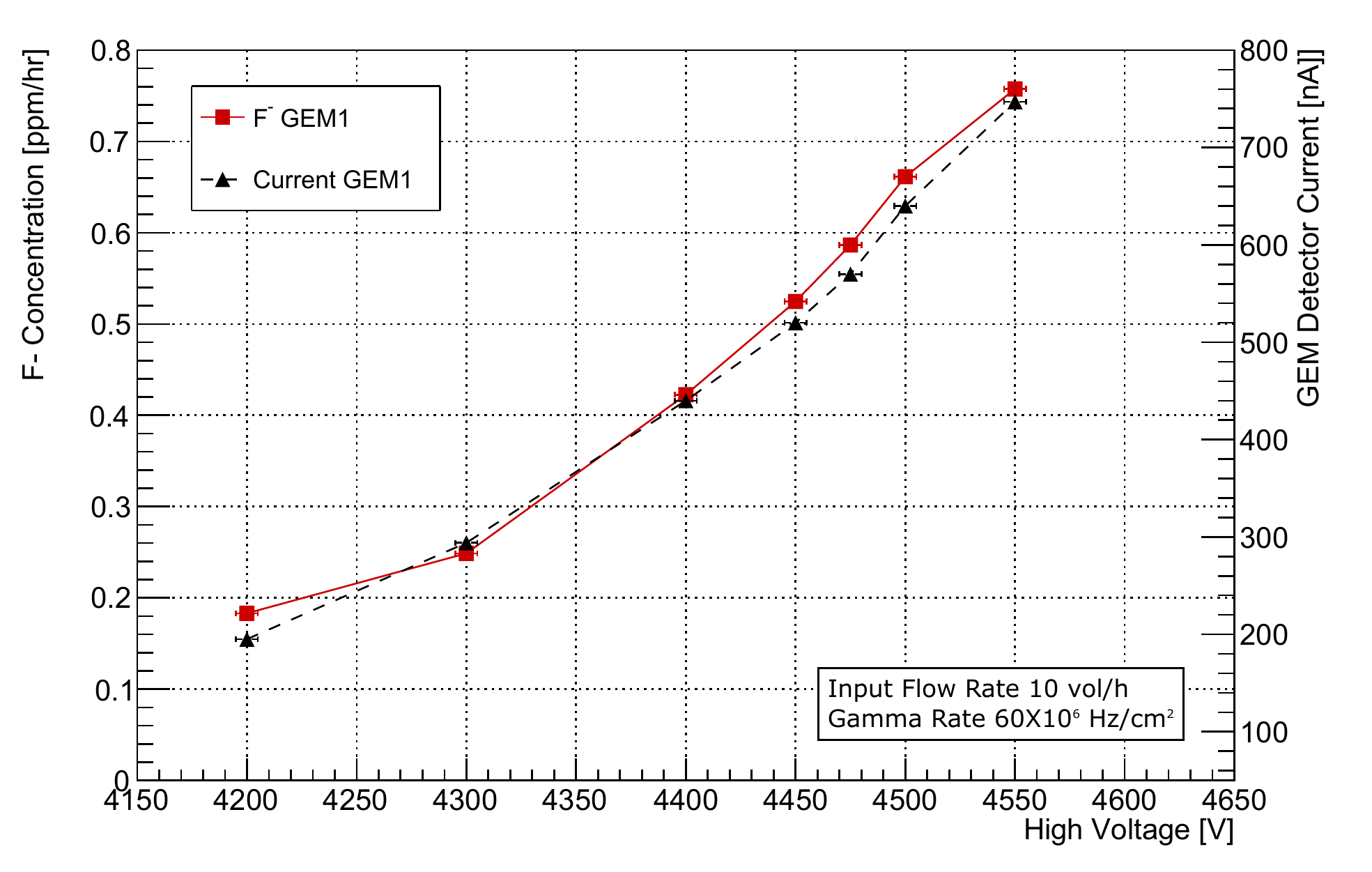}
	\caption{On the left, effect of O$_2$ pollution on the amplification gain of a Triple-GEM detector \cite{gasgem}. On the right, production rate of Fluoride impurities in a Triple-GEM operated with CF$_4$ (40\%) for increasing detector current \cite{fluoride}.}
	\label{gasimp}
\end{figure}
\\\\
Figure \ref{gas3} shows a simplified layout of a gas recirculation system where the increased level of complexity is clearly visible.
\begin{figure}[h!]
	\centering
	\includegraphics[width=8cm]{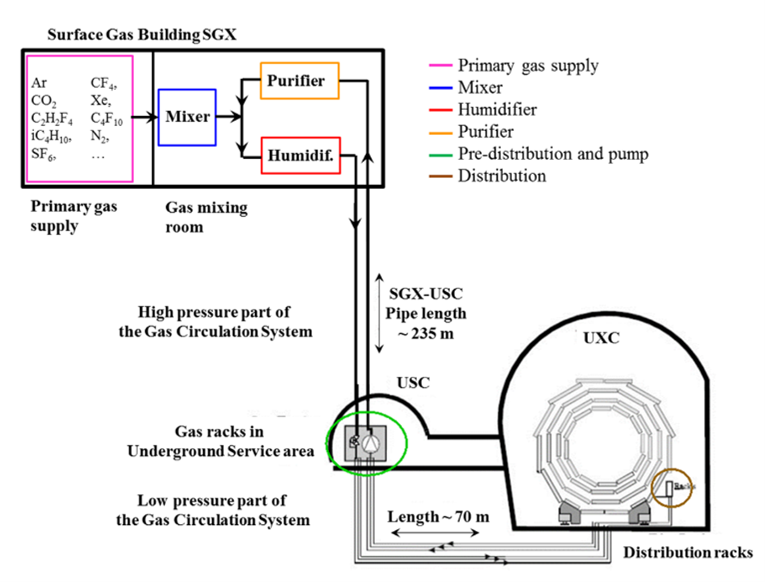}
	\caption{Schematic view of a gas recirculation system where all modules needed for operation are represented.}
	\label{gas3}
\end{figure}

\subsection{Gas recuperation systems}
\begin{figure}[h!]
	\centering
	\includegraphics[width=10cm]{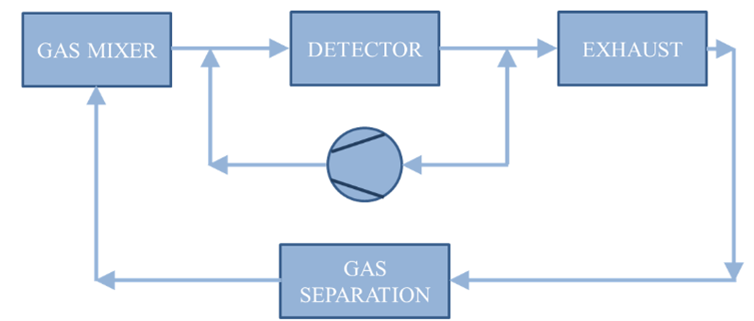}
	\caption{Schematic view of a gas recirculation system equipped with gas separation in the exhaust. Expensive or pollutant gas mixture components are separated from the exhausted gas of the recirculation system and re-used in the gas mixing module.}
	\label{gas4}
\end{figure}

In some cases, depending on the material used during construction, particle detectors can be gas tight but at the same time permeable to impurities that cannot be filtered (i.e. N$_2$). In these circumstances the gas recirculation rate is limited by the possibility to filter the impurities that accumulate in the gas stream. However, in principle the mixture can be recuperated from the exhaust of the gas recirculation system and sent to dedicated separation plant able to extract a specific mixture component (Figure \ref{gas4}). 
\\\\
The advantage related to gas recuperation plants is clearly represented by the possibility to achieve a further reduction in the gas emission (i.e. detector operational cost). The disadvantages are related to the introduction of a second level of complexity in addition to the gas recirculation system. Moreover, dedicated R\&D studies are required for the design of the recuperation plant and similarly for the definition and tuning of the mixture monitoring tools.  
\\\\
The CMS-CSC gas system is a typical example of this approach: both for requirements and complexity aspects related to operation. The CMS CSC muon detector uses a three components gas mixture made of Ar, CO$_2$ and CF$_4$ (40/50/10) The detector is tight (only 60 l/h of gas mixture are lost due to leak and gas analysis compared to a total detector volume of 90 m$^3$ and a circulation flow of 6.6 m$^3$/h). However, it has been discovered that the gas mixture is contaminated by air due to gas diffusion mechanism. Oxygen and water concentration can be kept under control using standard purification modules but unfortunately N$_2$ cannot be easily filtered by a standard purification module. In order to overcome this issue without making compromises on the gas mixture quality, a CF$_4$ recuperation plant has been developed based on warm gas separation. The extra-complexity introduced to monitor the CMS-CSC gas systems after the addition of the CF$_4$ recuperation plant is mainly related to the operation, maintenance, tuning of the plant and to a general reinforcement in the mixture analysis. Indeed, in addition to the standard gas analysis module (common to all the detector of the same experiment) a dedicated infrared analyser has been installed (continuous monitoring of CO$_2$ and CF$_4$), gas chromatographic analyses are performed one-two times per week on the gas mixture from the gas mixer and a new gas monitoring system based on single wire detectors has been implemented.

%%%%%%%%%%%%%%%%%%%%%%%%%%%%%%%%%%%%%%%%%%

%  If you would like to use BibTEX for the bibliography, please feel free to do so.  It is not required.

%  To use BibTeX,

%    1.  uncomment the following two lines, 
%    2.  comment out everything below from  \begin{thebibliography}{99}   to \end{thebibliography).
%    3.  create the file  myreferences.bib, and process this file in the usual way

%\bibliographystyle{JHEP}
%\bibliography{myreferences}  % file myreferences.bib

%%%%%%%%%%%%%%%%%%%%%%%%%%%%%%%%%%%%%%%%%

\newpage

%\bigskip 

%\medskip
\label{chap:intro}

%\medskip

%\Address{}

%\medskip

 %\begin{Abstract}
%\noindent 
%\en%d{Abstract}

%\snowmass

\def\thefootnote{\fnsymbol{footnote}}
\setcounter{footnote}{0}
%
%GDs will remain the primary choice for Muon tracking and triggering at future facilities whenever cost-effective, large-area coverage with low material budget and high detection efficiency is required.
%Moreover, muon systems are often designed to provide a precise momentum measurement, usually in combination with an inner tracker. 
%\chapter{Introduction}

\section*{Acknowledgements}
We are indebted with the CERN MPT workshop (in particular R. de Oliveira and his group) for ideas, discussions and the construction of the detectors; with the RD51 Collaboration for support with the tests at the Gas Detector Development (GDD) Laboratory and for the test-beam at CERN; with the team of the piM1 Beam facility for their support for the test beam at PSI; with the GIF++ facility team for their support during the beam and irradiation test at GIF++.

We would like to thank the organisers of Snowmass 2021 (US Community Study on the Future of Particle Physics) and in particular the conveners of "IF5: Micropattern gasesous detector working group".

\end{document}